\documentclass[prd,nofootinbib,reprint,preprintnumbers,showkeys]{revtex4-2}
\usepackage{xcolor}
\usepackage{amsmath,amsfonts,bm}
\usepackage{graphicx, epstopdf,scrextend}
\usepackage{mathrsfs,mathtools}
\usepackage{enumitem}

\definecolor{mygreen}{rgb}{0,0.6,0}
\definecolor{myorange}{rgb}{0.8,0.5,0}
\definecolor{darkgreen}{rgb}{0,0.4,0} 
\definecolor{darkblue}{rgb}{0,0,0.6} 
\usepackage[colorlinks=true,linkcolor=darkblue,citecolor=darkgreen]{hyperref}

\usepackage{tikz}
\usepackage[compat=1.1.0]{tikz-feynman}
\usepackage{pgfplots,pgfplotstable}

\usepgfplotslibrary{fillbetween}
\usetikzlibrary{intersections,arrows,bending}
\pgfplotsset{compat=newest}

\pgfplotsset{every axis/.style={
    width=12cm,
    height=10cm,
    grid=both,
    scaled ticks=false,
    yticklabel style={/pgf/number format/.cd, fixed,precision=5}
  }
}

\tikzfeynmanset{
  every empty square dot@@/.style={
    /tikz/shape=rectangle,
    /tikz/fill=none
  },
  every empty square dot/.style={/tikzfeynman/every empty square dot@@/.append style={#1}},
  empty square dot/.style={
    /tikzfeynman/every dot@@,
    /tikzfeynman/every empty square dot@@,
  },
  every ellipse blob@@/.style={
    /tikz/shape=ellipse,
    /tikz/graphs/as={},
    /tikz/draw,
    /tikz/fill=none,
    /tikz/outer sep=0.5\pgflinewidth,
    /tikz/inner sep=0pt,
    /tikz/minimum width=0.75cm,
    /tikz/minimum height=1.75cm,
    /tikz/pattern=north west lines,
  },
  every ellipse blob/.style={/tikzfeynman/every ellipse blob@@/.append style={#1}},
  ellipse blob/.style={
    /tikzfeynman/every ellipse blob@@,
  },
  cut/size/.store in=\tikzfeynman@cut@size,                                                                      
  cut/size=3pt,                                                                                                        
  cut/style/.store in=\tikzfeynman@cut@style,                                                                    
  cut/style={},                                                                                                        
  cut/.style args={[#1]#2}{                                                                                          
    /tikz/decoration={                                                                                                       
      markings,                                                                                                              
      mark=at position #2 with {
        \tikzfeynmanset{every cut/.cd,#1}
        \draw [line width=0.5mm, name path=l1, \tikzfeynman@cut@style] (-0.6pt, -\tikzfeynman@cut@size) -- (-0.6pt, \tikzfeynman@cut@size);
        \draw [line width=0.5mm, name path=l2, \tikzfeynman@cut@style] (0.6pt, \tikzfeynman@cut@size) -- (0.6pt, -\tikzfeynman@cut@size);
        \tikzfillbetween[of=l1 and l2]{white};
      }, 
    },                                                                                                                       
    /tikz/postaction={                                                                                                       
      /tikz/decorate=true,                                                                                                   
    },                                                                                                                       
  },
  cut/.default={[]0.5},
  arrow size=1.1pt,
  with arrow/.style={
    /tikz/decoration={
      markings,
      mark=at position #1 with {
        \node[
        transform shape,
        xshift=-0.5mm,
        fill,
        inner sep=\tikzfeynman@arrow@size,
        draw=none,
        isosceles triangle
        ] { };
      },
    },
    /tikz/postaction={
      /tikz/decorate=true,
    },
  },
  with reversed arrow/.style={
    /tikz/decoration={
      markings,
      mark=at position #1 with {
        \node[
        transform shape,
        xshift=-0.5mm,
        rotate=180,
        fill,
        inner sep=\tikzfeynman@arrow@size,
        draw=none,
        isosceles triangle
        ] { };
      },
    },
    /tikz/postaction={
      /tikz/decorate=true,
    },
  },
    with empty arrow/.style={
    /tikz/decoration={
      markings,
      mark=at position #1 with {
        \arrow[xshift=2mm]{Latex[width=2.5mm, length=2.5mm,fill=white]}
      },
    },
    /tikz/postaction={
      /tikz/decorate=true,
    },
  },
  every gluon@@/.style={
    /tikz/draw=none,
    /tikz/decoration={
      name=none,
      coil,
      aspect=-0.5,
      mirror,
      segment length=1mm,
      pre length=1mm,
      post length=1mm
    },
    /tikz/postaction={
      /tikz/draw,
      /tikz/decorate=true,
    },
  },
  every gluon/.style={/tikzfeynman/every gluon@@/.append style={#1}},
  gluon/.style={
    /tikzfeynman/every gluon@@,
  },
  momentum/arrow distance=2mm,
  momentum/arrow shorten=0.7,
  reversed momentum'@@/.style args={[#1]#2}{
    /tikz/preaction={
      /tikz/decoration={
        show path construction,
        moveto code={},
        lineto code={
          \tikzfeynmanset{momentum/.cd,#1}
          \path (\tikzinputsegmentlast) -- (\tikzinputsegmentfirst)
          coordinate [pos=\tikzfeynman@momentum@arrow@shorten] (tf@m@1)
          coordinate [pos=1 - \tikzfeynman@momentum@arrow@shorten] (tf@m@2);
          \draw [-Stealth, \tikzfeynman@momentum@arrow@style]
          ($(tf@m@1)!\tikzfeynman@momentum@arrow@distance!90:(tf@m@2)$)
          -- ($(tf@m@2)!\tikzfeynman@momentum@arrow@distance!-90:(tf@m@1)$)
          node [pos=0.5,
          auto,
          outer sep=\tikzfeynman@momentum@label@distance,
          \tikzfeynman@momentum@label@style] {#2};
        },
        curveto code={
          \tikzfeynmanset{momentum/.cd,#1}
          \path (\tikzinputsegmentlast)
          .. controls (\tikzinputsegmentsupportb) and (\tikzinputsegmentsupporta)
          .. (\tikzinputsegmentfirst)
          { \foreach \i in {1, ..., 50} {
              coordinate [pos=\tikzfeynman@momentum@arrow@shorten 
              + (1-2 * \tikzfeynman@momentum@arrow@shorten)*\i/50] (tf@m@\i) } };
          \draw [-Stealth, \tikzfeynman@momentum@arrow@style]
          ($(tf@m@1)!\tikzfeynman@momentum@arrow@distance!90:(tf@m@2)$)
          foreach \i [count=\j from 3] in {2, ..., 24} {
            -- ($(tf@m@\i)!\tikzfeynman@momentum@arrow@distance!90:(tf@m@\j)$)
          }
          -- ($(tf@m@25)!\tikzfeynman@momentum@arrow@distance!90:(tf@m@26)$)
          node [pos=0.5,
          auto,
          outer sep=\tikzfeynman@momentum@label@distance,
          \tikzfeynman@momentum@arrow@style] {#2}
          foreach \i [count=\j from 27] in {26, ..., 49} {
            -- ($(tf@m@\i)!\tikzfeynman@momentum@arrow@distance!90:(tf@m@\j)$)
          }
          -- ($(tf@m@50)!\tikzfeynman@momentum@arrow@distance!-90:(tf@m@49)$);
        },
        closepath code={
          \tikzfeynmanset{momentum/.cd,#1}
          \path (\tikzinputsegmentlast) -- (\tikzinputsegmentfirst)
          coordinate [pos=\tikzfeynman@momentum@arrow@shorten] (tf@m@1)
          coordinate [pos=1 - \tikzfeynman@momentum@arrow@shorten] (tf@m@2);
          \draw [-Stealth, \tikzfeynman@momentum@arrow@style]
          ($(tf@m@1)!\tikzfeynman@momentum@arrow@distance!90:(tf@m@2)$)
          -- ($(tf@m@2)!\tikzfeynman@momentum@arrow@distance!-90:(tf@m@1)$)
          node [pos=0.5, auto,
          outer sep=\tikzfeynman@momentum@label@distance,
          \tikzfeynman@momentum@label@style] {#2};
        },
      },
      /tikz/decorate=true,
    },
  },
}

\newcommand{\as}{\alpha_{\mathrm{s}}}

\newcommand{\LA}{\mathrm{A}}
\newcommand{\LB}{\mathrm{B}}
\newcommand{\LC}{\mathrm{C}}

\newcommand{\LF}{\mathrm{F}}

\newcommand{\LL}{\mathrm{L}}

\newcommand{\LP}{\mathrm{P}}
\newcommand{\LR}{\mathrm{R}}

\newcommand{\LS}{\mathrm{S}}

\newcommand{\LT}{\mathrm{T}}
\newcommand{\La}{\mathrm{a}}
\newcommand{\Lb}{\mathrm{b}}
\newcommand{\Lc}{\mathrm{c}}

\newcommand{\Lf}{\mathrm{f}}
\newcommand{\Lg}{\mathrm{g}}

\newcommand{\Lq}{\mathrm{q}}

\newcommand{\Ls}{\mathrm{s}}
\newcommand{\LU}{\mathrm{U}}

\newcommand{\MSbar}{\overline{\mathrm{MS}}}

\newcommand{\cA}{\mathcal{A}}

\newcommand{\cD}{\mathcal{D}}

\newcommand{\cL}{\mathcal{L}}
\newcommand{\cN}{\mathcal{N}}
\newcommand{\cO}{\mathcal{O}}

\newcommand{\cR}{\mathcal{R}}
\newcommand{\cS}{\mathcal{S}}

\newbox\charbox
\newbox\slabox
\def\s#1{{      
        \setbox\charbox=\hbox{$#1$}
        \setbox\slabox=\hbox{$/$}
        \dimen\charbox=\ht\slabox
        \advance\dimen\charbox by -\dp\slabox
        \advance\dimen\charbox by -\ht\charbox
        \advance\dimen\charbox by \dp\charbox
        \divide\dimen\charbox by 2
        \raise-\dimen\charbox\hbox to \wd\charbox{\hss/\hss}
        \llap{$#1$}
}}

\definecolor{red}{rgb}{1,0,0}

\def\mi{{\mathrm i}}

\def\ket#1{\big|{#1}\big\rangle}

\def\brax#1{\big\langle{#1}}   
\def\<>#1{\big\langle{#1}\big\rangle}
\def\[]#1{\big[{#1}\big]}

\pgfplotsset{every axis/.style={
    width=8.2cm,
    height=8.2cm,
    grid=both,
    scaled ticks=false,
    yticklabel style={/pgf/number format/.cd, fixed,precision=5}
  }
}

\newif\ifusefigs
\usefigstrue

\begin{document}

\title{Gauge choice for organizing infrared singularities in QCD}

\author{Zolt\'an Nagy}

\affiliation{
 Deutsches Elektronen-Synchrotron DESY, 
 Notkestr.\ 85, 22607 Hamburg, Germany
}

\email{Zoltan.Nagy@desy.de}

\author{Davison E.\ Soper}

\affiliation{
Institute for Fundamental Science,
University of Oregon,
Eugene, OR  97403-5203, USA
}

\email{soper@uoregon.edu}

\begin{abstract}
We explore the features of interpolating gauge for QCD. This gauge, defined by Doust and by Baulieu and Zwanziger, interpolates between Feynman gauge or Lorenz gauge and Coulomb gauge. We argue that it could be useful for defining the splitting functions for a parton shower beyond order $\as$ or for defining the infrared subtraction terms for higher order perturbative calculations.
\end{abstract}

\date{9 November 2023}

\preprint{DESY-23-052}

\maketitle


\makeatletter
\let\toc@pre\relax
\let\toc@post\relax
\makeatother 

\section{Introduction}
\label{sec:introduction}

It is an unsolved problem to specify an algorithm for a parton shower in which the splitting functions are defined at order $\as^2$ or beyond. The splitting functions can be based on the soft and collinear singularities of quantum chromodynamics (QCD) \cite{NSAllOrder}. Thus what one needs is to translate the singularities of Feynman graphs into functions from which the parton splitting functions are constructed. This is not a trivial project beyond leading order in $\as$ because one has both real emissions and virtual exchanges and both soft and collinear singularities and combinations of these. Thus one seeks a method that constructs the needed singular functions directly from Feynman graphs, without dealing with exceptions and special cases.

The construction of subtractions for the calculation of perturbative cross sections at next-to-next-to-leading order (NNLO)  and beyond presents similar problems. Here there are appropriate algorithms \cite{Czakon:2013goa, Cascioli:2014yka, Currie:2013dwa, Czakon:2019tmo, Grazzini:2017mhc, Boughezal:2015ded, DelDuca:2016ily, Chawdhry:2019bji, Anastasiou:2015vya, Duhr:2019kwi, Duhr:2020seh, Cieri:2018oms, Dulat:2018bfe, Dreyer:2018qbw, Dreyer:2016oyx}, but there is substantial ongoing effort to systematize and simplify these algorithms \cite{BabisCalcs, AnastasiouSterman}, {\em cf.} \cite{Nagy:2003qn, Nagy:2006xy}. 

These considerations lead to a certain difficulty. The Feynman diagrams are simplest if one uses a covariant gauge, particularly Feynman gauge. However, in Feynman gauge the treatment of collinear singularities is far from simple. Consider, for instance, a Feynman amplitude in which a quark with momentum $p-q$ couples to a gluon with momentum $q$, becoming a final state quark with momentum $p$ with $p^2 = 0$. Such an amplitude is singular when $q$ becomes collinear with $p$, so that the denominator of the quark propagator, $1/[(p-q)^2+ \mi 0]$ and the denominator of the gluon propagator, $1/[q^2 + \mi 0]$, both vanish. This creates a collinear singularity. In Feynman gauge, the leading term in the gluon propagator in the collinear limit becomes $p_\mu g^{\mu\nu} \propto q_\mu g^{\mu\nu} = q^\nu$. This gluon can then couple to any other line in the Feynman diagram, either an external, on-shell line or a virtual line. Any such connection retains the leading collinear singularity. One can deal with this surfeit of singularities using Ward identities, as we outline in Appendix \ref{sec:FeynmanGauge}. However at higher perturbative orders one can have multiple exchanged gluons with momenta collinear with different external parton momenta. These gluons can couple anywhere in the graph, including to each other. This can lead to the exceptions and special cases that one would like to avoid.

This argument suggests the use of a physical gauge, for instance an axial gauge $n\cdot A(x) = 0$ for some fixed vector $n$. In such a gauge, gluons never carry longitudinal polarizations $\varepsilon ^\nu(q) \propto q^\nu$, so the problems associated with longitudinally polarized gluons disappear. However, one then must deal with gauge-definition singularities $1/q\cdot n$, which need to be regulated somehow.

In this paper, we explore the use of a gauge defined by Doust \cite{Doust} and Baulieu and Zwanziger \cite{BaulieuZwanziger}. (We follow the construction of Ref.~\cite{BaulieuZwanziger}, although we choose what we think is a more transparent notation.) This gauge interpolates between a covariant gauge and Coulomb gauge. Accordingly, following Refs.~\cite{Doust, BaulieuZwanziger}, we will call it {\em interpolating gauge}. The gauge choice depends on a parameter $\xi$, where $\xi = 1$ corresponds interpolating from Feynman gauge and $\xi = 0$ corresponds to interpolating from Lorenz gauge. We mostly choose $\xi = 1$, corresponding to starting from Feynman gauge. The gauge definition also depends on a four vector $n$ that defines the time axis of a preferred reference frame. Finally, it depends on a parameter $v$, where $v = 1$ gives the starting covariant gauge and $v\to \infty$ gives Coulomb gauge. 

The definition of interpolating gauge is very simple. The gauge fixing term in the Lagrangian for a standard covariant gauge with gauge parameter $\xi$ is
\begin{equation}
\label{eq:LGF1}
\cL_\mathrm{GF}(x) = -\frac{1}{2\xi}\, (\partial_\mu A^\mu_a(x))
(\partial_\nu A^\nu_a(x))
\;.
\end{equation}
Using a reference frame in which $n = (1,0,0,0)$, the gauge fixing term for interpolating gauge is
\begin{equation}
\begin{split}
\label{eq:LGF2}
\cL_\mathrm{GF}(x) ={}& -\frac{v^2}{2\xi}\, 
\left(\frac{1}{v^2}\,\partial_0 A^0_a(x) - \sum_{i=1}^3\partial_i A^i_a(x)\right)
\\&\times
\left(\frac{1}{v^2}\,\partial_0 A^0_a(x) - \sum_{j=1}^3\partial_j A^j_a(x)\right)
\;.
\end{split}
\end{equation}
If we choose $v = 1$, we have the standard covariant gauge. With $v > 1$, we have the same general form of $\cL_\mathrm{GF}(x)$ except that the relative normalizations of the $\partial_0 A^0_a$ and $\partial_i A^i_a$ terms is modified.

The intent of Refs.~\cite{Doust, BaulieuZwanziger} was to better define Coulomb gauge. One might think that it would be ideal to use Coulomb gauge to help manage infrared singularities. It is, after all, a physical gauge in the sense that only transversely polarized gluons propagate. However, using Coulomb gauge requires taking limits $v \to \infty$. As we will see, interpolating gauge with any finite value of $v$ with $v > 1$ is physical enough for the purposes that we have in mind. We could, for instance, choose $v = 2$. Because of its useful features, we might call interpolating gauge a quasi-physical gauge. 

The definition of interpolating gauge depends on a parameter $\xi$ and a parameter $v$ with $v \ge 1$. Additionally, interpolating gauge, like Coulomb gauge, depends on a four vector $n$ with $n^2 = 1$ that defines the time axis in a preferred reference frame. We define an analogue $h^{\mu\nu}$ of the metric tensor $g^{\mu\nu}$. In a reference frame in which the components of $n^\mu$ are
\begin{equation}
n^\mu = (1,0,0,0)
\;,
\end{equation}
the components of $h^{\mu\nu}$ are
\begin{equation}
\label{eq:hmunumatrix}
h^{\mu\nu} = 
\begin{pmatrix}
1/v^2 & 0 & 0 & 0 \\
0 & -1 & 0 & 0 \\
0 & 0 & -1 & 0 \\
0 & 0 & 0 & -1 \\
\end{pmatrix}
\;.
\end{equation}
For any vector $q$ we define an associated vector $\tilde q$ by
\begin{equation}
\tilde q^\mu = h^\mu_\nu\, q^\nu
\;.
\end{equation}
Also
\begin{equation}
\tilde \partial_\mu = h_\mu^\nu\, \partial_\nu
\;.
\end{equation}
We think of $h^{\mu\nu}$ as being a modified metric tensor because in the gauge fixing Lagrangian we replace $\partial_\mu A_a^\mu  = \partial_\mu g^\mu_\nu A_a^\nu$ by $\tilde \partial_\mu A_a^\mu  = \partial_\mu h^\mu_\nu A_a^\nu$.

One might also consider gauge choices that interpolate between a covariant gauge and other physical gauges, such as the axial gauge defined by $n\cdot A = 0$. In Ref.~\cite{BaulieuZwanziger}, one can consider gauges with choices of $h^{\mu\nu}$ that are different from Eq.~(\ref{eq:hmunumatrix}). In this paper, we analyze the gauge defined by $\cL_\mathrm{GF}(x)$ in Eq.~(\ref{eq:LGF2}). We have two reasons for this preference. First, it uses a timelike vector $n$, which can be chosen as the direction of the total momentum for electron-positron annihilation and, for hadron-hadron collisions, as the direction of the total momentum of either the incoming hadrons or of the colliding partons at the Born level of the process considered. For hadron-hadron collisions, one could use a lightlike vector $n$ in the direction of one of the incoming partons, but this choice is not as useful for the description of the other incoming parton. Once one has chosen to use a timelike vector $n$, one still has the choice of a gauge fixing term in the Lagrangian. We believe that the choice in Eq.~(\ref{eq:LGF2}) is favored by its simplicity.

In the sections that follow, we define and analyze interpolating gauge in some detail. In the remainder of this Introduction, we very briefly review what the gluon propagator in interpolating gauge is and what advantages it might offer for calculations.

As explained in Sec.~\ref{sec:interpolatinggauge}, the gluon propagator in interpolating gauge is
\begin{equation}
\begin{split}
\label{eq:gluonpropagator2}
\mi D^{\mu\nu}(q) ={}&
\frac{\mi}{q^2+\mi 0}
  \Bigg[
    - g^{\mu\nu} 
    + \frac{q^\mu\,\tilde{q}^\nu + \tilde{q}^\mu\,q^\nu}
    {q\cdot\tilde{q} + \mi 0}
    \\&
    - \left(1+\frac{1}{v^2}\right)\frac{q^\mu\,q^\nu}
    {q\cdot\tilde{q} + \mi 0}\, 
 \\&
    - \frac{\xi-1}{v^2}\, 
    \frac{q^2\ q^\mu\,q^\nu}{(q\cdot\tilde{q}+ \mi 0)^2}
  \Bigg]
  \;.
\end{split}
\end{equation}
The ghost propagator is
\begin{equation}
\label{eq:ghostpropagator}
\mi D_\mathrm{ghost}(q) 
= \frac{\mi}{q \cdot \tilde q +\mi 0}
\;.
\end{equation}
There is a new denominator $q \cdot \tilde q$ here but it is not ambiguous how to define the singularity: it is $1/(q\cdot\tilde q + \mi 0)$.

We can understand the gluon propagator better by decomposing it into two parts:
\begin{equation}
D^{\mu\nu}(q) = D^{\mu\nu}_\LT(q) + D^{\mu\nu}_\LL(q)
\;.
\end{equation}

In a reference frame in which $n = (1,0,0,0)$, the components of $D^{\mu\nu}_\LT(q)$ are
\begin{equation}
\begin{split}
\label{eq:DmunuT}
D^{00}_{\LT}(q) ={}& 0
\;,
\\
D^{0i}_{\LT}(q) ={}& D^{i0}_{\LT}(q) = 0
\;,
\\
D^{ij}_{\LT}(q) ={}& 
    \frac{1}{q^2+\mi 0}
    \left\{\delta^{ij} - \frac{q^i q^j}{\vec q^{\,2}}\right\}
  \;.
\end{split}
\end{equation}
That is,
\begin{equation}
\label{eq:DTexpansion}
D^{\mu\nu}_{\LT}(q) =  \frac{1}{q^2+\mi 0}
\sum_{s=1,2} \varepsilon^\mu(q,s)\,\varepsilon^\nu(q,s)
\;,
\end{equation}
where the polarization vectors are the two real valued solutions\footnote{One often choses complex valued polarization vectors so that one can represent gluons with definite helicities. However, one then needs to distinguish between $\varepsilon$ and $\varepsilon^*$. Since this complicates the notation, we use real valued polarization vectors in this paper.} of
\begin{equation}
\begin{split}
\varepsilon(q,s)\cdot n ={}& 0
\;,
\\
\varepsilon(q,s)\cdot q ={}& 0
\;,
\end{split}
\end{equation}
normalized to 
\begin{equation}
\varepsilon(q,s) \cdot \varepsilon(q,s') = -\delta_{s,s'}
\;.
\end{equation}
Thus the T gluons are massless bosons with transverse polarizations.

The difficulty with collinear singularities that occurs in Feynman gauge is not present for T gluons because when $q = \lambda p$ we have $p_\mu D^{\mu\nu}(q) \propto q_\mu D^{\mu\nu}(q)$. Then $q_\mu D^{\mu\nu}(q) \propto - q_\mu g^{\mu\nu} = - q^\nu$ in Feynman gauge is replaced by $q_\mu \sum_s\varepsilon^\mu(q,s) \varepsilon^\nu(q,s)$, but $q_\mu \varepsilon^\mu(q,s) = 0$.

In a reference frame in which $n = (1,0,0,0)$, the components of the propagator for L gluons with the gauge parameter $\xi$ set to $\xi = 1$ are
\begin{equation}
\begin{split}
\label{eq:DmunuLstart}
D^{00}_{\LL}(q) ={}& 
- \frac{1}{q\cdot \tilde q +\mi 0}
\;,
\\
D^{0i}_{\LL}(q) ={}& D^{i0}_{\LL}(q) = 0
\;,
\\
D^{ij}_{\LL}(k) ={}& 
 \frac{1}{v^2}\,\frac{1}{q\cdot \tilde q +\mi 0}
  \frac{q^i q^j}{\vec q^{\,2}}
\;.
\end{split}
\end{equation}
This describes bosons with polarization vectors proportional to either $(1,\vec 0)$ or $(0,\vec q)$.

What is remarkable is that the L gluons are on shell not when $q^2 = 0$ but when $q\cdot\tilde q = 0$.\footnote{This tree level on-shell condition is modified at higher orders in $\as$.} Since  $q\cdot \tilde q = (q^0)^2/v^2 - \vec q^{\,2}$, the condition for on-shell propagation is
\begin{equation}
q^0 = \pm v  |\vec q\,|
\;.
\end{equation}
For a boson propagating in the $z$-direction with positive energy $\omega = v  |\vec q\,|$, the wave function in space-time is proportional to
\begin{equation}
e^{-\mi(\omega t - |\vec q\,| z)}
= e^{-\mi|\vec q\,|(v t -  z)}
\;.
\end{equation}
That is, the wave propagates with velocity $v$. Since we take $v > 1$, the L gluons are tachyons.

When we construct a cross section using interpolating gauge, the initial or final state partons should include quarks and T gluons, but not ghosts or L gluons. That is, we have an S matrix amplitude $\cS_i$ and a conjugate amplitude $\cS^\dagger_j$ with quarks and T gluons as external, on-shell particles. The Feynman rule factor for an external T gluon is 
\footnote{If there is a self-energy graph connected to the external line, one needs a limiting procedure with $q^2 \to 0$ according to the Lehmann-Symanzik-Zimmerman (LSZ) prescription. See Sec.~\ref{sec:dLSZdri}.}
\begin{equation}
(2\pi) \delta_+(q^2) \sum_s \varepsilon^\mu(q,s) \varepsilon^\nu(q,s)
\;,
\end{equation}
where $\delta_+(q^2)$ is $\delta(q^2)$ times a factor $\theta(q\cdot n > 0)$. 

The Feynman diagrams used to construct the S matrix for incoming and outgoing T gluons involve also virtual L gluons. One might be concerned that this S matrix in interpolating gauge differs from the S matrix in Feynman gauge or Lorenz gauge. However, as we will see in Sec.~\ref{sec:dSdri}, the theory obeys identities derived from Becchi-Rouet-Stora-Tyutin (BRST) symmetry that imply that the S matrix is independent of $v$ and $\xi$ and also independent of $n$. Thus the S matrix in interpolating gauge is the same as the S matrix in one of the covariant gauges.

An instructive way to write $D^{\mu\nu}_{\LL}(q)$ in Eq.~(\ref{eq:DmunuLstart}) is
\begin{equation}
\begin{split}
D^{\mu\nu}_{\LL}(q)
={}&
\frac{q^\mu q^\nu - q\cdot n\, (q^\mu n^\nu + n^\mu q^\nu)}
{[(q^0)^2  - v^2 \vec q^{\,2} + \mi 0]\ |\vec q\,|^2}
+\frac{n^\mu n^\nu}{|\vec q\,|^2}
\;.
\end{split}
\end{equation}
If we take $v\to \infty$ with fixed $q$, the first term vanishes. This leaves just $n^\mu n^\nu/|\vec q\,|^2$, which is the Coulomb potential. We conclude that interpolating gauge with $\xi = 1$ interpolates between Feynman gauge and Coulomb gauge.

What happens if one inserts a virtual L-gluon line with momentum $q$ into an amplitude, coupled to an external quark or an external T gluon with momentum $p$ with $p^2 = 0$? Then there are propagators with momenta $p-q$ and $q$. In Feynman gauge, this leads to a collinear singularity when $q = \lambda p$ since $(p-q)^2 = (1-\lambda)\, p^2$ and $q^2 = \lambda^2 p^2$ both vanish in the collinear limit. In interpolating gauge, $(p-q)^2 = (1-\lambda)\, p^2$ still vanishes, but the denominator for the L-gluon propagator is $q\cdot \tilde q + \mi 0$. In the collinear limit, this becomes 
\begin{equation}
\begin{split}
\lambda^2 p \cdot \tilde p ={}& 
\frac{\lambda^2}{v^2}\, p^2 
-\lambda^2 \left(1 - \frac{1}{v^2}\right) |\vec p\,|^2
\\
={}& -\lambda^2 \left(1 - \frac{1}{v^2}\right) |\vec p\,|^2
\;,
\end{split}
\end{equation}
which does not vanish. For this reason, exchanging an L gluon between two external partons can create a soft ($\vec q \to 0$) singularity but does not create a collinear singularity. We will see this in a more detailed calculation in Sec.~\ref{sec:exchange}.

In this Introduction, we have outlined very briefly why interpolating gauge might be useful, in spite of the complexity of the gluon propagator in this gauge. In Sec.~\ref{sec:interpolatinggauge}, we derive the propagators and vertices in interpolating gauge from the functional integral formulation of the theory. In Sec.~\ref{sec:TandLgluons}, we examine the decomposition of the gluon propagator into T and L parts in a little more detail than was presented above. In Sec.~\ref{sec:exchange}, we show how no collinear singularities arise from the exchange of a gluon between two external partons in interpolating gauge. In Sec.~\ref{sec:brst}, we examine BRST symmetry in this gauge.  Sec.~\ref{sec:renorm} explores the renormalization program. We examine the gluon self-energy function in Sec.~\ref{sec:gluonselfenergy} and the quark self-energy function in Sec.~\ref{sec:quarkselfenergy}. We assemble results about the infrared poles of the S matrix in Sec.~\ref{sec:Spoles}. Finally, Sec.~\ref{sec:conclusions} presents some conclusions. There are four appendices.

\section{Definition of interpolating gauge}
\label{sec:interpolatinggauge}

In this section, we use the functional integral approach to define SU(3) gauge theory in interpolating gauge, leading to the Feynman rules that one can use for calculations. The important step is the introduction of the gauge fixing function in Eq.~(\ref{eq:gaugefixing}) below. The rest of the analysis follows rather standard textbook methods, but we provide this analysis in order to present a self-contained derivation in a consistent notation.\footnote{We follow the convention of Schwartz \cite{SchwartzBook} for the sign of $g$. Much of the analysis is along the lines of that in Sterman \cite{StermanBook}.}

\subsection{Momenta and the tensor $h$}
\label{sec:hdef}

As sketched in the Introduction, we let $n$ be a timelike vector with $n^2 = 1$ that defines the time direction in a preferred reference frame that we often use. We use $n$ to define tensors $P^{\mu\nu}_\pm$ that project onto the direction along $n$ and the directions orthogonal to $n$:
\begin{equation}
\begin{split}
\label{eq:Pplusminus}
P^{\mu\nu}_+ ={}& n^\mu n^\nu
\;,
\\
P^{\mu\nu}_- ={}& g^{\mu\nu} - n^\mu n^\nu
\;.
\end{split}
\end{equation}
Then $g^{\mu\nu} = P^{\mu\nu}_+ +  P^{\mu\nu}_-$. The tensors $P_\pm$ act as projection operators:
\begin{equation}
\begin{split}
P^\mu_{\pm\,\alpha} P^{\alpha\nu}_\mp ={}& 0
\;,
\\
P^\mu_{\pm\,\alpha} P^{\alpha\nu}_\pm ={}& P^{\mu\nu}_\pm
\;.
\end{split}
\end{equation}
We let $v$ be a fixed parameter with $v \ge 1$ and define an analogue $h^{\mu\nu}$ of the metric tensor $g^{\mu\nu}$ by
\begin{equation}
\label{eq:hdef}
h^{\mu\nu} = \frac{1}{v^2}\,P^{\mu\nu}_+ + P^{\mu\nu}_-
\;.
\end{equation}

Using the definition of $h^{\mu\nu}$ and the properties of the projection tensors $P_\pm$, one derives the useful identity
\begin{equation}
\label{eq:hsquared}
h^\mu_\alpha h^\alpha_\nu = - \frac{1}{v^2}\,g^\mu_\nu 
+ \frac{v^2 + 1}{v^2}\,h^\mu_\nu 
\;.
\end{equation}

For any momentum $q$ we define a transformed momentum
\begin{equation}
\label{eq:tildedef}
\tilde q^\mu = h^\mu_\nu q^\nu
\;.
\end{equation}
This gives us
\begin{equation}
\begin{split}
\label{eq:qdottildeq}
\tilde q \cdot q ={}& q^2 - \left(1-\frac{1}{v^2}\right)
(q\cdot n)^2
\;,
\end{split}
\end{equation}
so that $q\cdot\tilde q < q^2$ when $v > 1$. Using Eq.~(\ref{eq:hsquared}), we also obtain
\begin{equation}
\label{eq:tildeqsq}
\tilde q^2 = -\frac{1}{v^2}\,q^2 +  \frac{v^2 + 1}{v^2} \,
q\cdot \tilde q
\;.
\end{equation}
%

\subsection{Functional integral definition of the gauge}
\label{sec:functionalintegral}

We use a functional integral over quark fields with flavor $f$, $\psi_f(x)$, and gauge boson fields, $A^\mu_a(x)$. We use the covariant derivative acting on quark fields
\begin{equation}
D^\mu(A) = \partial^\mu  - \mi g A^\mu_a(x) t_a
\;.
\end{equation}
For the covariant derivative acting on octet color fields, we make the color indices explicit:
\begin{equation}
D^\mu_{ac}(A) = \partial^\mu\delta_{ac}  + g A^\mu_b(x) f_{abc}
\;.
\end{equation}
We also use the field operator
\begin{equation}
F_a^{\mu\nu} = \partial^\mu A_a^\nu
- \partial^\nu A_a^\mu
+ g f_{abc}A^\mu_b(x) A^\nu_c(x)
\;.
\end{equation}
The gauge invariant Lagrangian is
\begin{equation}
\begin{split}
\cL(x) ={}& - \frac{1}{4}\,F_a^{\mu\nu}(x)F_{a,\mu\nu}(x)
\\ & + \sum_f \bar\psi_f(x) 
\left[\mi D^\mu(A)\,\gamma_\mu - m_f\right] \psi_f(x)
\;.
\end{split}
\end{equation}

We begin with the functional integral
\begin{equation}
\begin{split}
I_F ={}& \cN_0  
\int\!\cD \psi\ \cD \bar\psi\ \cD A\ 
\exp\!\left(\mi\int\!d^4 x\ \cL(x)\right)
\\&\times
F[A,\bar \psi, \psi]
\;.
\end{split}
\end{equation}
The function $F[A,\bar \psi, \psi]$ consists of gauge invariant combinations of the quark and gluon field operators. The normalization factor $\cN_0$ is not important for the construction because one considers $I_F$ divided by $I_1$ with $F[A,\bar \psi, \psi] = 1$. The functional integral over $A$ is not well defined at this stage because it implicitly includes an integral over the gauge group. In order to factor out an integral over the gauge group, we insert a functional integral
\begin{equation}
\label{eq:omegaintegral}
1 = \cN_1 \int \cD \omega\, 
\exp\!\left(\mi
\sum_c \int\!d^4x\ \frac{-1}{2a}\, \omega_c(x)^2
\right)
\;.
\end{equation}
Here $\omega_c(x)$ is a scalar field with a color index $c$ and $a$ is a constant parameter. We define another gauge parameter $\xi$ that we can use in place of $a$ by 
\begin{equation}
\xi = a v^2
\;.
\end{equation}
We also insert
\begin{equation}
1 = \int\! \cD \alpha\  \delta(G[A_\alpha])\,
\det\!\left[\frac{\delta G[A_\alpha]}{\delta\alpha}\right]
\;,
\end{equation}
where $G$ is the gauge fixing function defined below in Eq.~(\ref{eq:gaugefixing}). Here $\alpha_c(x)$ parameterizes a finite gauge transformation, under which a quark field transforms according to  
\begin{equation}
\psi_f(x) \to \psi_{f,\alpha}(x) = \exp(\mi g \alpha_c(x) t_c)\,\psi_f(x)
\;.
\end{equation}
Here the matrices $t_c$ are the generators of the fundamental representation of SU(3). In the adjoint representation of SU(3), the transformation matrix is
\begin{equation}
U(x) = \exp(\mi g \alpha_c(x) T_c)
\;.
\end{equation}
Then $A_\alpha$ is $A$ transformed by the gauge transformation,
\begin{equation}
\begin{split}
(A_\alpha)^\mu_a(x)T_a ={}& U(x) A^\mu_a(x)T_a U(x)^{-1}
\\&
-\frac{\mi}{g}\,[\partial^\mu U(x)]U(x)^{-1}
\;.
\end{split}
\end{equation}

The gauge fixing function $G[A_\alpha]$ is a function of a color index $c$ and a space-time position $x$ defined by
\begin{equation}
\label{eq:gaugefixing}
G[A]_c(x) = \tilde \partial_\mu A_c^\mu(x) - \omega_c(x)
\;.
\end{equation}
The delta function $\delta(G[A_\alpha])$ sets $G[A]_c(x)$ equal to zero for each $c$ and each $x$. This gauge fixing function replaces the function $\partial_\mu A_c^\mu(x) - \omega_c(x)$ that leads to a covariant gauge. This gives us
\begin{equation}
\begin{split}
I_\LF ={}& \cN_2  
\int\! \cD \alpha\, \cD \psi\, \cD \bar\psi\, \cD A\, \cD \omega\
\exp\!\left(\mi\int\!d^4 x\ \cL(x)\right)
\\&\times
\exp\!\left(\mi
\sum_c \int\!d^4x\ \frac{-1}{2a}\, \omega_c(x)^2
\right)
\\&\times
\delta(G[A_\alpha])\,
\det\!\left[\frac{\delta G[A_\alpha]}{\delta\alpha}\right] 
F[A,\bar \psi, \psi]
\;.
\end{split}
\end{equation}

We need the determinant $\delta G[A_\alpha]/\delta\alpha$ of the functional derivative of $G[A_\alpha]$ with respect to the gauge transformation $\alpha$. For this purpose, we consider a small variation $\delta \alpha$ in the gauge transformation. The corresponding variation in $(A_{\alpha})^\mu_a(x)$ is
\begin{equation}
\begin{split}
\delta (A_{\alpha})^\mu_a(x) ={}&   D(A_\alpha)^\mu_{ac}
\delta\alpha_c(x)
\;.
\end{split}
\end{equation}
Thus the variation in $G[A_\alpha]_a(x)$ is
\begin{equation}
\begin{split}
\delta G[A_{\alpha}]_a(x) ={}&  \tilde \partial_\mu D(A_\alpha)^\mu_{ac}
\delta\alpha_c(x)
\;.
\end{split}
\end{equation}
This gives us the functional derivative
\begin{equation}
\frac{\delta G[A_\alpha]_a(x)}{\delta \alpha_c(y)}
= \tilde \partial_\mu D(A_\alpha)^\mu_{ac} \delta(x-y)
\;.
\end{equation}

Before going further, we change integration variable from $A$ to $A_\alpha$. The Lagrangian and $F[A,\bar \psi, \psi]$ do not change when expressed as functions of $A_\alpha$ since these functions are gauge invariant. Next, we simply rename $A_\alpha$ as $A$. This gives us
\begin{equation}
\begin{split}
I_F ={}& \cN_2  
\int\! \cD \alpha\,
\cD \psi\, \cD \bar\psi\,
\cD A\,  \cD \omega\,
\exp\!\left(\mi\int\!d^4 x\ \cL(x)\right)
\\&\times
\exp\!\left(\mi
\sum_c \int\!d^4x\ \frac{-1}{2a} \omega_c(x)^2
\right)
\\&\times
\delta(G[A])\,
\det\!\left[\frac{\delta G[A]}{\delta\alpha}\right] 
F[A,\bar \psi, \psi]
\;.
\end{split}
\end{equation}
Now nothing depends on the gauge transformation $\alpha$, so we can simply absorb the integration $\int \cD\alpha$ into the normalization constant. Additionally, we can perform the integration over $\omega$ against $\delta(G[A])$, thus setting $\omega_c(x)$ to $\tilde \partial_\mu A^\mu_c(x)$. This gives us 
\begin{equation}
\begin{split}
I_F ={}& \cN_3  
\int\!\cD \psi\, \cD \bar\psi\,\cD A\,
\exp\!\left(\mi\int\!d^4 x\ \cL(x)\right)
\\&\times
\exp\!\left(\mi
\sum_c \int\!d^4x\ \frac{-1}{2a} (\tilde \partial_\mu A^\mu_c(x))^2
\right)
\\&\times
\det\!\left[\frac{\delta G[A]}{\delta\alpha}\right]
F[A,\bar \psi, \psi]
\;.
\end{split}
\end{equation}

The final step is to write the functional determinant as an integral over Grassmann fields $\eta_a(x)$ and $\bar\eta_a(x)$, the ghost and antighost fields:
\begin{equation}
\begin{split}
\det\!&\left[\frac{\delta G[A]}{\delta\alpha}\right]
\\
={}& \cN_4 
\int\!\cD \eta\, \cD \bar \eta\,
\exp\!\left(
-\mi \int\!d^4 x\ \bar \eta_a(x)\,
\tilde \partial_\mu D(A)^\mu_{ac}\,
\eta_c(x)
\right)
.
\end{split}
\end{equation}
This gives us
\begin{equation}
\begin{split}
I_F ={}& \cN   
\int\!\cD \psi\, \cD \bar\psi\,\cD A\,\cD \eta\, \cD \bar \eta\,
\exp\!\left(\mi\int\!d^4 x\ \cL(x)\right)
\\&\times
\exp\!\left(\mi
\sum_c \int\!d^4x\ \frac{-1}{2a} (\tilde \partial_\mu A^\mu_c(x))^2
\right)
\\&\times
\exp\!\left(
-\mi \int\!d^4 x\ \bar \eta_a(x)\,
\tilde \partial_\mu D(A)^\mu_{ac}\,
\eta_c(x)
\right)
\\&\times
F[A,\bar \psi, \psi]
\;.
\end{split}
\end{equation}
We now have a functional integral with the usual gauge invariant Lagrangian $\cL(x)$, a gauge fixing Lagrangian
\begin{equation}
\label{eq:LGF}
\cL_\mathrm{GF}(x) = -\frac{1}{2a}\, (\tilde \partial_\mu A^\mu_a(x))
(\tilde \partial_\nu A^\nu_a(x))
\end{equation}
and ghost fields with a ghost Lagrangian
\begin{equation}
\label{eq:Lghost}
\cL_\mathrm{ghost}(x) =
-\bar \eta_a(x)\,
\tilde \partial_\mu D(A)^\mu_{ac}\,
\eta_c(x)
\;.
\end{equation}
This is just the same as with a covariant gauge except that $\tilde \partial_\mu$ replaces $\partial_\mu$.

\subsection{Propagators}
\label{sec:propagators}

The terms in the Lagrangian that are quadratic in the gauge field are
\begin{equation}
\begin{split}
\cL_{A^2}(x) ={}& \frac{1}{2}
\Big\{
(\partial_\mu A_a^\nu)(\partial_\nu A_a^\mu)
- (\partial_\mu A_a^\nu)(\partial^\mu A_{a\,\nu})
\\&
-\frac{1}{a}(\tilde\partial_\mu A_a^\mu)(\tilde\partial_\nu A_a^\nu)
\Big\}
\;.
\end{split}
\end{equation}
In momentum space, this becomes a two-point vertex $-\mi \delta_{ab} \Gamma_\mathrm{tree}^{\mu\nu}$ for a gluon with momentum $q$,
\begin{equation}
\Gamma_\mathrm{tree}^{\mu\nu} = 
q^\mu q^\nu - q^2 g^{\mu\nu}
- \frac{1}{a}\,\tilde q^\mu \tilde q^\nu
\;.
\end{equation}
The tree-level gluon propagator is $\mi \delta_{cd} D^{\mu\nu}(q)$ where $D^\mu_\alpha \Gamma_{\mathrm{tree}}^{\alpha \nu} = g^{\mu\nu}$:
\begin{equation}
\begin{split}
\label{eq:gluonpropagator}
D^{\mu\nu}(q) ={}&
\frac{1}{q^2+\mi 0}
  \Big[
    - g^{\mu\nu} 
    + \frac{q^\mu\,\tilde{q}^\nu + \tilde{q}^\mu\,q^\nu}
    {q\cdot\tilde{q}+ \mi 0}
\\&\quad
    - \frac{\tilde{q}^2+a\, q^2}{(q\cdot\tilde{q}+ \mi 0)^2}\, 
    q^\mu\,q^\nu
  \Big]
  \;.
\end{split}
\end{equation}
We can use Eq.~(\ref{eq:tildeqsq}) for $\tilde q^2$ and use the gauge parameter $\xi = a v^2$ instead of $a$. Then the gluon propagator is expressed as in Eq.~(\ref{eq:gluonpropagator2}):
\begin{equation}
\begin{split}
\label{eq:gluonpropagator2bis}
D^{\mu\nu}(q) ={}&
\frac{1}{q^2+\mi 0}
  \Bigg[
    - g^{\mu\nu} 
    + \frac{q^\mu\,\tilde{q}^\nu + \tilde{q}^\mu\,q^\nu}
    {q\cdot\tilde{q} + \mi 0}
    \\&
    - \left(1+\frac{1}{v^2}\right)\frac{q^\mu\,q^\nu}
    {q\cdot\tilde{q} + \mi 0}  \Bigg] 
 \\&
    - \frac{\xi-1}{v^2}\, 
    \frac{q^\mu\,q^\nu}{(q\cdot\tilde{q}+ \mi 0)^2}
  \;.
\end{split}
\end{equation}
Evidently, this is simplest if we choose $\xi = 1$. That will be our favored choice. We will return to properties of $D^{\mu\nu}$ in Sec.~\ref{sec:TandLgluons}.

In the case that $v = 1$, we have $\tilde q = q$ and
\begin{equation}
\begin{split}
\label{eq:gluonpropagatorv1}
D^{\mu\nu}(q) ={}&
\frac{1}{q^2+\mi 0}
  \Bigg[
    - g^{\mu\nu} 
    + (1-\xi)\,\frac{q^\mu\, q^\nu}
    {q^2 + \mi 0}
    \Bigg]
  \;.
\end{split}
\end{equation}
This is the usual covariant gauge with gauge parameter $\xi$, with $\xi = 1$ giving Feynman gauge. In the case that $v\to \infty$ with any fixed $q$ and with any fixed $\xi$, we have $\tilde q^\mu \to P_-^{\mu\alpha} q_\alpha$. Then the limit does not depend on $\xi$ and is
\begin{equation}
\begin{split}
D^{\mu\nu}(q) \to {}&
-\frac{1}{q^2 + \mi 0}\left[P_-^{\mu\nu}
+ \frac{P_-^{\mu\alpha}q_\alpha\,P_-^{\nu\beta}q_\beta}
{q_\alpha P_-^{\alpha\beta} q_\beta}
\right]
\\&
- \frac{n^\mu n^\nu}{q_\alpha P_-^{\alpha\beta} q_\beta}
\;.
\end{split}
\end{equation}
This is the propagator in Coulomb gauge, expressed in covariant form. The second term is the Coulomb potential,
\begin{equation}
D^{00} = \frac{1}{|\vec q\,|^2}
\end{equation}
in a frame in which $n^\mu = (1,0,0,0)$.

For the ghost propagator, we need the inverse of the part of the ghost Lagrangian (\ref{eq:Lghost}) that does not include $A_b^\mu(x)$. The ghost propagator is $\delta_{ab}\,\mi D^\mathrm{ghost}(q)$ with
\begin{equation}
\label{eq:ghostpropagatorbis}
D^\mathrm{ghost}(q) 
= \frac{1}{q \cdot \tilde q +\mi 0}
\;.
\end{equation}

The quark propagator $\mi D^\mathrm{quark}$ takes the familiar form derived from the $\bar \psi \ \psi$ part of the Lagrangian,
\begin{equation}
D^\mathrm{quark}(q) = \frac{\s{q} + m_f}{q^2 - m_f^2 + \mi 0}
\;.
\end{equation}
%

\subsection{Vertices}
\label{sec:vertices}

\begin{figure}[t]
    \begin{tikzpicture}[baseline={([yshift=-.5ex]current bounding box.center)}]
      \begin{feynman}
        \vertex [dot] (a) {};
        \vertex [above=of a](b) {$a,\mu$};
        \vertex [right=of a](c) {$i,\alpha$};
        \vertex [left=of a](d) {$i',\alpha'$};
      
        \diagram*{
          (a) -- [gluon](b),
          (c) -- [fermion](a) -- [fermion](d),
        };	
      \end{feynman}
    \end{tikzpicture}
\caption{
Quark-gluon vertex.
\label{fig:qqg}}
\end{figure}
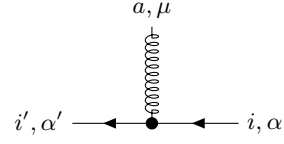

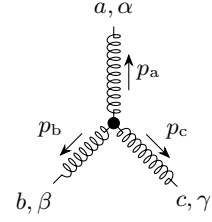
\begin{figure}[t]
     \begin{tikzpicture}[baseline={([yshift=-.5ex]current bounding box.center)}]
      \begin{feynman}
        \vertex [dot] (a) {};
        \vertex [above=of a](b) {$a,\alpha$};
        \vertex [below left=of a](c) {$b,\beta$};
        \vertex [below right=of a](d) {$c,\gamma$};
       
        \diagram*{
          (a) -- [gluon, rmomentum={$p_\La$}](b),
          (a) -- [gluon, rmomentum={$p_\Lb$}](c),
          (a) -- [gluon, rmomentum'={$p_\Lc$}](d),
        };	
      \end{feynman}
    \end{tikzpicture}
\caption{
Triple-gluon vertex.
\label{fig:ggg}}
\end{figure}

\begin{figure}[t]
      \begin{tikzpicture}[baseline={([yshift=-.5ex]current bounding box.center)}]
        \begin{feynman}
          \vertex [dot] (o) {};
          \vertex [above left=of o](a) {$\alpha, a$};
          \vertex [above right=of o](b) {$\delta, d$};
          \vertex [below right=of o](c) {$\gamma, c$};
          \vertex [below left=of o](d) {$\beta, b$};
          \diagram*{
            (o) -- [gluon](a),
            (o) -- [gluon](b),
            (o) -- [gluon](c),
            (o) -- [gluon](d),
          };	
        \end{feynman}
      \end{tikzpicture}
\caption{
Four gluon vertex
\label{fig:gggg}}
\end{figure}
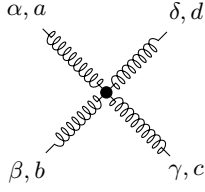

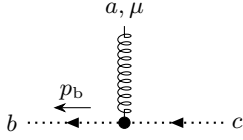
\begin{figure}[t]
 \begin{tikzpicture}[baseline={([yshift=-.5ex]current bounding box.center)}]
      \begin{feynman}
        \vertex [dot] (a) {};
        \vertex [above=of a](b) {$a,\mu$};
        \vertex [right=of a](c) {$c$};
        \vertex [left=of a](d) {$b$};
        \vertex at ($(a)!0.5!(d) + (2cm, 1cm)$) (c2);
        
        \diagram*{
          (a) -- [gluon](b),
          (c) -- [ghost, with arrow=0.5](a) 
          -- [ghost, with arrow=0.5, rmomentum={$p_\Lb$}](d),
        };	
      \end{feynman}
    \end{tikzpicture}
\caption{
Ghost-gluon vertex.
\label{fig:ghostgluon}}
\end{figure}

The Feynman rules for the vertices can be read off from the Lagrangian. The quark-gluon vertex is
\begin{equation}
\Gamma^{a,i' i}_{\mu,\alpha'\alpha}(q,k,p) = \mi g\, 
(\gamma_\mu)_{\alpha'\alpha} (t_a)_{i' i}
\;,
\end{equation}
as illustrated in Fig.~\ref{fig:qqg}. The triple-gluon vertex (with momenta leaving the vertex) is given by
\begin{equation}
\begin{split}
\label{eq:threegluonvertex}
\Gamma^{a b c}_{\alpha \beta \gamma}&(p_\La, p_\Lb, p_\Lc) 
\\={}&
-g f_{abc}
\big\{ 
g_{\alpha\beta}(p_\La - p_\Lb)_\gamma
+ g_{\beta\gamma}(p_\Lb - p_\Lc)_\alpha
\\&
+ g_{\gamma\alpha} (p_\Lc - p_\La)_\beta
\big\}
\;,
\end{split}
\end{equation}
as illustrated in Fig.~\ref{fig:ggg}. The four gluon vertex is
\begin{equation}
\begin{split}
\label{eq:fourgluonvertex}
\Gamma^{abcd}_{\alpha \beta \gamma \delta} ={}&
- \mi g^2 f_{\bar a a b} f_{\bar a c d}
\big\{ 
g_{\alpha \gamma} g_{\beta \delta}
- g_{\alpha \delta} g_{\beta \gamma}
\big\}
\\&
- \mi g^2  f_{\bar a a d} f_{\bar a b c}
\big\{ 
g_{\alpha \beta} g_{\gamma \delta}
- g_{\alpha \gamma} g_{\beta \delta}
\big\}
\\&
- \mi g^2 f_{\bar a a c} f_{\bar a b d}
\big\{ 
g_{\alpha \beta} g_{\gamma \delta}
- g_{\alpha \delta} g_{\beta \gamma}
\big\}
\;,
\end{split}
\end{equation}
as illustrated in Fig.~\ref{fig:gggg}. The ghost-gluon vertex is
\begin{equation}
\label{eq:ghostgluonvertex}
\Gamma^{abc}_\mu(p_\La,p_\Lb,p_\Lc) =  g f_{abc}\, \tilde p_{\Lb,\mu}
\;,
\end{equation}
as illustrated in Fig.~\ref{fig:ghostgluon}. The momentum of the outgoing ghost line is $p_\Lb$.

\section{T gluons and L gluons}
\label{sec:TandLgluons}

We have seen that the propagator for quarks contains a pole $1/(q^2 + \mi 0) = 1/(q^\alpha q^\beta g_{\alpha \beta} + \mi 0)$. On the other hand, the propagator for ghosts contains a pole $1/(q\cdot \tilde q + \mi 0) = 1/(q^\alpha q^\beta h_{\alpha \beta} + \mi 0)$. This implies that in coordinate space the propagator for quarks is singular on the light cone, $x^\mu x^\nu g_{\mu\nu} = 0$ while the propagator for ghosts is singular on the surface $x^\mu x^\nu h^{-1}_{\mu\nu} = 0$. On this surface, $x^2 < 0$. That is, between interactions, ghosts propagate faster than the speed of light. In a frame in which $n = (1,0,0,0)$, we have $x^\mu x^\nu h^{-1}_{\mu\nu} = v^2 t^2 - \vec x^{\,2}$, so the ghosts propagate with speed $v$ in this frame.

For gluons, the situation is a little more complicated. With our preferred choice $\xi = 1$, the gluon propagator, Eq.~(\ref{eq:gluonpropagator2bis}), is
\begin{equation}
\begin{split}
\label{eq:gluonpropagator3}
D^{\mu\nu}(q) ={}&
\frac{1}{q^2+\mi 0}
  \Bigg[
    - g^{\mu\nu} 
    + \frac{q^\mu\,\tilde{q}^\nu + \tilde{q}^\mu\,q^\nu}
    {q\cdot\tilde{q} + \mi 0}
    \\&
    - \left(1+\frac{1}{v^2}\right)\frac{q^\mu\,q^\nu}
    {q\cdot\tilde{q} + \mi 0}\, 
      \Bigg]
  \;.
\end{split}
\end{equation}
This propagator contains products of poles $1/(q^2 + \mi 0)$ and $1/(q\cdot \tilde q + \mi 0)$. We can simplify it by writing it as a sum of a propagator $D_\LT^{\mu\nu}(q)$ with only $1/(q^2 + \mi 0)$ poles and  a propagator $D_\LL^{\mu\nu}(q)$ with only $1/(q\cdot\tilde q + \mi 0)$ poles. We use the projection tensor $P_-^{\alpha \beta}$, Eq.~(\ref{eq:Pplusminus}) and obtain factors $q\cdot P_- \cdot q = q_\alpha P_-^{\alpha\beta} q_\beta$ in denominators. We find
\begin{equation}
\label{eq:gluondecompositionC}
D^{\mu\nu}(q)
= D^{\mu\nu}_{\LT}(q) + D^{\mu\nu}_{\LL}(q)
\;,
\end{equation}
where
\begin{equation}
\begin{split}
\label{eq:DmunuTandL}
D^{\mu\nu}_{\LT}(q) ={}& 
\frac{P^{\mu\alpha}_- P^{\nu\beta}_-}{q^2 + \mi 0}
\left\{
-g_{\alpha\beta} + \frac{q_\alpha q_\beta}{q\cdot P_- \cdot q}
\right\}
\;,
\\
D^{\mu\nu}_{\LL}(q) ={}& 
-\frac{1}{q\cdot\tilde q + \mi 0}
\left\{
\frac{P^{\mu\alpha}_- q_\alpha P^{\nu\beta}_-q_\beta}
{v^2\,q \cdot  P_- \cdot  q}
+ n^\mu n^\nu
\right\}
\;.
\end{split}
\end{equation}
The components of $D^{\mu\nu}_\LT(q)$ and $D^{\mu\nu}_\LL(q)$ in a frame in which $n = (1,0 ,0,0)$ are given in Eqs.~(\ref{eq:DmunuT}) and (\ref{eq:DmunuLstart}). The factor $q\cdot P_- \cdot q = -\vec q^{\,2}$ appears in denominators. This quantity is always negative or zero, so it does not need a $+\mi 0$ prescription. To prove Eq.~(\ref{eq:DmunuTandL}), one manipulates the components of $D^{\mu\nu}(q)$ in this frame. 

The propagator $D^{\mu\nu}_\LT(q)$ contains a pole $1/(q^2 + \mi 0)$ but no pole $1/(q\cdot\tilde q + \mi 0)$. It has another important property: it is entirely transverse. That is 
\begin{equation}
q_\mu D^{\mu\nu}_{\LT}(q) = n_\mu D^{\mu\nu}_{\LT}(q) = 0
\,.
\end{equation}
Thus we can expand it according to Eq.~(\ref{eq:DTexpansion}),
\begin{equation}
\label{eq:DTexpansionbis}
D^{\mu\nu}_{\LT}(q) =  \frac{1}{q^2+\mi 0}
\sum_{s=1,2} \varepsilon^\mu(q,s)\varepsilon^\nu(q,s)
\;,
\end{equation}
using polarization vectors $\varepsilon^\mu(q,s)$ that are real valued functions of an index $s \in \{1,2\}$ and the part of $q$ that is orthogonal to $n$, $P_-^{\mu\alpha} q_\alpha$. The polarization vectors are solutions of $q\cdot \varepsilon(q,s) = n\cdot \varepsilon(q,s) = 0$ that are orthogonal to each other and normalized to $\varepsilon ^2 = -1$. These are the same polarization vectors that one uses in Coulomb gauge.

The propagator $D^{\mu\nu}_\LL(q)$ contains a pole $1/(q\cdot\tilde q + \mi 0)$ but no pole $1/(q^2 + \mi 0)$. It has another important property: it is entirely longitudinal:
\begin{equation}
\label{eq:DLlongitudinal}
\varepsilon_\mu(q,s) D^{\mu\nu}_{\LL}(q) = 0
\,.
\end{equation}
The two terms in $D^{\mu\nu}_{\LL}(q)$ in Eq.~(\ref{eq:DmunuTandL}) correspond to two additional choices of polarization vectors $\varepsilon$. In the first term, $n\cdot \varepsilon = 0$ and $\varepsilon$ is proportional to the part of $q$ that is orthogonal to $n$. In the second term, $\varepsilon$ is proportional to $n$.

\section{Exchange of a soft gluon}
\label{sec:exchange}

Consider a graph for an S matrix element involving a parton with momentum $l$ with $l^2 = 0$ and another parton with momentum $k$ with $k^2 = 0$. These partons could be either quarks, antiquarks, or T gluons. We suppose that these are final state partons, so $l^0 > 0$ and $k^0 > 0$. The partons exchange a virtual gluon with momentum $q$, so that before the exchange the parton momenta are $l-q$ and $k+q$. The amplitude for this exchange is singular in the limit that the exchanged gluon is soft, $q \to 0$. In this limit, the part of the S matrix describing the exchange can be approximated by the eikonal approximation,
\begin{equation}
\begin{split}
\label{eq:veikonal}
\cA^\mathrm{eik} ={}&
\frac{-\mi\,4\pi\as}{(q \cdot l - \mi 0)(q \cdot k + \mi 0)}\,
l_\mu k_\nu D^{\mu\nu}(\vec q)
\;.
\end{split}
\end{equation}
This multiplies a color factor $\bm T_l \cdot \bm T_k$.
 
If we were to calculate $\cA^\mathrm{eik}$ using Feynman gauge, we would find that in addition to the soft singularity, the graph has singularities when $q$ is collinear to $l$ or $-k$. In the presence of these collinear singularities, the eikonal approximation is not adequate. We would need to subtract the collinear singularities from the amplitude and then add them back using Ward identities to sum their contributions over all ways of attaching the gluon to the rest of the graph. This use of Ward identities to organize the infrared singularities is fairly simple when only one gluon in a loop can have a momentum that is soft or collinear with the momentum of an external parton. It is not so simple when more than one gluon can be collinear with an external parton momentum or soft \cite{BabisCalcs, AnastasiouSterman}. We will discuss this in the simple case in Appendix~\ref{sec:FeynmanGauge}.

In this section, we examine this soft gluon exchange in interpolating gauge with $\xi = 1$. We analyze $\cA^\mathrm{eik}$ for T and L gluon exchanges separately. We will see quite directly that there is a soft singularity but no collinear singularities for either T or L gluon exchange.

We examine $\cA^\mathrm{eik}$ in a reference frame in which $n = (1,0,0,0)$. We will want to integrate over $q$, so we define
\begin{equation}
V = 
\mu^{2\epsilon}\int\!\frac{d^{4-2\epsilon} q}{(2\pi)^{4-2\epsilon}}\ 
\theta(|\vec q\,| < Q)\,
\cA^\mathrm{eik}
\;.
\end{equation}
Here we have inserted a factor $\theta(|\vec q\,| < Q)$ with an arbitrary value of $Q$ to provide an ultraviolet cutoff because we have assumed that $q$ is very small in order to obtain the eikonal approximation, but in doing so we have created an artificial UV divergence.

Consider first the exchange of a T gluon. We have
\begin{equation}
\begin{split}
V_\LT ={}&
\mu^{2\epsilon}\int\!\frac{d^{4-2\epsilon} q}{(2\pi)^{4-2\epsilon}}\,
\theta(|\vec q\,| < Q)
\left[\vec l\cdot \vec k - \frac{\vec l\cdot \vec q \ \vec k \cdot \vec q}
{\vec q^{\,2}}\right]
\\&\times
\frac{-\mi\,4\pi\as}{(q \cdot l - \mi 0)(q \cdot k + \mi 0)}
\frac{1}{q^2 + \mi 0}
\;.
\end{split}
\end{equation}
We can now perform the $q^0$ integration by closing the contour in one half plane or the other. There are two contributions initially because $1/(q^2 + \mi 0)$ has two poles. When we combine the terms, we separate the integrand into terms even and odd under $\vec q \to - \vec q$ and throw away the odd part. There is a singular denominator that we separate into even and odd parts using
\begin{equation}
\begin{split}
\frac{1}{\vec q \cdot \vec l /|\vec l\,| 
- \vec q \cdot \vec k/|\vec k\,| + \mi 0} 
={}& \frac{1}{\left[\vec q \cdot \vec l /|\vec l\,| - \vec q \cdot \vec k/|\vec k\,|\right]_\LP} 
\\&
-\mi \pi 
\delta\!\left(\frac{\vec q \cdot \vec l }{|\vec l\,|}
- \frac{\vec q \cdot \vec k }{|\vec k\,|}\right)
.
\end{split}
\end{equation}
The first term is odd, the second is even. After combining the terms, the result is
\begin{equation}
\begin{split}
\label{eq:VT}
V_\LT ={}& 
-\mu^{2\epsilon}\int\!\frac{d^{3-2\epsilon} \vec q}{(2\pi)^{3-2\epsilon}}\,
\theta(|\vec q| < Q)\,
\frac{2\pi\as}{|\vec q|^3}
\\&\times
\left[\vec u_l\cdot \vec u_k - \vec u_q\cdot \vec u_l \ \vec u_q \cdot \vec u_k
\right] 
\\&\times
\Bigg\{
\frac{1 + \vec u_q \cdot \vec u_l\ \vec u_q \cdot \vec u_k}
{\big[1 - \left(\vec u_q \cdot \vec u_l\right)^2\big]
\big[1 - \left(\vec u_q \cdot \vec u_k\right)^2\big]}
\\&\quad
- \mi \pi 
\delta\!\left(\vec u_q \cdot \vec u_l 
- \vec u_q \cdot \vec u_k\right)\,
\frac{2}
{1 - \left(\vec u_q \cdot \vec u_l\right)^2}
\Bigg\}
,
\end{split}
\end{equation}
where $\vec u_l = \vec l/|\vec l\,|$ is a unit vector in the direction of $\vec l$, $\vec u_k = \vec k/|\vec k\,|$, and $\vec u_q = \vec q/|\vec q\,|$.

The integrand is singular in the soft limit $|\vec q\,| \to 0$. This singularity will produce a pole $1/\epsilon$. The first term in braces -- the real term -- is also singular in the collinear limits $\vec u_q \to \vec u_l$ and $\vec u_q \to \vec u_k$. These collinear singularities would be strong enough to produce another $1/\epsilon$. However, in these collinear limits, the factor $[\vec u_l\cdot \vec u_k - \vec u_q\cdot \vec u_l \ \vec u_q \cdot \vec u_k]$ vanishes. This eliminates the collinear poles.\footnote{If $\vec u_q = \vec u_l + \delta\vec u_q$, there is a factor $\delta \vec u_q^{\,2}$ in the denominator. There is a factor $\vec u_k \cdot \delta\vec u_q$ in the numerator. The left-over integrable singularity cancels because $\vec u_k \cdot \delta\vec u_q$ is an odd function of $\delta\vec u_q$.} This cancellation can be traced back to the fact that $q\cdot \varepsilon(q,s) = 0$.

We have not at this point examined the possibility that exchange of a T gluon with momentum $q$ between a parton with momentum $l$ and another parton could be singular when $q$ is collinear to $l$ but is {\em not} soft, $q \sim x\,l$ with a finite coefficient $x$. In this case, we cannot use the eikonal approximation. However, the amplitude is then proportional to a factor $J_\mu \varepsilon^\mu(q,s)$, where $J^\mu$ is the current function for the parton. In the collinear limit, we find that $J_\mu \propto q_\mu$, so that we find a factor $q \cdot \varepsilon(q,s) = 0$ that cancels the collinear singularity. (See Appendix~\ref{sec:FeynmanGauge}.) 

Consider next the exchange of an L gluon. We have, again in the frame in which $n = (1,0,0,0)$,
\begin{equation}
\begin{split}
V_\LL ={}&
\mu^{2\epsilon}\!\int\!\frac{d^{4-2\epsilon} q}{(2\pi)^{4-2\epsilon}}\,
\theta(|\vec q\,| < Q)
\\&\times
\left[v^2 |\vec l| |\vec k| - \frac{\vec l\cdot \vec q \ \vec k \cdot \vec q}
{\vec q^{\,2}}\right]
\\&\times
\frac{\mi\,4\pi\as}{(q \cdot l - \mi 0)(q \cdot k + \mi 0)}
\frac{1}{v^2(q\cdot \tilde q + \mi 0)}
\;.
\end{split}
\end{equation}
We perform the $q^0$ integration and collect the resulting terms that are even under $\vec q \to - \vec q$. This gives
\begin{equation}
\begin{split}
\label{eq:VL}
V_\LL ={}& 
\mu^{2\epsilon}\int\!\frac{d^{3-2\epsilon} \vec q}{(2\pi)^{3-2\epsilon}}\,
\theta(|\vec q| < Q)\,
\frac{2\pi\as}{|\vec q|^3}
\\&\times
\left[v^2 - \vec u_q\cdot \vec u_l \ \vec u_q \cdot \vec u_k
\right] 
\\&\times
\Bigg\{
\frac{v^2 + \vec u_q \cdot \vec u_l\ \vec u_q \cdot \vec u_k}
{\big[v^2 - \left(\vec u_q \cdot \vec u_l\right)^2\big]
\big[v^2 - \left(\vec u_q \cdot \vec u_k\right)^2\big]}
\\&\quad
- \mi \pi 
\delta\!\left(\vec u_q \cdot \vec u_l 
- \vec u_q \cdot \vec u_k\right)\,
\frac{2}
{v^2 - \left(\vec u_q \cdot \vec u_l\right)^2}
\Bigg\}
\;.
\end{split}
\end{equation}
Just as with $V_\LT$, there is a soft singularity from $|\vec q|\to 0$ that will produce a pole, $1/\epsilon$. In contrast to the case with $V_\LT$, the numerator factor $[v^2 - \vec u_q\cdot \vec u_l \ \vec u_q \cdot \vec u_k]$ does not vanish in the collinear limits $\vec u_q \to \vec u_l$ or $\vec u_q \to \vec u_k$. However, again in contrast to the case with $V_\LT$, as long as $v^2 > 1$, the denominator factor $1/[v^2 - (\vec u_q \cdot \vec u_l)^2]$ is not singular when $\vec u_q \to \vec u_l$ and $1/[v^2 - (\vec u_q \cdot \vec u_k)^2]$ is not singular when $\vec u_q \to \vec u_k$. Thus there are no collinear singularities that need to be cancelled. This non-appearance of collinear singularities can be traced back to the fact that if $l \ne 0$, $q \ne 0$, and $v^2 > 0$, it is kinematically not possible to have $l^2 = 0$, $(l-q)^2 = 0$, and $q\cdot\tilde q = 0$ at the same time.

We can perform the $|\vec q\,|$ integration in Eqs.~(\ref{eq:VT}) and (\ref{eq:VL}) to give an infrared (IR) pole:
\begin{equation}
\mu^{2\epsilon}\int\!\frac{d|\vec q\,|}{(2\pi)^{-2\epsilon}}\,
\frac{\theta(|\vec q| < Q)}{|\vec q\,|^{1+2\epsilon}}
= -\frac{1}{2\epsilon} \left( 1  + \cO(\epsilon)\right)
.
\end{equation}
We can then perform integrations over the angles of $\vec q$ with $\epsilon = 0$. This gives us the infrared pole in $V = V_\LT + V_\LL$ for an exchange between partons $l$ and $k$:
\begin{equation}
\begin{split}
\label{eq:Vsoftpole}
V_{lk} ={}& 
\frac{\as}{4\pi}\,\frac{2}{\epsilon}
\Bigg\{\!
- \log\!\left(\frac{1-\vec u_l\cdot \vec u_k}{2}\!\right)
- \mi\pi\, \theta(p_l\cdot p_k\! >\! 0)
\\ &
+ \log\!\left(\frac{v-1}{v+1}\right)
-\frac{v-1}{v}
+ \cO(\epsilon)
\Bigg\}
\;.
\end{split}
\end{equation}
The $-\mi\pi$ contribution in Eq.~(\ref{eq:Vsoftpole}) appears in the case of a gluon exchange between two final state partons and also in the case of a gluon exchange between two initial state partons. In a cross section, these exchanges cancel (at order $\as$) between exchanges in the ket amplitude and in the conjugate bra amplitude. There is no $\mi \pi$ contribution arising from an exchange between an initial state parton and a final state parton. For this reason, we have supplied a factor to indicate that the $-\mi \pi$ contribution is present only when $p_l\cdot p_k > 0$. We can rewrite this in an instructive form, including the $-\mi\pi$ term, as
\begin{equation}
\begin{split}
\label{eq:Vsoftpole2}
V_{lk} ={}& 
\frac{\as}{4\pi}\,\frac{2}{\epsilon}
\Bigg\{\!
- \log\!\left(\frac{- 2\, p_l\cdot p_k + \mi 0}{4\, p_l\cdot n\ p_k\cdot n}\!\right)
\\ &
+ \log\!\left(\frac{v-1}{v+1}\right)
-\frac{v-1}{v}
+ \cO(\epsilon)
\Bigg\}
\;.
\end{split}
\end{equation}

The log of $v-1$ in Eq.~(\ref{eq:Vsoftpole2}) is an indication that if we were to choose Feynman gauge, $v = 1$, a collinear pole would appear. With $v-1 > 0$ we have
\begin{equation}
\frac{1}{\epsilon}\left[1 - (v-1)^{-\epsilon}\right]
= \log(v-1) + \cO(\epsilon)
\;.
\end{equation}
But if we take $(v-1) \to 0$ with $\epsilon$ fixed with $\epsilon < 0$, we have $(v-1)^{-\epsilon} \to 0$. Then
\begin{equation}
\frac{1}{\epsilon}\left[1 - (v-1)^{-\epsilon}\right]
\to \frac{1}{\epsilon}
\;.
\end{equation}

The effect of the $v$ and $n$ dependent factors in $V$ can be better understood by considering the role of color in the exchange. We have calculated the singular contribution to the S matrix from exchanging a soft gluon between partons with labels $l$ and $k$. This exchange comes with a color factor $\bm T_l \cdot \bm T_k = T_l^a T_k^a$, where $T_l^a$ is the color generator matrix for coupling a gluon with color $a$ to the parton line with index $l$. Summing over $l$ and $k$, the total contribution is
\begin{equation}
\begin{split}
S_\mathrm{exch} ={}& \frac{1}{2}\sum_l \sum_{k \ne l} V_{lk}
\bm T_l \cdot \bm T_k
+ \cO(\epsilon^0)
\;.
\end{split}
\end{equation}
Using Eq.~(\ref{eq:Vsoftpole2}) and adding and subtracting a logarithm of the renormalization scale, this is
\begin{equation}
\begin{split}
S_\mathrm{exch} ={}& \frac{1}{2}\sum_l \sum_{k \ne l} 
\bm T_l  \cdot  \bm T_k\,
\frac{\as}{4\pi}\,\frac{2}{\epsilon}
\\&\times
\Bigg\{ 
- \log\!\left( \frac{-2p_l\cdot p_k\!+\!\mi 0}{\mu^2} \right)
\\ &
-\frac{1}{2}\log\!\left(\frac{\mu^2}{4\, (p_l\cdot n)^2}\!\right)
-\frac{1}{2}\log\!\left(\frac{\mu^2}{4\, (p_k\cdot n)^2}\!\right)
\\&
+ \log\!\left(\frac{v-1}{v+1}\right)
-\frac{v-1}{v}
+ \cO(\epsilon)
\Bigg\}
\;.
\end{split}
\end{equation}
Color conservation gives us $\sum_k \bm T_k = \sum_l \bm T_l = 0$ when the sums include all index values, including $k = l$. Thus
\begin{equation}
\begin{split}
\label{eq:Sexchange}
S_\mathrm{exch} ={}& -\sum_l \sum_{k \ne l} 
\bm T_l \cdot \bm T_k
\frac{\as}{4\pi}\,\frac{S_\epsilon}{\epsilon}
\log\!\left(\!\frac{-2p_l\!\cdot\! p_k\!+\!\mi 0}{\mu^2}\!\right)
\\ &
+ \sum_l \bm T_l^2\,
\frac{\as}{4\pi}\,\frac{S_\epsilon}{\epsilon}
\Bigg\{\frac{v-1}{v}
- \log\!\left(\frac{v-1}{v+1}\right)
\\&\quad
+ \log\!\left(\frac{\mu^2}{4\, (p_l\cdot n)^2}\!\right)
\Bigg\}
+ \cO(\epsilon)
\;.
\end{split}
\end{equation}
The factor $\bm T_l^2 = \bm T_l \cdot \bm T_l$ is either $C_\LF$ or $C_\LA$ depending on whether parton $l$ is a quark or a gluon. We will combine this result with the results from self-energy graphs in Sec.~\ref{sec:Spoles}.

\section{BRST symmetry}
\label{sec:brst}

The definition of interpolating gauge depends on two gauge parameters, which we can take to be $v$ and $a$. (We often use parameters $v$ and $\xi = a v^2$, but in this section it is more convenient to use $v$ and $a$.) How do Green functions depend on $v$ and $a$? To find out, we can use BRST symmetry \cite{brs1, brs2, tyutin}. We will also use BRST symmetry to derive the form of a standard Ward identity in the case of interpolating gauge. This analysis is complementary to the analysis of BRST symmetry in Refs.~\cite{BaulieuZwanziger, Baulieu1985}. 

\subsection{The fields and the Lagrangian}
\label{sec:brstL}

For the purposes of this section, we replace the ghost fields $\eta_a(x)$ and $\bar\eta_a(x)$ by fields $c_a(x)$ and $\bar c_a(x)$ with a slightly different normalization:
\begin{equation}
\begin{split}
c_a(x) ={}& \eta_a(x)
\;,
\\
\bar c_a(x) ={}& \sqrt{a}\,\bar\eta_a(x)
\;.
\end{split}
\end{equation}

We seek to find how Green functions depend on the gauge parameters $v$ and $a$. The dependence of the Lagrangian on the gauge parameters resides in the gauge fixing term $\cL_\mathrm{GF}(x)$, Eq.~(\ref{eq:LGF}), and the ghost term $\cL_\mathrm{ghost}(x)$, Eq.~(\ref{eq:Lghost}), in the Lagrangian. We can write these terms in a compact form using the matrix
\begin{equation}
H^\mu_\alpha = \frac{1}{\sqrt a}\,h^\mu_\alpha
\;.
\end{equation}
The gauge dependent parts of the Lagrangian are
\begin{equation}
\begin{split}
\label{eq:gaugedependentL}
\cL_\mathrm{GF}(x) ={}&
-\frac{1}{2}\,[H^\nu_\alpha\partial_\nu A^\alpha_a(x)]
[H^\mu_\beta\partial_\mu A^\beta_a(x)]
\;,
\\
\cL_\mathrm{ghost}(x) ={}&
- \bar c_a(x) 
H^\mu_\alpha \partial_\mu D(A)^\alpha_{ac}
c_c(x)
\;.
\end{split}
\end{equation}
Now we can define derivatives of $H^\mu_\alpha$ with respect to the parameters that we use to define the gauge. Let us name the parameters $r_i$, $i = 1,\dots,6$, according to
\begin{equation}
\begin{split}
\label{eq:ridef}
r_1 ={}& a \;,
\\
r_2 ={}& v \;,
\\
r_{\beta + 3} ={}& n^\beta \;, \qquad \beta = 0,\dots,3
\;.
\end{split}
\end{equation}
Then we define a matrix $X$ by
\begin{equation}
\frac{\partial}{\partial r_i} H^\mu_\alpha(r) = X^\mu_{i,\alpha}(r)
\;.
\end{equation}
Concretely, this is
\begin{equation}
\begin{split}
\frac{\partial}{\partial a} H^\mu_\alpha \equiv{}& X^\mu_{1,\alpha}
= - \frac{1}{2 a^{3/2}}\, h^\mu_\alpha
\;,
\\
\frac{\partial}{\partial v} H^\mu_\alpha \equiv{}& X^\mu_{2,\alpha}
= - \frac{2}{v^3\sqrt a}\, n^\mu n_\alpha
\;,
\\
\frac{\partial}{\partial n^\beta} H^\mu_\alpha \equiv{}& X^\mu_{\beta + 3,\alpha}
\\ ={}& - \frac{1}{\sqrt a}\, \left(1-\frac{1}{v^2}\right)
\left(g^\mu_\beta n_\alpha + n^\mu g_{\alpha \beta}\right)\;.
\end{split}
\end{equation}
The derivatives of the Lagrangian are
\begin{equation}
\begin{split}
\label{eq:dLdri}
\frac{\partial\cL}{\partial r_i} ={}& 
- [H^\nu_\alpha\partial_\nu A^\alpha_a(x)]
[X^\mu_{i,\beta}\partial_\mu A^\beta_a(x)]
\\&
- \bar c_a(x) 
X^\mu_{i,\alpha} \partial_\mu D(A)^\alpha_{ac}
c_c(x)
\;.
\end{split}
\end{equation}
%

\subsection{The BRST transformation}
\label{sec:brstdef}

The theory in interpolating gauge has an exact symmetry under a BRST transformation \cite{brs1,brs2,tyutin}. For each field $\phi \in \{A,\psi,\bar \psi, c,\bar c\}$, the BRST transformation is 
\begin{equation}
\phi \to  \phi +  \theta\,\delta_\mathrm{brst} \phi
\;,
\end{equation}
where $\theta$ is a variable that anticommutes with itself and with $\psi_f(c)$, $\bar \psi_f(x)$, $c_a(x)$ and $\bar c_a(x)$. If we apply a BRST transformation to a product of fields, we use
\begin{equation}
\phi_1 \phi_2 \cdots \phi_n \to \phi_1 \phi_2 \cdots \phi_n
+ \theta\, \delta_\mathrm{brst}(\phi_1 \phi_2 \cdots \phi_n )
\;,
\end{equation}
where
\begin{equation}
\begin{split}
\theta\, \delta_\mathrm{brst}(\phi_1 \phi_2 \cdots \phi_n ) ={}& \theta \delta_\mathrm{brst} (\phi_1) \phi_2\cdots \phi_n
\\&
+ \phi_1 \theta \delta_\mathrm{brst} (\phi_2) \cdots \phi_n
\\&
+\cdots + \phi_1 \phi_2 \cdots \theta \delta_\mathrm{brst} (\phi_n)
\;.
\end{split}
\end{equation}
Then
\begin{equation}
\begin{split}
\delta_\mathrm{brst}(\phi_1 \phi_2 \cdots \phi_n)
={}& \delta_\mathrm{brst} (\phi_1) \phi_2\cdots \phi_n
\\&
+ (-1)^{n_2} \phi_1 \delta_\mathrm{brst} (\phi_2) \cdots \phi_n
+ \cdots
\\&
+ (-1)^{n_n} \phi_1 \phi_2 \cdots \delta_\mathrm{brst} (\phi_n)
\;.
\end{split}
\end{equation}
The signs are $(-1)^{n_j}$ where $n_j$ is the number of fields $\psi$, $\bar \psi$, $c$ and $\bar c$ that are to the left of the transformed field $\phi_j$.

With this notation, the BRST transformation for the individual fields are
\begin{equation}
\begin{split}
\delta_\mathrm{brst}A^\mu_a(x) ={}& D_{ac}^\mu(A(x))\, c_c(x) 
\\
={}& \delta_{ac}\partial^\mu c_c(x)
+ g f_{abc}\, A^\mu_b(x)\, c_c(x)
\;,
\\
\delta_\mathrm{brst}\psi_f(x) ={}& \mi g\,  c_a(x) t_a \psi_f(x)
\;,
\\
\delta_\mathrm{brst}\bar\psi_f(x) ={}& \mi g\, \bar\psi_f(x) c_a(x) t_a
\;,
\\
\delta_\mathrm{brst}c_a(x) ={}& -\frac{g}{2}\,f_{abc}\,
c_b(x)\, c_c(x)
\;,
\\
\delta_\mathrm{brst}\bar c_a(x) ={}& 
- H^\mu_\alpha\, \partial_\mu A^\alpha_a(x)
\;.
\end{split}
\end{equation}
%

\subsection{Dependence of Green functions on the gauge parameters}
\label{sec:dGdri}

Consider a Green function in interpolating gauge,
\begin{equation}
\begin{split}
G ={}&
\Big\langle 
A^{\mu_1}_{a_1}(x_1) \psi_f(x_2) \bar\psi_f(x_3) \cdots
\Big\rangle 
\\ ={}& 
\cN \int\!\cD \psi\, \cD \bar\psi\,\cD A\,\cD \eta\, \cD \bar \eta\ 
e^{\mi S} 
\\&\times
A^{\mu_1}_{a_1}(x_1) \psi_f(x_2) \bar\psi_f(x_3) \cdots
\;.
\end{split}
\end{equation}
We have indicated a gluon field, a quark field, and an antiquark field, but there could be more of these fields. Ghost and antighost fields are not included. The Green function is given as a functional integral over the quark, antiquark, gluon, ghost and antighost fields, weighted by $\exp(\mi S)$, where $S$ is the action. The normalization $\cN$ sets $\left\langle 1\right\rangle = 1$. 

The action depends on the gauge parameters $r_i$, Eq.~(\ref{eq:ridef}).   Accounting for a possible variation of the normalization factor, the derivative of the Green function with respect to one of the $r_i$ is
\begin{align}
\label{eq:dGdvsq}
\frac{\partial G}{\partial r_i}
 ={}& 
\left\langle 
\int\! dx\ 
\mi\frac{\partial \cL(x)}{\partial r_i}\
A^{\mu_1}_{a_1}(x_1) \psi_f(x_2) \bar\psi_f(x_3) \cdots
\right\rangle
\notag
\\& 
- \left\langle 
A^{\mu_1}_{a_1}(x_1) \psi_f(x_2) \bar\psi_f(x_3) \cdots
\right\rangle
\notag
\\&\quad\times
\left\langle 
\int\! dx\ 
\mi\frac{\partial \cL(x)}{\partial r_i}\
\right\rangle
\;.
\end{align}
Here $\partial \cL(x)/\partial r_i$ is given by Eq.~(\ref{eq:dLdri}).

We can use BRST invariance to find $\partial G/\partial r_i$. We note that $\partial \cL(x)/\partial r_i$ is the BRST variation of another quantity:
\begin{equation}
\frac{\partial\cL(x)}{\partial r_i}
= \delta_\mathrm{brst}\cR_{i}(x)
\;,
\end{equation}
where
\begin{equation}
\begin{split}
\cR_{i}(x) ={}& \bar c_a(x) X^\mu_{i,\alpha} \partial_\mu  A^\alpha_a(x)
\;.
\end{split}
\end{equation}
We consider the BRST variation of
\begin{equation}
G_0 =
\mi\int\! dx
\left\langle 
\cR_i(x)\,
A^{\mu_1}_{a_1}(x_1) \psi_f(x_2) \bar\psi_f(x_3) \cdots
\right\rangle
.
\end{equation}
Since Green functions are invariant under BRST transformations, we have
\begin{align}
0 ={}& 
\mi\int\! dx
\left\langle 
\frac{\partial \cL(x)}{dr_i}\,
A^{\mu_1}_{a_1}(x_1) \psi_f(x_2) \bar\psi_f(x_3) \cdots
\right\rangle
\notag
\\&
- \mi\int\! dx\,
\Big\langle 
\cR_i(x)
\left[\delta_\mathrm{brst} A^{\mu_1}_{a_1}(x_1)\right] 
\psi_f(x_2) \bar\psi_f(x_3) \cdots
\Big\rangle
\notag
\\&
- \mi\int\! dx\,
\Big\langle 
\cR_i(x)
A^{\mu_1}_{a_1}(x_1) 
\left[\delta_\mathrm{brst} \psi_f(x_2)\right] 
\bar\psi_f(x_3) \cdots
\Big\rangle
\notag
\\&
+ \mi\int\! dx\,
\Big\langle 
\cR_i(x) 
A^{\mu_1}_{a_1}(x_1) 
\psi_f(x_2) 
\left[\delta_\mathrm{brst} \bar\psi_f(x_3)\right] \cdots
\Big\rangle
\notag
\\&
+ \cdots \;.
\end{align}
This also gives us
\begin{equation}
\begin{split}
\bigg\langle  
\int\! dx\ 
\mi 
\frac{\partial \cL(x)}{\partial r_i}
\bigg\rangle 
={}&
0\;.
\end{split}
\end{equation}
Thus the second term in Eq.~(\ref{eq:dGdvsq}), describing a possible shift in the normalization, vanishes.

\begin{figure*}[t]
\begin{equation*}
  \label{eq:brs-amplitude1}
  \begin{split}
    \frac{\partial}{\partial r_i}\;
    \begin{tikzpicture}[baseline=(b.base)]
      \begin{feynman}[]
        \vertex[blob, minimum size=1.2cm] (b) {};
        \vertex [dot, label={[right] $\mu_1,a_1$}] at ($(b) + (60:2.0cm)$) (c) {};
        \vertex [dot, label={[right] $\alpha$}] at ($(b) + (15:2.0cm)$) (d) {};
        \vertex [dot, label={[right] $\beta$}] at ($(b) + (-30:2.0cm)$) (e) {};
        \diagram*{
          (b)--[gluon, rmomentum'={$p_1$}](c);
          (b)--[fermion, rmomentum'={[label style={above}]$p_2$}](d);
          (b)--[anti fermion,rmomentum'={[label style={above right}]$p_3$}](e);
        };
        \vertex [] at ($(b) + (-80:1.cm)$) (o1) {$\cdot$};
        \vertex [] at ($(b) + (-95:1.0cm)$) (o2) {$\cdot$};
        \vertex [] at ($(b) + (-110:1.0cm)$) (o3) {$\cdot$};
        \vertex [] at ($(b) + (-125:1.0cm)$) (o4) {$\cdot$};
       \end{feynman}
    \end{tikzpicture}
    \; = {}&\;
    \begin{tikzpicture}[baseline=(a.base)]
      \begin{feynman}[]
        \vertex[blob, minimum size=1.2cm] (b) {};
        \vertex [empty square dot, label={[left] $R_i$}] at ($(b) + (180:2cm)$)  (a) {};
        \vertex [dot, label={[right] $a_1,\mu_1$}] at ($(b) + (60:2.0cm)$) (c) {};
        \vertex [dot, label={[right] $\alpha$}] at ($(b) + (-5:2.0cm)$) (d) {};
        \vertex [dot, label={[right] $\beta$}] at ($(b) + (-50:2.0cm)$) (e) {};
        \diagram*{
          (a)--[ghost, quarter left, with arrow=0.5, rmomentum'={$\ell$}] (b)
          --[gluon, quarter left, rmomentum'={$\ell$}](a);
          (b)--[ghost,  with empty arrow=0.5,rmomentum'={$p_1$}](c);
          (b)--[fermion, rmomentum'={[label style={above}]$p_2$}](d);
          (b)--[anti fermion,  rmomentum'={[label style={above right}]$p_3$}](e);
        };
        \vertex [] at ($(b) + (-70:1cm)$) (o1) {$\cdot$};
        \vertex [] at ($(b) + (-85:1cm)$) (o2) {$\cdot$};
        \vertex [] at ($(b) + (-100:1cm)$) (o3) {$\cdot$};
        \vertex [] at ($(b) + (-115:1cm)$) (o4) {$\cdot$};
      \end{feynman}
    \end{tikzpicture}
    \;+\;
    \begin{tikzpicture}[baseline=(a.base)]
      \begin{feynman}[]
        \vertex[blob, minimum size=1.2cm] (b) {};
        \vertex [empty square dot, label={[left] $R_i$}] at ($(b) + (180:2cm)$)  (a) {};
        \vertex [empty dot, label={[right]$\mu_1, a_1$}] at ($(b) + (60:2cm)$) (c) {};
        \vertex [dot, label={[right] $\alpha$}] at ($(b) + (-5:2cm)$) (d) {};
        \vertex [dot, label={[right] $\beta$}] at ($(b) + (-50:2cm)$) (e) {};
        \diagram*{
          (a)--[ghost, quarter left, with arrow=0.5, rmomentum'={$ \ell$}] 
          (b)--[gluon, quarter left, rmomentum'={$\ell$}](a);
          (b)--[ghost, quarter left, with arrow=0.5, rmomentum'={$ k$}](c);
          (b)--[gluon, rmomentum={[label style={right}]$ p_1-k$}](c);
          (b)--[fermion, rmomentum'={[label style={above}]$ p_2$}](d);
          (b)--[anti fermion, rmomentum'={[label style={above right}]$ p_3$}](e);
        };
        \vertex [] at ($(b) + (-70:1cm)$) (o1) {$\cdot$};
        \vertex [] at ($(b) + (-85:1cm)$) (o2) {$\cdot$};
        \vertex [] at ($(b) + (-100:1cm)$) (o3) {$\cdot$};
        \vertex [] at ($(b) + (-115:1cm)$) (o4) {$\cdot$};
      \end{feynman}
    \end{tikzpicture}
    \\
   & +\;
   \begin{tikzpicture}[baseline=(a.base)]
     \begin{feynman}[]
       \vertex[blob, minimum size=1.2cm] (b) {};
       \vertex [empty square dot, label={[left] $R_i$}] 
       at ($(b) + (180:2cm)$)  (a) {};
       \vertex [dot, label={[right] $\mu_1, a_1$}] at ($(b) + (60:2cm)$) (c) {};
       \vertex [empty dot, label={[right] $\alpha$}]  at ($(b) + (0:2cm)$) (d) {};
       \vertex [dot, label={[right] $\beta$}] at ($(b) + (-60:2cm)$) (e) {};
       \diagram*{
         (a)--[ghost, quarter left, with arrow=0.5, rmomentum'={$ \ell$}] (b)
         --[gluon, quarter left, rmomentum'={$ \ell$}](a);
         (b)--[ghost, quarter left, with arrow=0.5, rmomentum'={$ k$}](d);
         (b)--[gluon, rmomentum'={[label style={}]$ p_1$}](c);
         (b)--[fermion, rmomentum={[label style={below}]$ p_2-k$}](d);
         (b)--[anti fermion, rmomentum={[label style={left}]$ \,p_3$}](e);
       };
       \vertex [] at ($(b) + (-75:1cm)$) (o1) {$\cdot$};
       \vertex [] at ($(b) + (-90:1cm)$) (o2) {$\cdot$};
       \vertex [] at ($(b) + (-105:1cm)$) (o3) {$\cdot$};
       \vertex [] at ($(b) + (-120:1cm)$) (o3) {$\cdot$};
     \end{feynman}
   \end{tikzpicture}
   \;+\;
   \begin{tikzpicture}[baseline=(a.base)]
     \begin{feynman}[]
       \vertex[blob, minimum size=1.2cm] (b) {};
       \vertex [empty square dot, label={[left] $ R_i$}] 
       at ($(b) + (180:2.0cm)$)  (a) {};
       \vertex [dot, label={[right] $\mu_1, a_1$}] at ($(b) + (60:2cm)$) (c) {};
       \vertex [dot, label={[right] $\alpha$}] at ($(b) + (15:2cm)$) (d) {};
       \vertex [empty dot, label={[right] $\beta$}] at ($(b) + (-45:2cm)$) (e) {};
       \diagram*{
         (a)--[ghost, quarter left, with arrow=0.5, rmomentum'={$ \ell$}] 
         (b)--[gluon, quarter left, 
         rmomentum'={[arrow shorten=0.7,arrow distance=2mm]$ \ell$}](a);
         (b)--[ghost, quarter left, with arrow=0.5, rmomentum'={$ k$}](e);
         (b)--[gluon, rmomentum'={[label style={}]$p_1$}](c);
         (b)--[fermion, rmomentum'={[label style={above}]$p_2$}](d);
         (b)--[anti fermion, rmomentum={[label style={below left}]$ p_3-k$}](e);
       };
       \vertex [] at ($(b) + (-70:1cm)$) (o1) {$\cdot$};
       \vertex [] at ($(b) + (-85:1cm)$) (o2) {$\cdot$};
       \vertex [] at ($(b) + (-100:1cm)$) (o3) {$\cdot$};
       \vertex [] at ($(b) + (-115:1cm)$) (o4) {$\cdot$};
     \end{feynman}
   \end{tikzpicture}
   \\
   &+\;\cdots\;\;.
 \end{split}
\end{equation*}
\caption{
Identity for the derivative of a full Green function with respect to gauge parameter $r_i$, Eq.~(\ref{eq:dGdriidentity}). The shaded circle represents a full Green function for gluon, quark, and antiquark fields. The rule for the square two point vertex is given in Eq.~(\ref{eq:Rvertex}). The lines with dots at their ends are propagators. In the first term on the right-hand side of the identity, the open arrow on the ghost propagator represents a factor of $p_1^{\mu_1}$. The rules for the vertices represented by open circles are given in Eqs.~(\ref{eq:compositevertexq}), (\ref{eq:compositevertexqbar}), and (\ref{eq:compositevertexg}).
\label{fig:dGdri}}
\end{figure*}
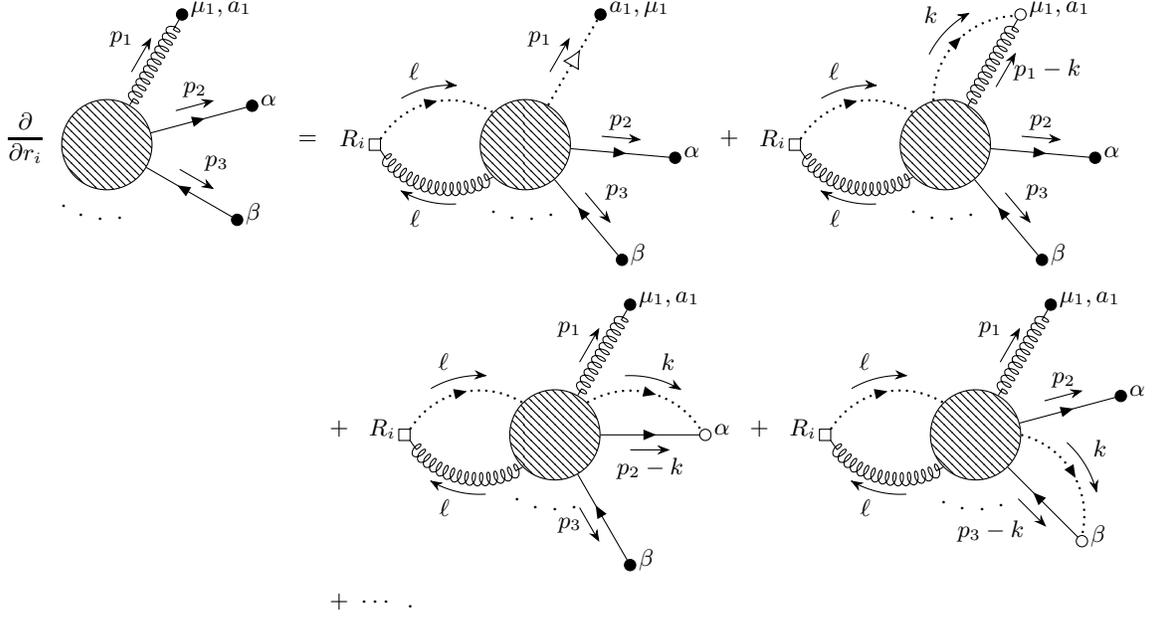

\begin{widetext} 
This gives us
\begin{equation}
\begin{split}
\label{eq:dGdriidentity}
\frac{\partial}{\partial r_i} 
\Big\langle 
A^{\mu_1}_{a_1}(x_1) \psi_f(x_2) \bar\psi_f(x_3) \cdots
\Big\rangle
={}& \int\! dx\,
\Big\langle 
\cR_i(x)
\left[\mi D_{a_1 c}^\mu(A(x_1))\, \eta_c(x_1)\right] 
\psi_f(x_2) \bar\psi_f(x_3) \cdots
\Big\rangle
\\&
- g\int\! dx\,
\Big\langle 
\cR_i(x) 
A^{\mu_1}_{a_1}(x_1) 
\left[\eta_a(x_2) t_a \psi_f(x_2)\right] 
\bar\psi_f(x_3) \cdots
\Big\rangle
\\&
+ g \int\! dx\,
\Big\langle 
\cR_i(x)
A^{\mu_1}_{a_1}(x_1) 
\psi_f(x_2) 
\left[\bar\psi_f(x_3) \eta_a(x_3) t_a\right] \cdots
\Big\rangle
+ \cdots \;.
\end{split}
\end{equation}
\end{widetext} 

This equation is illustrated in Fig.~\ref{fig:dGdri}. The factors $\cR_i(x)$, written in terms of $X^\mu_\alpha$ and $a$ and the antighost field $\bar \eta_(x)$, are
\begin{equation}
\begin{split}
\label{eq:Rvertexinx}
\cR_{i}(x) ={}& \bar \eta_a(x) \sqrt{a}\,X^\mu_{i,\alpha} \partial_\mu  A^\alpha_a(x)
\;.
\end{split}
\end{equation}
These operators destroy a gluon and create a ghost with the same color index or else destroy an antighost and create a gluon. We need these operators integrated over $x$, so in momentum space the operators $\int\!d^4x\, \cR_{i}(x)$ conserve momentum. In momentum space, if the momentum of the gluon entering the $\cR$ vertex is $\ell$ and the gluon polarization is $\alpha$, then the momentum of the ghost leaving the vertex is also $\ell$ and the value of the vertex is
\begin{equation}
\label{eq:Rvertex}
R_{i,\alpha}(\ell) = -\mi \ell_\mu \sqrt{a}\, X^\mu_{i,\alpha}
\;,
\end{equation}
as illustrated by the square two-point vertices in Fig.~\ref{fig:dGdri}. On the right-hand side of Eq.~(\ref{eq:dGdriidentity}), there is a contribution for each external parton. 

A quark line with momentum $p_2$ in the original Green function is replaced with a vertex at which a quark line with momentum $p_2 - k$ meets a ghost with momentum $k$ and color $a$. The vertex is simply 
\begin{equation}
\label{eq:compositevertexq}
V_\Lq(p_2,p_2-k,k,a) = - g\, t_a
\;.
\end{equation}
There is no propagator for the quark after this vertex. 

An antiquark line with momentum $p_3$ in the original Green function is replaced with a vertex at which an antiquark line with momentum $p_3 - k$ meets a ghost with momentum $k$ and color $a$. The vertex is
\begin{equation}
\label{eq:compositevertexqbar}
V_{\bar \Lq}(p_3,p_3-k,k,a) = + g\, t_a
\;,
\end{equation}
with no subsequent antiquark propagator.

\begin{figure*}[t]
\begin{equation*}
  \begin{split}
    \frac{\tilde\ell_\mu}{a}\;
    \begin{tikzpicture}[baseline=(b.base)]
      \begin{feynman}[]
        \vertex[blob, minimum size=1.2cm] (b) {};
        \vertex [dot, label={[above] $c,\mu$}] at ($(b) + (180:2cm)$)  (a) {};
        \vertex [dot, label={[right] $\mu_1,a_1$}] at ($(b) + (60:2.0cm)$) (c) {};
        \vertex [dot, label={[right] $\alpha$}] at ($(b) + (15:2.0cm)$) (d) {};
        \vertex [dot, label={[right] $\beta$}]  at ($(b) + (-30:2.0cm)$) (e) {};
        \diagram*{
          (a)--[gluon, rmomentum'={$\ell$}](b)--[gluon, rmomentum'={$p_1$}](c);
          (b)--[fermion, rmomentum'={[label style={above}]$p_2$}](d);
          (b)--[anti fermion,rmomentum'={[label style={above right}]$p_3$}](e);
        };
        \vertex [] at ($(b) + (-80:1.cm)$) (o1) {$\cdot$};
        \vertex [] at ($(b) + (-95:1.0cm)$) (o2) {$\cdot$};
        \vertex [] at ($(b) + (-110:1.0cm)$) (o3) {$\cdot$};
        \vertex [] at ($(b) + (-125:1.0cm)$) (o4) {$\cdot$};
       \end{feynman}
    \end{tikzpicture}
    \; = {}&\;
    \begin{tikzpicture}[baseline=(a.base)]
      \begin{feynman}[]
        \vertex[blob, minimum size=1.2cm] (b) {};
        \vertex [dot, label={[above] $c$}] at ($(b) + (180:2cm)$)  (a) {};
        \vertex [dot, label={[right] $a_1,\mu_1$}] at ($(b) + (60:2.0cm)$) (c) {};
        \vertex [dot, label={[right] $\alpha$}] at ($(b) + (-5:2.0cm)$) (d) {};
        \vertex [dot, label={[right] $\beta$}] at ($(b) + (-50:2.0cm)$) (e) {};
        \diagram*{
          (a)--[ghost, with arrow=0.5, rmomentum'={$\ell$}] (b);
          (b)--[ghost,  with empty arrow=0.5,rmomentum'={$p_1$}](c);
          (b)--[fermion, rmomentum'={[label style={above}]$p_2$}](d);
          (b)--[anti fermion, rmomentum'={[label style={above right}]$p_3$}](e);
        };
        \vertex [] at ($(b) + (-70:1cm)$) (o1) {$\cdot$};
        \vertex [] at ($(b) + (-85:1cm)$) (o2) {$\cdot$};
        \vertex [] at ($(b) + (-100:1cm)$) (o3) {$\cdot$};
        \vertex [] at ($(b) + (-115:1cm)$) (o4) {$\cdot$};
      \end{feynman}
    \end{tikzpicture}
    \;+\;
    \begin{tikzpicture}[baseline=(a.base)]
      \begin{feynman}[]
        \vertex[blob, minimum size=1.2cm] (b) {};
        \vertex [dot, label={[above] $c$}]               
        at ($(b) + (180:2cm)$) (a) {};
        \vertex [empty dot, label={[right]$\mu_1, a_1$}] 
        at ($(b) + (60:2cm)$)  (c) {};
        \vertex [dot, label={[right] $\alpha$}]          
        at ($(b) + (-5:2cm)$)  (d) {};
        \vertex [dot, label={[right] $\beta$}]           
        at ($(b) + (-50:2cm)$) (e) {};
        \diagram*{
          (a)--[ghost, with arrow=0.5, rmomentum'={$ \ell$}] (b);
          (b)--[ghost, quarter left, with arrow=0.5, rmomentum'={$ k$}](c);
          (b)--[gluon, rmomentum={[label style={right}]$ p_1-k$}](c);
          (b)--[fermion, rmomentum'={[label style={above}]$ p_2$}](d);
          (b)--[anti fermion, rmomentum'={[label style={above right}]$ p_3$}](e);
        };
        \vertex [] at ($(b) + (-70:1cm)$) (o1) {$\cdot$};
        \vertex [] at ($(b) + (-85:1cm)$) (o2) {$\cdot$};
        \vertex [] at ($(b) + (-100:1cm)$) (o3) {$\cdot$};
        \vertex [] at ($(b) + (-115:1cm)$) (o4) {$\cdot$};
      \end{feynman}
    \end{tikzpicture}
    \\
   & +\;
    \begin{tikzpicture}[baseline=(a.base)]
      \begin{feynman}[]
        \vertex[blob, minimum size=1.2cm] (b) {};
        \vertex [dot, label={[above] $c$}]            
        at ($(b) + (180:2cm)$)  (a) {};
        \vertex [dot, label={[right] $\mu_1, a_1$}]   
        at ($(b) + (60:2cm)$) (c) {};
        \vertex [empty dot, label={[right] $\alpha$}] 
        at ($(b) + (0:2cm)$) (d) {};
        \vertex [dot, label={[right] $\beta$}]        
        at ($(b) + (-60:2cm)$) (e) {};
        \diagram*{
          (a)--[ghost, with arrow=0.5, rmomentum'={$\ell$}] (b);
          (b)--[ghost, quarter left, with arrow=0.5, rmomentum'={$ k$}](d);
          (b)--[gluon, rmomentum'={[label style={}]$ p_1$}](c);
          (b)--[fermion, rmomentum={[label style={below}]$ p_2-k$}](d);
          (b)--[anti fermion, rmomentum={[label style={left}]$ \,p_3$}](e);
        };
        \vertex [] at ($(b) + (-75:1cm)$) (o1) {$\cdot$};
        \vertex [] at ($(b) + (-90:1cm)$) (o2) {$\cdot$};
        \vertex [] at ($(b) + (-105:1cm)$) (o3) {$\cdot$};
        \vertex [] at ($(b) + (-120:1cm)$) (o3) {$\cdot$};
      \end{feynman}
    \end{tikzpicture}
    \;+\;
    \begin{tikzpicture}[baseline=(a.base)]
      \begin{feynman}[]
        \vertex[blob, minimum size=1.2cm] (b) {};
        \vertex [dot, label={[above] $c$}]            
        at ($(b) + (180:2.0cm)$)  (a) {};
        \vertex [dot, label={[right] $\mu_1, a_1$}]  
        at ($(b) + (60:2cm)$) (c) {};
        \vertex [dot, label={[right] $\alpha$}]      
        at ($(b) + (15:2cm)$) (d) {};
        \vertex [empty dot, label={[right] $\beta$}] 
        at ($(b) + (-45:2cm)$) (e) {};
        \diagram*{
          (a)--[ghost, with arrow=0.5, rmomentum'={$ \ell$}] (b);  
          (b)--[ghost, quarter left, with arrow=0.5, rmomentum'={$ k$}](e);
          (b)--[gluon, rmomentum'={[label style={}]$p_1$}](c);
          (b)--[fermion, rmomentum'={[label style={above}]$p_2$}](d);
          (b)--[anti fermion, rmomentum={[label style={below left}]$ p_3-k$}](e);
        };
        \vertex [] at ($(b) + (-70:1cm)$) (o1) {$\cdot$};
        \vertex [] at ($(b) + (-85:1cm)$) (o2) {$\cdot$};
        \vertex [] at ($(b) + (-100:1cm)$) (o3) {$\cdot$};
        \vertex [] at ($(b) + (-115:1cm)$) (o4) {$\cdot$};
      \end{feynman}
    \end{tikzpicture}
    \\
    &+\;\cdots\;\;.
  \end{split}
\end{equation*}
\caption{
Identity (\ref{eq:STDidentity}) for a Green function in which one external gluon line carrying momentum $\ell$ and polarization $\mu$ is multiplied by $\tilde\ell_\mu/a$. On the right hand side of the identity, the gluon line is replaced by a ghost line. The other graphical notations are the same as in Fig.~\ref{fig:dGdri}.
\label{fig:otherWardid}}
\end{figure*}
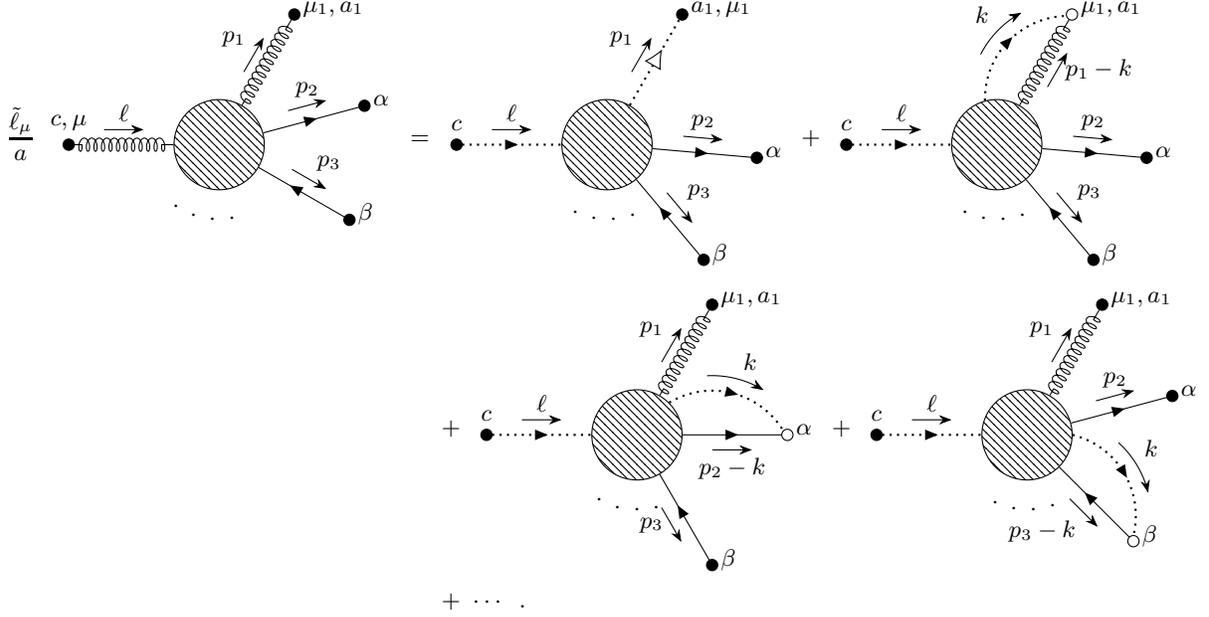

For a gluon line with momentum $p_1$,  polarization $\mu_1$, and color $a_1$ in the original graph, there are two terms, according to the two terms in 
\begin{equation}
\begin{split}
\mi D_{a_1 c}^{\mu_1}(A(x_1))\, \eta_c(x_1)
={}& \mi\, \partial^{\mu_1} \eta_{a_1}(x_1)
\\&
+ \mi\,g f_{a_1 bc} A^{\mu_1}_b(x_1) \eta_c(x_1)
\;.
\end{split}
\end{equation}
In the first term, the gluon is simply replaced by a ghost with the same color index and we multiply by $p_1^{\mu_1}$. In the second term, the gluon is replaced with a vertex at which a gluon line with momentum $p_1 - k$, polarization $\mu_1$, and color $b$ meets a ghost with momentum $k$ and color $c$. The vertex is  
\begin{equation}
\label{eq:compositevertexg}
V_{\Lg}(p_1,p_1-k,k;a_1,b,c) =  \mi\,g f_{a_1 bc}
\end{equation}
with no subsequent propagator.

\subsection{Another Ward identity}
\label{sec:StdWard}

We can use BRST invariance to derive another identity for Green functions. This identity is quite standard, except that with interpolating gauge the standard $\partial_\mu A_a^\mu(x)$ is replaced by $\tilde\partial_\mu A_a^\mu(x)$. We start with the Green function
\begin{equation}
G_0 =
-\left\langle 
\bar \eta_c(x)\,
A^{\mu_1}_{a_1}(x_1) \psi_f(x_2) \bar\psi_f(x_3) \cdots
\right\rangle
.
\end{equation}
The BRST variation of $\bar \eta_c(x)$ is
\begin{equation}
\delta_\mathrm{brst}\bar \eta_c(x) = - \frac{1}{a}\,\tilde\partial_\mu A^\mu_c(x)
\;.
\end{equation}
The BRST variation of $G_0$ vanishes, so
\begin{widetext} 
\begin{equation}
\begin{split}
\label{eq:STDidentity}
\frac{1}{a}
\Big\langle 
[-\mi\,\tilde\partial_\mu A^\mu_c(x)]\,
A^{\mu_1}_{a_1}(x_1) \psi_f(x_2) \bar\psi_f(x_3) \cdots
\Big\rangle
={}& 
\Big\langle 
\bar\eta_c(x)
\left[\mi D_{a_1 a}^\mu(A(x_1))\, c_a(x_1)\right] 
\psi_f(x_2) \bar\psi_f(x_3) \cdots
\Big\rangle
\\&
- g\,
\Big\langle 
\bar\eta_c(x)
A^{\mu_1}_{a_1}(x_1) 
\left[\eta_a(x_2) t_a \psi_f(x_2)\right] 
\bar\psi_f(x_3) \cdots
\Big\rangle
\\&
+ g\,
\Big\langle 
\bar\eta_c(x)
A^{\mu_1}_{a_1}(x_1) 
\psi_f(x_2) 
\left[\bar\psi_f(x_3) \eta_a(x_3) t_a\right] \cdots
\Big\rangle
+ \cdots \;.
\end{split}
\end{equation}

\end{widetext} 
In this identity, we consider a Green function with gluon, quark, and antiquark lines as in the previous subsection and with one additional gluon line with polarization $\mu$ and color index $c$ that carries momentum $\ell$ into the graph.  We multiply the Green function by $(1/a)\, \tilde \ell_\mu$. This gives a sum of Green functions in which the gluon is replaced by a ghost with momentum $\ell$ and color index $c$. In each of these Green functions, the ghost line interacts with one of the external parton lines just as in the identity of the previous subsection. This identity is depicted in Fig.~\ref{fig:otherWardid}.

In a covariant gauge, with $v = 1$, $\xi = a$, the additional gluon line is multiplied by $\ell_\mu$ instead of $\tilde \ell_\mu$. Then this is an identity that is described in many field theory textbooks \cite{StermanBook}.

\subsection{Dependence of the LSZ factors on the $r_i$}
\label{sec:dLSZdri}

In the following subsection, we will examine how the S matrix depends on the parameters $r_i$ defined in Eq.~(\ref{eq:ridef}). As a step in this endeavor, we first consider the factors $R_\Lg$ and $R_\Lq$ that appear in the LSZ reduction formula that relates the S matrix to Green functions.

The LSZ factor for gluons is defined from the full gluon propagator $\delta_{ab}G^{\mu\nu}(p)$. Here $\mu,\nu$ are the Lorentz indices for the two ends of the gluon line and $a,b$ are the color indices. We multiply $G$ by polarization vectors $\varepsilon_\mu(p,s)$ and $\varepsilon_\nu(p,s)$ for the transversely polarized gluons, with $\varepsilon^2 = -1$. The polarization vectors are functions of the part of the momentum $p$ that is orthogonal to $n$, as discussed in Sec.~\ref{sec:TandLgluons}. The LSZ factor $R_\Lg$ is defined by
\begin{equation}
\label{eq:Rgdef}
\varepsilon_\mu(p,s)\,G^{\mu\nu}(p)\,\varepsilon_\nu(p,s)
\sim \frac{\mi R_\Lg(p)}{p^2 + \mi 0}
\;.
\end{equation}
We start with $p^2 \ne 0$ but then we take the limit $p^2 \to 0$ in Eq.~(\ref{eq:Rgdef}). We here use $A\sim B$ to mean that $A/B \to 1$ in the limit $p^2 \to 0$. Thus $\mi R_\Lg(p)$ is the residue of the pole at $p^2 = 0$. Even with $p^2 = 0$, $R_\Lg$ can depend on $(p\cdot n)^2$. For this reason, the notation indicates that $R_\Lg$ depends on $p$.

In Eq.~(\ref{eq:Rgdef}), we work in the renormalized theory in $4 - 2\epsilon$ dimensions, with $\epsilon < 0$ to control integrations that are otherwise divergent in the infrared. When we take $\epsilon \to 0$, infrared poles $1/\epsilon^n$ appear, as we will see in Sec.~\ref{sec:gluonselfenergy}.

We rewrite Eq.~(\ref{eq:Rgdef}) in the form
\begin{equation}
\label{eq:Rgdef2}
\mi p^2\,R_\Lg(p) \sim
p^2 \varepsilon_\mu(p,s)\,G^{\mu\nu}(p)\,\varepsilon_\nu(p,s)\,p^2
\;.
\end{equation}
Here the factors of $p^2$ on the right hand side of the equation are proportional to inverse tree level propagators for T gluons.

In order to study the dependence of $R_\Lg$ on the gauge parameters, we first investigate the structure of the full gluon propagator $G^{\mu\nu}$.
In Sec.~\ref{sec:TandLgluons}, we decomposed the tree level gluon propagator  into separate contributions for T gluons and L gluons:
\begin{equation}
D^{\mu\nu}(p) = D_\LT^{\mu\nu}(p) + D_\LL^{\mu\nu}(p)
\;,
\end{equation}
with
\begin{equation}
\begin{split}
D_\LT^{\mu\nu}(p) ={}& \frac{P_\LT^{\mu\nu}(p)}{p^2 + \mi 0}
\;,
\\
D_\LL^{\mu\nu}(p) ={}& \frac{P_\LL^{\mu\nu}(p)}{p\cdot \tilde p + \mi 0}
\;.
\end{split}
\end{equation}
The numerator functions are
\begin{equation}
\begin{split}
P_\LT^{\mu\nu}(p) ={}& \sum_{s=1,2} \varepsilon^\mu(p,s)\,\varepsilon^\nu(p,s)
\\={}&
- g^{\mu\nu} + n^\mu n^\nu 
\\&
- 
\frac{(p^\mu - p\cdot n\, n^\mu)(p^\nu - p\cdot n\, n^\nu)}
{(p\cdot n)^2 - p^2}
\;,
\\
P_\LL^{\mu\nu}(p) ={}& - n^\mu n^\nu
+ \frac{(p^\mu\! - p\cdot n\,  n^\mu)(p^\nu\! - p\cdot n\,  n^\nu)}
{v^2 [(p\cdot n)^2 - p^2]}
\\&
- \frac{\xi-1}{v^2}\, 
    \frac{p^\mu\,p^\nu}{p\cdot\tilde{p} + \mi 0}
\;.
\end{split}
\end{equation}

We can also decompose the gluon self-energy into
\begin{equation}
\begin{split}
\Pi^{\mu\nu}(p) ={}& \Pi_\LT^{\mu\nu}(p)
+
\Pi_\LL^{\mu\nu}(p)
\;,
\end{split}
\end{equation}
where $\Pi_\LT^{\mu\nu}(p)$ has the structure
\begin{equation}
\Pi_\LT^{\mu\nu}(p) = P_\LT^{\mu\nu}(p)\, \pi_\LT(p)
\end{equation}
and where $\Pi_\LL^{\mu\nu}(p)$ can be decomposed into terms proportional to $p^\mu p^\nu$, $n^\mu n^\nu$ and $p^\mu n^\nu + n^\mu p^\nu$. With this definition, $\Pi_\LL^{\mu\nu}(p)$ does {\em not} contain any terms proportional to $g^{\mu\nu}$. Those terms belong in $\Pi_\LT^{\mu\nu}(p)$.

Now, the full propagator obeys the Dyson equation
\begin{equation}
G^\mu_\nu(p) = D^\mu_\nu(p)
+ G^\mu_\alpha(p) \Pi^\alpha_\beta(p) D^\beta_\nu(p)
\;.
\end{equation}
This has the perturbative solution, with dots denoting contractions in the Lorentz indices,
\begin{equation}
\begin{split}
G ={}& D + D\cdot\Pi\cdot D + 
D\cdot\Pi\cdot D\cdot\Pi\cdot D + \cdots
\;.
\end{split}
\end{equation}
In each term, we substitute $D = D_\LT + D_\LL$ and $\Pi = \Pi_\LT + \Pi_\LL$. We note that
\begin{equation}
\begin{split}
P_\LT^{\mu\nu}(p)\,p_\nu ={}& \,p_\nu P_\LT^{\nu\mu}(p) = 0
\;,
\\
P_\LT^{\mu\nu}(p)\,n_\nu ={}& \,n_\nu P_\LT^{\nu\mu}(p) = 0
\;,
\end{split}
\end{equation}
so that
\begin{equation}
\begin{split}
D_\LT\cdot\Pi_\LL ={}& \Pi_\LL \cdot D_\LT  = 0
\;,
\\
\Pi_\LT\cdot D_\LL ={}& D_\LL \cdot \Pi_\LT  = 0
\;.
\end{split}
\end{equation}
Because of this, the structure of the gluon propagator simplifies:
\begin{equation}
G^{\mu\nu}(p) = G_\LT^{\mu\nu}(p) + G_\LL^{\mu\nu}(p)
\;,
\end{equation}
with
\begin{equation}
\begin{split}
\label{eq:GTGL}
G_\LT ={}& D_\LT + D_\LT\cdot\Pi_\LT\cdot D_\LT 
\\&+ 
D_\LT\cdot\Pi_\LT\cdot D_\LT\cdot\Pi_\LT\cdot D_\LT + \cdots
\;,
\\
G_\LL ={}& D_\LL + D_\LL\cdot\Pi_\LL\cdot D_\LL 
\\&+ 
D_L\cdot\Pi_\LL\cdot D_\LL\cdot\Pi_\LL\cdot D_\LL + \cdots
\;.
\end{split}
\end{equation}
The propagator $G_\LT^{\mu\nu}(p)$ for T gluons has poles at $p^2 = 0$ but no poles at $p\cdot \tilde p = 0$. The propagator $G_\LL^{\mu\nu}(p)$ for L gluons, expanded perturbatively, has poles at $p\cdot \tilde p = 0$ but no poles at $p^2 = 0$. In Eq.~(\ref{eq:Rgdef2}), there is a factor $p^2$ and we take the limit $p^2 \to 0$. Thus $G_\LT^{\mu\nu}(p)$ contributes to $R_\Lg$ but $G_\LL^{\mu\nu}(p)$ does not.

The residue $R_\Lg$ defined in Eq.~(\ref{eq:Rgdef2}) depends on the gauge parameters $r_i$ defined in Eq.~(\ref{eq:ridef}). To find its derivative with respect to $r_i$, we can differentiate $G^{\mu\nu}(p)$ with respect to $r_i$. Four of the $r_i$ are the components of the vector $n$ that defines the preferred reference frame used to specify the gauge. For these parameters, we should take note that the polarization  vectors $\varepsilon(p,s)$ depend on $n$ since they obey $n\cdot \varepsilon(p,s)=0$. However, this dependence does not matter in Eq.~(\ref{eq:Rgdef2}). To see why, differentiate $p\cdot \varepsilon = 0$ and $n\cdot \varepsilon = 0$ to obtain
\begin{equation}
\begin{split}
p_\nu\frac{\partial \varepsilon^\nu(p,s)}{\partial n^\alpha} ={}& 0
\;,
\\
n_\nu\frac{\partial \varepsilon^\nu(p,s)}{\partial n^\alpha} 
={}& -  \varepsilon_\alpha(p,s)
\;.
\end{split}
\end{equation}
We can achieve this with
\begin{equation}
\begin{split}
\label{eq:dvarepsilondn}
\frac{\partial \varepsilon^\nu(p,s)}{\partial n^\alpha} ={}& 
- \frac{p\cdot n\, p^\nu - p^2 n^\nu}{(p\cdot n)^2 - p^2}\ \varepsilon_\alpha(p,s)
\;.
\end{split}
\end{equation}
More generally, with $n \to n' = n + \delta n$, one can transform $\varepsilon$ to $\varepsilon' = \varepsilon + \delta\varepsilon$ using Eq.~(\ref{eq:dvarepsilondn}), then rotate $\varepsilon'$ in the plane orthogonal to $p$ and $n'$ by an angle $\delta\phi$. We define the transform of $\varepsilon$ not to include this extra rotation.

Now suppose that we differentiate $\varepsilon_\nu(p,s)$ in Eq.~(\ref{eq:Rgdef2}) with respect to $n_\alpha$. In any graph that contributes to $G_\LT^{\mu\nu}(p)$ in Eq.~(\ref{eq:Rgdef2}), there is a factor $D_\LT^{\beta \nu}(p)$ that multiplies $\varepsilon_\nu(p,s)$. However,
\begin{equation}
\label{eq:dvarepsilondngives0}
D_\LT^{\beta \nu}(p)\ 
\frac{\partial \varepsilon_\nu(p,s)}{\partial n^\alpha} = 0
\end{equation}
because $D_\LT^{\beta \nu}(p) p_\nu = D_\LT^{\beta \nu}(p) n_\nu = 0$. We conclude that the dependence on $n$ of $\varepsilon_\nu(p,s)$ and $\varepsilon_\mu(p,s)$ in Eq.~(\ref{eq:Rgdef2}) does not affect the dependence of $R_\Lg$ on $n$.

\begin{figure*}[t]
\begin{equation*}
  \begin{split}
    \mi p^2\frac{\partial R_\Lg(p)}{\partial r_i}\sim {}& 
    \frac{\partial}{\partial r_i}
    \begin{tikzpicture}[baseline=(b.base)]
      \begin{feynman}[]
        \vertex[blob,minimum size=1.2cm] (b) {};
        \vertex [dot] at ($(b) + (180:1.7cm)$) (a) {};
        \vertex [dot] at ($(b) + (0:1.7cm)$) (c) {};
        \vertex [] at ($(b) + (180:2.5cm)$) (a1) {};
        \vertex [] at ($(b) + (0:2.5cm)$) (c1) {};
        \diagram*{
          (a1)--[Bracket-, gluon](a)--[gluon, rmomentum'={$p$}]
          (b) --[gluon, rmomentum'={$p$}](c)--[-Bracket,gluon](c1);
        };
       \end{feynman}
    \end{tikzpicture}
    \\
  	={}&
    \begin{tikzpicture}[baseline=(a.base)]
      \begin{feynman}[]
        \vertex[blob,fill=none, minimum size=1.2cm] (b) {$\Gamma$};
        \vertex [empty square dot, label={[left] $R_i$}] 
        at ($(b) + (180:1.5cm)$)  (a) {};
        \vertex [empty dot] at ($(b) + (60:1.5cm)$) (c) {};
        \vertex [] at ($(b) + (60:2.5cm)$) (d) {$ $};
        \vertex [blob, minimum size=1.2cm] at ($(b) + (0:2.25cm)$) (e) {};
        \vertex [dot] at ($(e) + (0:1.7cm)$) (f) {};
        \vertex [] at ($(e) + (0:2.5cm)$) (f1) {};
        \diagram*{
          (a)--[ghost, quarter left, with arrow=0.5, rmomentum'={$\ell$}] (b)
          --[gluon, quarter left, rmomentum'={$\ell$}](a);
          (b)--[ghost, quarter left, with arrow=0.5, rmomentum'={$k$}](c);
          (b)--[gluon, momentum={[label style={right}]$p+k$}](c)
          --[-Bracket,gluon, momentum={$p$}](d);
          (b)--[gluon, rmomentum={$p$}](e) 
          --[gluon, rmomentum'={$p$}](f)--[-Bracket,gluon](f1);
         };
    \end{feynman}
    \end{tikzpicture}
  	+
    \begin{tikzpicture}[baseline=(a.base)]
      \begin{feynman}[]
       \vertex[blob,fill=none, minimum size=1.2cm] (b) {$\Gamma$};
        \vertex [empty square dot, label={[right] $R_i$}] 
        at ($(b) + (1800:1.5cm)$)  (a) {};
        \vertex [empty dot] at ($(b) + (120:1.5cm)$) (c) {};
        \vertex [] at ($(b) + (120:2.5cm)$) (d) {$ $};
        \vertex [blob, minimum size=1.2cm] at ($(b) + (180:2.25cm)$) (e) {};
        \vertex [dot] at ($(e) + (180:1.7cm)$) (f) {};
        \vertex [] at ($(e) + (180:2.5cm)$) (f1) {};
        \diagram*{
          (a)--[ghost, quarter left, with arrow=0.5, rmomentum'={$\ell$}] (b)
          --[gluon, quarter left, rmomentum'={$\ell$}](a);
          (b)--[ghost, quarter left, with arrow=0.5, rmomentum'={$k$}](c);
          (b)--[gluon, rmomentum={[label style={right}]$p-k$}](c)
          --[-Bracket, gluon, rmomentum={$p$}](d);
          (b)--[gluon, momentum'={$p$}](e) 
          --[gluon, momentum'={$p$}](f)--[-Bracket,gluon](f1);
         };
    \end{feynman}
    \end{tikzpicture}
  \end{split}
\end{equation*}
\caption{
Illustration of Eq.~(\ref{eq:dRgdri}). The gluon lines terminated with ``$\;\scriptstyle{[}\;$'' or ``$\;\scriptstyle{]}\;$'' stand for $p^2\,\varepsilon(p,s)$. These factors do not vanish as long as $p^2 \ne 0$.
\label{fig:gpropagatoridentity}}
\end{figure*}
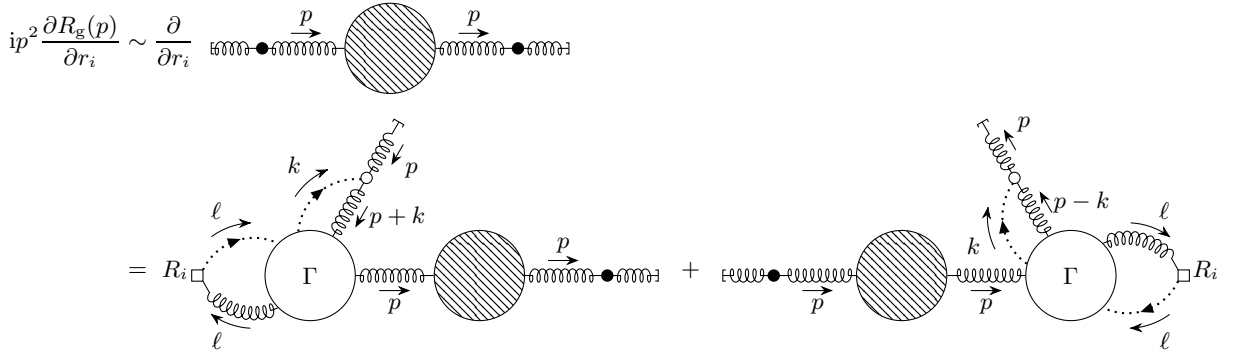

This analysis shows that the derivative of $R_\Lg(p)$ with respect to any of the gauge parameters $r_i$ is given by
\begin{equation}
\begin{split}
\label{eq:GtoRg}
\mi p^2\,\frac{\partial R_\Lg(p)}{\partial r_i} \sim{}& 
p^2 \varepsilon_\mu(p,s)\,
\frac{\partial G^{\mu\nu}(p)}{\partial r_i}\,\varepsilon_\nu(p,s)\,p^2
\;.
\end{split}
\end{equation}
To obtain $\partial G^{\mu\nu}(p)/\partial r_i$ we use the identity (\ref{eq:dGdriidentity}). There are now contributions from both ends of the gluon propagator. For each end, there is a term in $\partial G^{\mu\nu}(p)/\partial r_i$ represented as the first term in Fig.~\ref{fig:dGdri}. This term is proportional to $p^{\mu}$ for the left end and $p^{\nu}$ for the right end. This term does not contribute to Eq.~(\ref{eq:GtoRg}) because $p \cdot \varepsilon(p,s) = 0$. The remaining contribution for the right end has a vertex that combines a gluon with momentum $p-k$ and Lorentz index $\nu$ with a ghost with momentum $k$, with no attached gluon propagator. The remaining contribution for the left end has a vertex that combines a gluon with momentum $-p-k$ coming out of the graph and Lorentz index $\mu$ with a ghost with momentum $k$, with no attached gluon propagator. We can write these contributions as
\begin{equation}
\begin{split}
\label{eq:dRgdri}
\mi p^2\,\frac{\partial R_\Lg(p)}{\partial r_i} \sim{}& 
p^2  \varepsilon_\mu(p,s)
G^{\mu}_\alpha(p) \Gamma^{\alpha\nu}(p)
\varepsilon_\nu(p,s)\,p^2
\\ &\!\!\!\!\!\!
+ p^2  \varepsilon_\mu(p,s)
\Gamma^{\alpha\mu}(-p) G^{\nu}_\alpha(p) 
\varepsilon_\nu(p,s) p^2
.
\end{split}
\end{equation}
Here $\Gamma^{\alpha\nu}(p)$ contains the right hand gluon-ghost-gluon vertex. It is one particle irreducible: it has no cut that cuts a single gluon line. Thus $\Gamma^{\alpha\nu}(p)$ must contain the two particle gluon-ghost vertex $R$, Eq.~(\ref{eq:Rvertex}). There is still at least a tree level gluon propagator on the left. Summing over graphs, there is a complete gluon propagator $G^{\mu}_\alpha(p)$ on the left. Similarly $\Gamma^{\nu\alpha}(-p)$ contains the left hand gluon-ghost-gluon vertex. This equation is illustrated in Fig.~\ref{fig:gpropagatoridentity}.

The Lorentz structure of $\Gamma$ can be simplified using invariance under Lorentz transformations that leave $p$ and $n$ unchanged:
\begin{equation}
\begin{split}
\Gamma^{\alpha\nu}(p)\,\varepsilon_\nu(p,s) ={}&  
\varepsilon^\alpha(p,s)\,\Gamma_{\!\Lg}(p)
\;.
\end{split}
\end{equation}
Here $\Gamma_{\!\Lg}(p)$ is a scalar function of $p^2$ and $(p\cdot n)^2$. This gives us
\begin{equation}
\begin{split}
\frac{\partial R_\Lg(p)}{\partial r_i} \sim{}& 
-\mi p^2 \varepsilon_\mu(p,s)\,
G^{\mu\nu}(p) \,\varepsilon_\nu(p,s)
\,\Gamma_{\!\Lg}(p)
\\ &
-\mi p^2 \varepsilon_\mu(p,s)\,
G^{\mu\nu}(p) \,\varepsilon_\nu(p,s)
\,\Gamma_{\!\Lg}(p)
\;.
\end{split}
\end{equation}
Now we can take the $p^2 \to 0$ limit, using the definition Eq.~(\ref{eq:Rgdef}). This gives
\begin{equation}
\begin{split}
\frac{\partial R_\Lg(p)}{\partial r_i} ={}& 
2 R_\Lg(p)
\,\Gamma_{\!\Lg}(p)
\;.
\end{split}
\end{equation}

We now turn to the LSZ factor for quarks, which is defined from the full propagator $\delta_{ij}G^{\alpha\beta}(p)$ for quarks. Here $\alpha,\beta$ are the Dirac indices for the two ends of the quark line and $i,j$ are the color indices. In our analysis below, we do not write Dirac indices explicitly.  The LSZ residue factor $R_\Lq$ is defined by
\begin{equation}
\label{eq:Rqdef}
G(p) \sim \frac{\mi \s{p}}{p^2 + \mi 0}\ R_\Lq(p)
\end{equation}
in the limit $p^2 \to 0$. 

Our analysis will make use of Dirac spinors $u(p,s)$. Since we want to take a limit $p^2 \to 0$ with $p^2 \ne 0$ to start with, we should be careful with the definition. We define a lightlike momentum vector $p_0$ using a timelike reference vector $n_\Lq$, which is distinct from the gauge fixing vector $n$:
\begin{equation}
\label{eq:p0def}
p = p_0 + \frac{(1+\beta)\, p^2}{2 p \cdot n_\Lq}\,n_\Lq
\;.
\end{equation}
with
\begin{equation}
p_0^2 = 0
\;.
\end{equation}
Then
\begin{equation}
\beta = \frac{2}{1 + \sqrt{1- p^2 n_\Lq^2/(p\cdot n_\Lq)^2}} - 1
\;.
\end{equation}
That is, $\beta = \cO(p^2)$ as $p^2 \to 0$. We take $u(p,s) = u(p_0(p,n_\Lq),s)$ to be the usual solutions of the massless Dirac equation for momentum $p_0$: 
\begin{equation}
\begin{split}
\label{eq:upsdef}
\s{p}_0 u(p,s) ={}& 0
\;,
\\
\bar u(p,s) \gamma ^\mu u(p,s) ={}& 2 p_0^\mu
\;.
\end{split}
\end{equation}
With this definition, 
\begin{equation}
\bar u(p,s)\,\s{p}\,u(p,s) = (1 - \beta)\, p^2
\;.
\end{equation}

We can express the relation (\ref{eq:Rqdef}) between $G(p)$ and  $R_\Lq(p)$ using spinors:
\begin{equation}
\begin{split}
\label{eq:GtoRderivation}
\bar u(p,s)\,\s{p}\,G(p)\,\s{p}\,u(p,s) \sim{}& 
\frac{\mi R_\Lq(p)}{p^2 + \mi 0}\,\bar u(p,s)\,\s{p}\,\s{p}\,\s{p}\,u(p,s)
\\
={}& \mi R_\Lq \,\bar u(p,s)\,\s{p}\,u(p,s)
\\
\sim{}& \mi R_\Lq \,p^2
\;.
\end{split}
\end{equation}
Then our definition of the LSZ factor is
\begin{equation}
\label{eq:Rqdef2}
\mi p^2\,R_\Lq(p) \sim
\bar u(p,s)\,\s{p}\, G (p)\,\s{p}\, u(p,s)
\;. 
\end{equation}
%

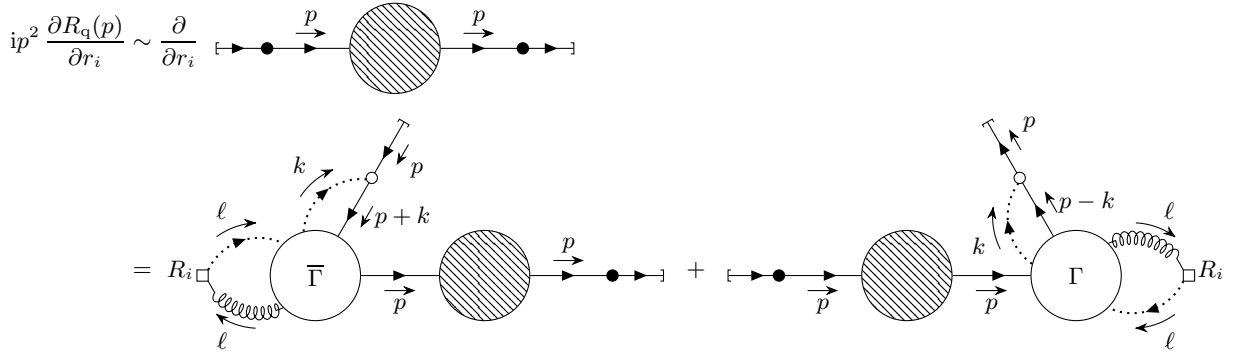
\begin{figure*}[t]
\begin{equation*}
  \begin{split}
  \mi p^2\,
    \frac{\partial R_\Lq(p)}{\partial r_i}\sim {}& 
    \frac{\partial}{\partial r_i}
    \begin{tikzpicture}[baseline=(b.base)]
      \begin{feynman}[]
        \vertex[blob,minimum size=1.2cm] (b) {};
        \vertex [dot] at ($(b) + (180:1.7cm)$) (a) {};
        \vertex [dot] at ($(b) + (0:1.7cm)$) (c) {};
        \vertex [] at ($(b) + (180:2.5cm)$) (a1) {};
        \vertex [] at ($(b) + (0:2.5cm)$) (c1) {};
        \diagram*{
          (a1)--[Bracket-, fermion](a)--[fermion, rmomentum'={$p$}](b) 
          --[fermion, rmomentum'={$p$}](c)--[-Bracket,fermion](c1);
        };
       \end{feynman}
    \end{tikzpicture}
    \\
  	={}&
    \begin{tikzpicture}[baseline=(a.base)]
      \begin{feynman}[]
        \vertex[blob,fill=none, minimum size=1.2cm] (b) {$\overline\Gamma$};
        \vertex [empty square dot, label={[left] $R_i$}] 
        at ($(b) + (180:1.5cm)$)  (a) {};
        \vertex [empty dot] at ($(b) + (60:1.5cm)$) (c) {};
        \vertex [] at ($(b) + (60:2.5cm)$) (d) {$ $};
        \vertex [blob, minimum size=1.2cm] at ($(b) + (0:2.25cm)$) (e) {};
        \vertex [dot] at ($(e) + (0:1.7cm)$) (f) {};
        \vertex [] at ($(e) + (0:2.5cm)$) (f1) {};
        \diagram*{
          (a)--[ghost, quarter left, with arrow=0.5, rmomentum'={$\ell$}] (b)
          --[gluon, quarter left, rmomentum'={$\ell$}](a);
          (b)--[ghost, quarter left, with arrow=0.5, rmomentum'={$k$}](c);
          (b)--[anti fermion, momentum={[label style={right}]$p+k$}](c)
          --[-Bracket,anti fermion, momentum={$p$}](d);
          (b)--[fermion, rmomentum={$p$}](e)
          --[fermion, rmomentum'={$p$}](f)--[-Bracket,fermion](f1);
         };
    \end{feynman}
    \end{tikzpicture}
  	+
    \begin{tikzpicture}[baseline=(a.base)]
      \begin{feynman}[]
       \vertex[blob,fill=none, minimum size=1.2cm] (b) {$\Gamma$};
        \vertex [empty square dot, label={[right] $R_i$}] 
        at ($(b) + (1800:1.5cm)$)  (a) {};
        \vertex [empty dot] at ($(b) + (120:1.5cm)$) (c) {};
        \vertex [] at ($(b) + (120:2.5cm)$) (d) {$ $};
        \vertex [blob, minimum size=1.2cm] at ($(b) + (180:2.25cm)$) (e) {};
        \vertex [dot] at ($(e) + (180:1.7cm)$) (f) {};
        \vertex [] at ($(e) + (180:2.5cm)$) (f1) {};
        \diagram*{
          (a)--[ghost, quarter left, with arrow=0.5, rmomentum'={$\ell$}] (b)
          --[gluon, quarter left, rmomentum'={$\ell$}](a);
          (b)--[ghost, quarter left, with arrow=0.5, rmomentum'={$k$}](c);
          (b)--[fermion, rmomentum={[label style={right}]$p-k$}](c)
          --[-Bracket, fermion, rmomentum={$p$}](d);
          (b)--[anti fermion, momentum'={$p$}](e)
          --[anti fermion, momentum'={$p$}](f)--[-Bracket,anti fermion](f1);
         };
    \end{feynman}
    \end{tikzpicture}
  \end{split}
\end{equation*}
\caption{
Illustration of Eq.~(\ref{eq:dRdri1}). The incoming quark line terminated with a ``$\;\scriptstyle{[}\;$''  stands for $\s{p}\, u(p,s)$ and the outgoing terminated quark line stands for $\bar u(p,s)\, \s{p}$. These factors do not vanish as long as $p^2 \ne 0$.
\label{fig:qpropagatoridentity}}
\end{figure*}

We can also determine how $R_\Lq$ depends on the gauge parameters $r_i$. We start with
\begin{equation}
\label{eq:dRqdri}
\frac{\partial G(p)}{\partial r_i}
\sim \frac{\mi \s{p}}{p^2 + \mi 0}\ 
\frac{\partial R_\Lq(p)}{\partial r_i}
\end{equation}
Then the steps in Eq.~(\ref{eq:GtoRderivation}) give us
\begin{equation}
\mi p^2\,
\frac{\partial R_\Lq(p) }{\partial r_i} \sim
\bar u(p,s)\,\s{p}\,
\frac{\partial G(p)}{\partial r_i}\,\s{p}\, u(p,s)
\;. 
\end{equation}
To obtain $\partial G(p)/\partial r_i$ we use the identity (\ref{eq:dGdriidentity}). There are contributions from both ends of the quark propagator. The contribution for the right end has a vertex that combines a quark with momentum $p-k$ and Lorentz index $\beta$ with a ghost with momentum $k$, with no attached gluon propagator. The contribution for the left end has a vertex that combines an incoming quark with momentum $p+k$ and Lorentz index $\alpha$ with a ghost that supplies momentum $k$, with no attached quark propagator. We can write these contributions as
\begin{equation}
\begin{split}
\label{eq:dRdri1}
\frac{\partial R_\Lq(p) }{\partial r_i} \sim{}&
\frac{-\mi}{p^2}\, \bar u(p,s)\s{p}\,
\big[ G(p)\,\Gamma(p) + \overline\Gamma(-p)\,G(p)
\big]
\\&\times
\s{p}\, u(p,s)
\;.
\end{split}
\end{equation}
This equation is illustrated in Fig.~\ref{fig:qpropagatoridentity}.

The Dirac structure of $\Gamma$ can be specified. We note that $\Gamma$ contains an even number of $\gamma$-matrices and is a function of $p$ and $n$. It thus has the form
\begin{equation}
\begin{split}
\Gamma ={}& \Gamma_{\!\Lq} \,1 
+ \frac{\Gamma'_{\!\Lq}}{p^2}\,\s{p}_0 \s{p}
\;.
\end{split}
\end{equation}
Here $\Gamma_{\!\Lq}(p)$ and $\Gamma_{\!\Lq}'(p)$ are a scalar functions of $p$ and $n$. 

Now, using $\s{p}_0\,u(p,s) = 0$,
\begin{equation}
\begin{split}
\Gamma\,\s{p}\,u(p,s) ={}& 
\Gamma_{\!\Lq}\, \s{p}\,u(p,s)
+ \Gamma'_{\!\Lq}\,\s{p}_0\,u(p,s)
\\
={}& \Gamma_{\!\Lq}\, \s{p}\,u(p,s)
\;.
\end{split}
\end{equation}
Similarly,
\begin{equation}
\begin{split}
\bar u(p,s)\,\s{p}\,\overline\Gamma  ={}& 
\Gamma_{\!\Lq}\, \bar u(p,s)\,\s{p}
\;.
\end{split}
\end{equation}

Now Eq.~(\ref{eq:dRdri1}) becomes
\begin{equation}
\begin{split}
\label{eq:dRdri2}
\frac{\partial R_\Lq(p) }{\partial r_i} \sim{}&
\frac{-\mi}{p^2}\, \bar u(p,s)\s{p}\,
G(p)
\s{p}\, u(p,s)\ 2 \Gamma_{\!\Lq}
\;.
\end{split}
\end{equation}
Using Eq.~(\ref{eq:Rqdef2}) then gives
\begin{equation}
\begin{split}
\frac{\partial R_\Lq(p) }{\partial r_i} ={}&
2 R_\Lq(p)\,\Gamma_{\!\Lq}(p)
\;.
\end{split}
\end{equation}
The same analysis applies to antiquarks, with the same function $\Gamma_\Lq$.

\subsection{Dependence of the S matrix on the $r_i$}
\label{sec:dSdri}

We have seen in subsection \ref{sec:dGdri} how a Green function for gluons, quarks, and antiquarks depends on the gauge parameters $r_i$ defined in Eq.~(\ref{eq:ridef}). We can use this result to find how the S matrix depends on the $r_i$. Let us call the Green function in momentum space $G(p_1,\mu_1,a_1; p_2,\alpha_2, i_2; p_3, \beta_3, j_3; \dots)$. Here the gluon with momentum $p_1$ has vector index $\mu_1$ and color $\bm 8$ index $a_1$; the quark with momentum $p_2$ has Dirac index $\alpha_2$ and color $\bm 3$ index $i_2$; the antiquark with momentum $p_3$ has Dirac index $\beta_3$ and color $\bar{\bm 3}$ index $j_3$. There can be more partons, indicted by the ellipsis. All momenta are defined to be outgoing.

From the Green function, we can construct the S matrix $S(p_1,s_1,a_1; p_2,s_2, i_2; p_3, s_3, j_3; \dots)$. We start with off-shell partons with momenta $p_j$, but we will then take a limit $p_j^2 \to 0$. The variables $s_i$ label the transverse polarizations of the gluons or the spins of the quarks or antiquarks. The S matrix is related to the Green function by
\begin{align}
\label{eq:Smatrix}
S(p_1,& s_1,a_1; p_2,s_2, i_2; p_3, s_3, j_3; \dots)
\notag\\
={}& G(p_1,\mu_1,a_1; p_2,\alpha_2, i_2; p_3, \beta_3, j_3; \dots)
\notag\\ & \times
\frac{(\mi p_1^2)\, \varepsilon_{\mu_1}(p_1,s_1)}{\sqrt{R_\Lg(p_1)}}\,
\frac{\overline u_{\alpha_2}(p_2,s_2)\,(-\mi \s{p}_2)}{\sqrt{R_\Lq(p_2)}}
\\&\times
\frac{(-\mi \s{p}_3)\, v_{\beta_3}(p_3,s_3)}{\sqrt{R_\Lq(p_3)}}
\times \cdots
\;.
\notag
\end{align}
This definition applies using the renormalized Green function $G$ and the factors $R$ in $4-2\epsilon$ dimensions, with infrared divergences regulated by keeping $\epsilon < 0$, as in Eq.~(\ref{eq:Rgdef}). For quarks and antiquarks, we have multiplied by an inverse tree level propagator, $-\mi \s{p}$. For each parton, we have multiplied by the appropriate polarization vector or spinor, with an implied sum over the polarization or spinor index. (This is for final state quarks and antiquarks. The choice is modified for initial state antiquarks or quarks.)  The polarization vectors and spinors are functions of the parts of the momenta $p_1$, $p_2$, and $p_3$ that are orthogonal to $n$, as in Sec.~\ref{sec:dLSZdri}. Thus $\overline u_{\alpha_2}(p_2,s_2) \s{p}_2$ and $\s{p}_3 v_{\alpha_2}(p_3,s_3)$ do not vanish as long as $p_2^2$ and $p_3^2$ are not zero. 

In $G$, the propagator for T gluons, $G_\LT$ (Eq.~(\ref{eq:GTGL})) multiplies $\varepsilon_{\mu_1}(p_1,s_1)$. The propagator for L gluons, $G_\LL$, does not appear because it lacks poles $1/p_1^2$. We will differentiate $S$ with respect to the gauge parameters $r_i$. Four of these parameters are the components of $n$. The polarization vector $\varepsilon_{\mu_1}(p_1,s_1)$ depends on $n$. However, the derivative of $\varepsilon_{\mu_1}(p_1,s_1)$ with respect to $n_\alpha$ gives zero when contracted with $G_\LT$, as we saw in Eq.~(\ref{eq:dvarepsilondngives0}). Thus we can treat $\varepsilon_{\mu_1}(p_1,s_1)$ as if it were independent of all of the $r_i$.

In Eq.~(\ref{eq:Smatrix}), we have started with the full Green function and divided by a factor $\sqrt{R}$ for each external leg. An alternative formulation, which we use later in this paper, is to divide by $R$ for each external leg, giving the Green function amputated on the external legs, then to multiply by $\sqrt{R}$.

We would now like to see how the S matrix depends on the gauge parameters. We differentiate $S$ with respect to the gauge parameter $r_i$. Here we have to differentiate both $G$ and each factor of $1/\sqrt R$. We get
\begin{widetext}
\begin{equation}
\begin{split}
\label{eq:dSdri1}
\frac{\partial S}{\partial r_i} ={}& 
\frac{\partial G}{\partial r_i}\,
\frac{(\mi p_1^2)\, \varepsilon_{\mu_1}(p_1,s_1)}{\sqrt{R_\Lg(p_1)}}\,
\frac{\overline u_{\alpha_2}(p_2,s_2)\,(-\mi \s{p}_2)}{\sqrt{R_\Lq(p_2)}}\,
\frac{(-\mi \s{p}_3)\, v_{\beta_3}(p_3,s_3)}{\sqrt{R_{\Lq(p_3)}}}
\times \cdots
\\ &
- G 
\left[
\frac{(\mi p_1^2)\, \varepsilon_{\mu_1}(p_1,s_1)}{\sqrt{R_\Lg(p_1)}}\,
\frac{\overline u_{\alpha_2}(p_2,s_2)\,(-\mi \s{p}_2)}{\sqrt{R_\Lq(p_2)}}\,
\frac{(-\mi \s{p}_3)\, v_{\beta_3}(p_3,s_3)}{\sqrt{R_\Lq(p_3)}}
\times \cdots
\right]
\\ &
\times
\left[
\frac{1}{2 R_\Lg(p_1)}\frac{\partial R_\Lg(p_1)}{\partial r_i}
+\frac{1}{2 R_\Lq(p_2)}\frac{\partial R_\Lq(p_2)}{\partial r_i}
+\frac{1}{2 R_\Lq(p_3)}\frac{\partial R_\Lq(p_3)}{\partial r_i}
+ \cdots
\right]
\;.
\end{split}
\end{equation}
\end{widetext}

Let us examine the term in Eq.~(\ref{eq:dSdri1}) proportional to $\partial G/\partial r_i$. There is a term in $\partial G/\partial r_i$, represented as the first term in Fig.~\ref{fig:dGdri}, that is proportional to $p_1^{\mu_1}$. This term does not contribute to Eq.~(\ref{eq:dSdri1}) because $p_1 \cdot \varepsilon(p_1,s_1) = 0$. 

In the next term in $\partial G/\partial r_i$, a gluon with momentum $p_1 - k$ joins a ghost with momentum $k$ at a vertex $V_\Lg$, Eq.~(\ref{eq:compositevertexg}). There is no subsequent propagator with a factor $1/p_1^2$. Since we multiply by $p_1^2$ in Eq.~(\ref{eq:dSdri1}), this contribution vanishes when we take $p_1^2 \to 0$ for many of the graphs that contribute to $\partial G/\partial r_i$. However, there are some graphs in which a gluon line carrying momentum $p_1$ enters a one-particle irreducible subgraph $\Gamma$ that involves the special gluon-ghost mixing vertex together with other interactions and finally creates a gluon and a ghost that combine in the vertex $V_\Lg$. This single gluon propagator has a factor $1/p_1^2$ that cancels the factor $p_1^2$ to give us a finite contribution when $p_1^2 \to 0$.  After analyzing the structure of the one-particle-irreducible graphs as in Sec.~\ref{sec:dLSZdri}, we obtain a contribution of the form
\begin{equation}
\label{eq:GitoGAgluon}
G\,
\frac{\mi p_1^2 \varepsilon_{\mu_1}(p_1,s_1)}{\sqrt{R_\Lg(p_1)}}\
\Gamma_{\!\Lg}
\;.
\end{equation}

In the next term in $\partial G/\partial r_i$, a quark with momentum $p_2 - k$ joins a ghost with momentum $k$ at a vertex $V_\Lq$, Eq.~(\ref{eq:compositevertexq}). There is no subsequent propagator with a factor $\s{p}_2/p_2^2$. Thus this contribution vanishes when we take $p_2^2 \to 0$ for many graphs. However, there are some graphs in which a quark line carrying momentum $p_2$ enters a one-particle-irreducible subgraph that involves the special gluon-ghost mixing vertex together with other interactions and finally creates a quark and a ghost that combine in the vertex $V_\Lq$. This single gluon propagator has a factor $\s{p}_2/p_2^2$ that gives us a finite contribution when $p_2^2 \to 0$. After analyzing the structure of the one-particle-irreducible graphs as in Sec.~\ref{sec:dLSZdri}, we obtain a contribution of the form
\begin{equation}
\label{eq:GitoGAquark}
G\,
\frac{\overline u_{\alpha_2}(p_2,s_2)\,(-\mi \s{p}_2)}{\sqrt{R_\Lq(p_2)}}\
\Gamma_{\!\Lq}
\;.
\end{equation}

In the next term in $\partial G/\partial r_i$, an antiquark with momentum $p_3 - k$ joins a ghost with momentum $k$ at a vertex $V_{\bar\Lq}$, Eq.~(\ref{eq:compositevertexqbar}). The surviving contributions come from one-particle-irreducible subgraphs coupled to a full Green function. As in the quark case, these give a contribution
\begin{equation}
\label{eq:GitoGAqbar}
G\,
\frac{(-\mi \s{p}_3)\,v_{\beta_3}(p_3,s_3)}{\sqrt{R_\Lq(p_3)}}\
\Gamma_{\!\Lq}
\;.
\end{equation}

Summing these contributions, we have
\begin{widetext}
\begin{equation}
\begin{split}
\label{eq:dSdri2}
\frac{\partial S}{\partial r_i} ={}& 
G 
\left[
\frac{(\mi p_1^2)\, \varepsilon_{\mu_1}(p_1,s_1)}{\sqrt{R_\Lg(p_1)}}\,
\frac{\overline u_{\alpha_2}(p_2,s_2)\,(-\mi \s{p}_2)}{\sqrt{R_\Lq(p_2)}}\,
\frac{(-\mi \s{p}_3)\, v_{\beta_3}(p_3,s_3)}{\sqrt{R_\Lq(p_3)}}
\times \cdots
\right]
\\ &
\times
\left[
\Gamma_{\!\Lg}
-\frac{1}{2 R_\Lg(p_1)}\frac{\partial R_\Lg(p_1)}{\partial r_i}
+ \Gamma_{\!\Lq}
-\frac{1}{2 R_\Lq(p_2)}\frac{\partial R_\Lq(p_2)}{\partial r_i}
+ \Gamma_{\!\Lq}
-\frac{1}{2 R_\Lq(p_3)}\frac{\partial R_\Lq(p_3)}{\partial r_i}
+ \cdots
\right]
\;.
\end{split}
\end{equation}
\end{widetext}

In Sec.~\ref{sec:dLSZdri}, we found the derivatives of $R_\Lg$ and $R_\Lq$ with respect to $r_i$:
\begin{equation}
\begin{split}
\Gamma_{\!\Lg}
={}& \frac{1}{2 R_\Lg}\frac{\partial R_\Lg}{\partial r_i}
\;,
\\
\Gamma_{\!\Lq}
={}& \frac{1}{2 R_\Lq}\frac{\partial R_\Lq}{\partial r_i}
\;.
\end{split}
\end{equation}
Inserting this into Eq.~(\ref{eq:dSdri2}), we find that the S matrix is invariant under changes of the gauge parameters,
\begin{equation}
\label{eq:dSdrifinal}
\frac{\partial S}{\partial r_i} = 0
\;.
\end{equation}
%

\section{Renormalization}
\label{sec:renorm}

Ref.~\cite{BaulieuZwanziger} provided an argument, based on BRST invariance, that QCD in interpolating gauge can be renormalized. In this section, we calculate the one-loop contributions to the renormalization factors $Z$.

To renormalize the theory in interpolating gauge, we need to renormalize the coupling $g$, the gauge parameters $v$ and $\xi$, and the field strengths. In each case, we relate a ``bare'' quantity, indicated with a subscript B, to a corresponding renormalized quantity by means of a factor $Z$ or $\sqrt Z$. In the case of the gluon field, $\sqrt Z$ is a matrix. Each of the $Z$ factors depend on $\as$ and the gauge parameters $v^2$ and $\xi$. We expand the $Z$ factors in powers of $\as$, beginning with 1 or the unit matrix at order $\as^0$. In this section, we examine the $\as^1$ contributions.

The field strength renormalizations take the form
\begin{equation}
\begin{split}
\label{eq:fieldstrengthZs}
A^\mu_a(x)_\LB ={}& [Z_A^{1/2}]^{\mu}_\nu\, A^\nu_a(x)
\;,\\
\psi_\alpha(x)_\LB ={}& Z_\psi^{1/2}\, \psi_\alpha(x)
\;,\\
\bar\psi_\alpha(x)_\LB ={}& Z_\psi^{1/2}\, \bar\psi_\alpha(x)
\;,\\
\eta_a(x)_\LB ={}& Z_\eta^{1/2}\, \eta_a(x)
\;,\\
\bar\eta_a(x)_\LB ={}& Z_\eta^{1/2}\, \bar\eta_a(x)
\;.
\end{split}
\end{equation}
The field strength renormalization factor for the gluon field needs to be a Lorentz tensor instead of just a scalar because the definition of the gauge uses a fixed vector $n$. The coupling and gauge parameters are renormalized as
\begin{equation}
\begin{split}
g_\LB ={}& Z_g\,g
\;,
\\
v^2_\LB ={}& Z_v\,v^2
\;,
\\
\xi_\LB ={}& Z_\xi\,\xi
\;.
\end{split}
\end{equation}
%

\subsection{The renormalization factors}
\label{sec:theZs}

We begin with a statement of the results for the renormalization factors $Z$. In the following subsection, we exhibit the calculation that leads to these results.

At order $\as$, the renormalization factor for the coupling is the same as in a covariant gauge,
\begin{equation}
\label{eq:Zgresult}
Z_g = 1 - \frac{\as}{4\pi}\,
\frac{S_\epsilon}{\epsilon}\, \gamma_\Lg + \cO(\as^2)
\;,
\end{equation}
where
\begin{equation}
\label{eq:gammag}
\gamma_\Lg = \frac{11}{6}\,C_A - \frac{2}{3}\,T_R n_f
\end{equation}
and $S_\epsilon$ is the standard coefficient of $1/\epsilon$ for $\MSbar$ renormalization,
\begin{equation}
\label{eq:epsilonMSbar}
S_\epsilon = \frac{(4\pi)^\epsilon}{\Gamma(1-\epsilon)}
\;.
\end{equation}

The quark field renormalization is
\begin{equation}
\begin{split}
\label{eq:Zqresult}
Z_\psi ={}& 1 -\frac{\as}{4\pi}\,\frac{S_\epsilon}{\epsilon}
\left[\frac{(v-1)^2}{v(v+1)}
+ \frac{\xi}{v}\right]C_\LF  + \cO(\as^2)
\;.
\end{split}
\end{equation}
The ghost field renormalization is
\begin{equation}
\label{eq:Zghostresult}
Z_\eta = 1 + \frac{\as}{4\pi}\, \frac{S_\epsilon}{\epsilon}
\left[\frac{16v^2+v+1}{12v(v+1)}-\frac{\xi}{4 v}\right] C_\LA
+ \cO(\as^2)
\;.
\end{equation}

The components of $A_a^\mu(x)$ along $n$ and orthogonal to $n$ renormalize differently. We find
\begin{equation}
\begin{split}
\label{eq:ZA}
Z_A^{\mu\nu} ={}& g^{\mu\nu} 
+ \frac{\as}{4\pi}\,\frac{S_\epsilon}{\epsilon}
\big[c_A(v,\xi)\, g^{\mu\nu} + \tilde{c}_A(v,\xi)\, h^{\mu\nu}\big]
\\&
+\cO(\as^2)
\;,
\end{split}
\end{equation}
where the coefficients are
\begin{equation}
\begin{split}
\label{eq:cAtildecA}
c_A(v,\xi) = {}& 
\left[\frac{22 v^3+35 v^2+20 v-1}{6v(v+1)^2}-\frac{\xi}{2 v}\right]C_\LA
\\&\quad
-\frac{4}{3}\,T_R n_\Lf
\;,\\
\tilde{c}_A(v,\xi) ={}& -\frac{4 v (2 v + 1) }{3(v+1)^2}\,C_\LA
\;.
\end{split}
\end{equation}
Using the projections $P_\pm$ along $n$ and orthogonal to $n$ defined in Eq.~(\ref{eq:Pplusminus}), $Z_A^{\mu\nu}$ is
\begin{equation}
\begin{split}
\label{eq:ZAalt}
Z_A^{\mu\nu} ={}& g^{\mu\nu} 
\\&
+ \frac{\as}{4\pi}\,\frac{S_\epsilon}{\epsilon}
\bigg[\!\left(c_A(v,\xi) + \frac{1}{v^2}\,\tilde{c}_A(v,\xi)\right)\!P_+^{\mu\nu}
\\&\quad
+ \big(c_A(v,\xi) + \tilde{c}_A(v,\xi)\big)P_-^{\mu\nu} \bigg]
+\cO(\as^2)
\;.
\end{split}
\end{equation}

The renormalization factors for the gauge parameters are determined from the renormalization of the gauge field:
\begin{equation}
\begin{split}
\label{eq:ZvZxi}
Z_v ={}& 1 - \frac{\as}{4\pi}\,\frac{S_\epsilon}{\epsilon}\,
\frac{v^2 - 1}{2v^2}\, \tilde{c}_A(v,\xi)  + \cO(\as^2)
\;,
\\
Z_\xi ={}& 1 + \frac{\as}{4\pi}\,\frac{S_\epsilon}{\epsilon}\,
\left[ c_A(v,\xi)
+ \frac{v^2+1}{2v^2}\, \tilde{c}_A(v,\xi)\right] 
\\&
+ \cO(\as^2)
\;.
\end{split}
\end{equation}
%

\begin{figure}[t]
    \begin{tikzpicture}[baseline=(a.base)]
      \begin{feynman}
        \vertex [dot] (a) {};
        \vertex [left=of a](c) {};
        \vertex at ($(a) + (2cm, 0cm)$)[dot] (d) {};
        \vertex [right=of d](e) {};  
        \diagram*{
          (a) -- [gluon, half left, momentum'={[arrow shorten=0.7]$p-q$}](d),
          (d) -- [gluon, half left, rmomentum'={[arrow shorten=0.7]$q$}](a),
          (c) -- [gluon, momentum'={$p$}](a), 
          (d) -- [gluon, momentum'={$p$}](e),
        };	
      \end{feynman}
    \end{tikzpicture}
    \begin{tikzpicture}[baseline=(a.base)]
      \begin{feynman}
        \vertex [dot] (a) {};
        \vertex at ($(a) - (2.5cm, 0cm)$)[](c) {};
        \vertex at ($(a) + (2.5cm, 0cm)$)[](e) {};  
        \diagram* {
          a -- [gluon, out=135, in=45, loop, min distance=3cm, 
          momentum'={[arrow shorten=0.7]$q$}] a;
          (c) -- [gluon, momentum={[arrow shorten=0.7]$p$}]
          (a) -- [gluon, momentum={[arrow shorten=0.7]$p$}] (e);
        };	
      \end{feynman}
    \end{tikzpicture}
    \begin{tikzpicture}[baseline=(a.base)]
      \begin{feynman}
        \vertex [dot] (a) {};
        \vertex [left=of a](c) {};
        \vertex at ($(a) + (2cm, 0cm)$)[dot] (d) {};
        \vertex [right=of d](e) {};  
        \diagram*{
          (d) -- [ghost, with arrow=0.5, half left, 
          rmomentum'={[arrow shorten=0.7]$q$}](a),
          (a) -- [ghost, with arrow=0.5, half left, 
          momentum'={[arrow shorten=0.7]$p-q$}](d),
          (c) -- [gluon, momentum'={$p$}](a), 
          (d) -- [gluon, momentum'={$p$}](e),
        };	
      \end{feynman}
    \end{tikzpicture}
    \begin{tikzpicture}[baseline=(a.base)]
      \begin{feynman}
        \vertex [dot] (a) {};
        \vertex [left=of a](c) {};
        \vertex at ($(a) + (2cm, 0cm)$)[dot] (d) {};
        \vertex [right=of d](e) {};  
        \diagram*{
          (a) -- [fermion, with arrow=0.5, half left, 
          momentum'={[arrow shorten=0.7]$p-q$}](d),
          (d) -- [fermion, with arrow=0.5, half left, 
          rmomentum'={[arrow shorten=0.7]$q$}](a),
          (c) -- [gluon, momentum'={$p$}](a), 
          (d) -- [gluon, momentum'={$p$}](e),
        };	
      \end{feynman}
    \end{tikzpicture}
\caption{
Gluon self-energy. There are four graphs: a gluon loop, a gluon tadpole, a ghost loop, and a quark loop.
\label{fig:GluonSE}}
\end{figure}
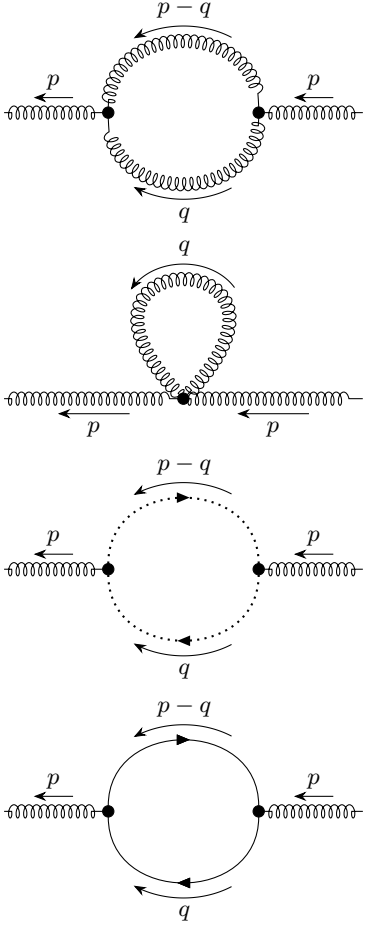

\subsection{Determination of the renormalization factors}
\label{sec:findingZs}

The renormalization factors $Z$  at order $\as$ can be determined by one-loop calculations. We illustrate this with the gluon propagator. Its renormalization is more complicated than in Feynman gauge because we are using a gauge condition that breaks manifest Lorentz invariance.

The inverse gluon propagator is $-\mi \Gamma^{\mu\nu}(p)$ with
\begin{equation}
\Gamma^{\mu\nu}(p) = 
\Gamma_\mathrm{tree}^{\mu\nu}(v,\xi;p)
- \Pi^{\mu\nu}(p)
\;.
\end{equation}
Here $\Pi^{\mu\nu}(p)$ is the gluon self-energy tensor, which we calculate at order $\as$ using the graphs in Fig.~\ref{fig:GluonSE}. The tree level inverse gluon propagator,
\begin{equation}
\label{eq:Gammatree0}
\Gamma_\mathrm{tree}^{\mu\nu}(v,\xi;p) = 
- g^{\mu\nu} p^2 + p^\mu p^\nu - \frac{v^2}{\xi}\,\tilde p^\mu \tilde p^\nu
\;,
\end{equation}
depends on the gauge parameters $v$ and $\xi$ through the factor $v^2/\xi$ and through the presence of $v$ in the definition (\ref{eq:hdef}) of $\tilde p^\mu = h^\mu_\nu p^\nu$.

The parameters $v^2$ and $\xi$ are renormalized and, in addition, $\Gamma^{\mu\nu}$ is renormalized according to
\begin{equation}
\Gamma^{\mu\nu} = [Z_A^{1/2}]^\mu_\alpha 
\Gamma_\LB^{\alpha\beta}[Z_A^{1/2}]^\nu_\beta 
\;,
\end{equation}
where
\begin{equation}
\Gamma^{\mu\nu}_\LB(p) = 
\Gamma_\mathrm{tree}^{\mu\nu}(v_\LB, \xi_\LB; p)
- \Pi_\LB^{\mu\nu}(p)
\;.
\end{equation}
We calculate $\Pi^{\mu\nu}_\LB$ in the bare theory from the diagrams in Fig.~\ref{fig:GluonSE}. However, in this calculation we substitute the renormalized versions of $g$, $v^2$, and $\xi$ for their bare versions since $\Pi^{\mu\nu}$ is already of order $\as$. Then $\Pi^{\mu\nu}_\LB$ thus calculated will contain ultraviolet (UV) poles $1/\epsilon$. The renormalization program will remove these poles.
 
We write $[Z_A^{1/2}]^{\mu\nu}$ to first order in the form
\begin{equation}
[Z_A^{1/2}]^{\mu\nu} = 
g^{\mu\nu} + \frac{1}{2} \delta Z_I\, g^{\mu\nu}
+ \frac{1}{2} \delta Z_h\, h^{\mu\nu} + \cO(\as^2)
\;.
\end{equation}
(The subscript $I$ on $\delta Z_I$ refers to the fact that the metric tensor acts as the identity operator on vectors: $g^\mu_\alpha p^\alpha = p^\mu$.) After using Eq.~(\ref{eq:hsquared}), we find for the counterterms from $Z_A$:
\begin{align}
\label{eq:Gammamunu}
\Gamma^{\mu\nu} ={}&
\Gamma^{\mu\nu}_\LB
-   (\delta Z_I g^{\mu\nu} + \delta Z_h h^{\mu\nu})p^2
\notag
\\&
+ \delta Z_I p^\mu p^\nu
+  \frac{\xi + 1}{2\xi}\delta Z_h
(p^\mu \tilde p^\nu + \tilde p^\mu p^\nu)
\\&
-\left[\frac{v^2}{\xi} \delta Z_I + \frac{v^2+1}{\xi} \delta Z_h\right]
\tilde p^\mu\tilde p^\nu 
+\cO(\as^2)
\;.
\notag
\end{align}

There are also counterterms from the renormalization of $v^2$ and $\xi$:
\begin{equation}
\begin{split}
v^2_\LB ={}& v^2 + \delta Z_v v^2 + \cO(\as)
\;,
\\
\xi_\LB ={}& \xi + \delta Z_\xi \xi + \cO(\as)
\;.
\end{split}
\end{equation}
\begin{widetext}
When we account for the direct appearance of $v^2$ and $\xi$ in Eq.~(\ref{eq:Gammatree0}) and the appearance of $v^2$ in $\tilde p^\lambda = h^\lambda_\alpha p^\alpha$, these relations give us
\begin{equation}
\begin{split}
\label{eq:GammaB}
\Gamma^{\mu\nu}_\LB ={}& \Gamma^{\mu\nu}_\mathrm{tree}(v_\LB,\xi_\LB; p) - \Pi^{\mu\nu}_\LB
\\={}& \Gamma^{\mu\nu}_\mathrm{tree}(v,\xi; p) - \Pi^{\mu\nu}_\LB
+
\frac{v^2}{\xi}\,\frac{1}{v^2-1}\,\delta Z_v
(p^\mu \tilde p^\nu + \tilde p^\mu p^\nu)
+ \frac{v^2}{\xi}\,
\left[\delta Z_\xi - \frac{v^2+1}{v^2-1}\,\delta Z_v\right]
\tilde p^\mu \tilde p^\nu
+\cO(\as^2)
\;.
\end{split}
\end{equation}
We substitute $\Gamma^{\mu\nu}_\LB$ from Eq.~(\ref{eq:GammaB}) into Eq.~(\ref{eq:Gammamunu}) for $\Gamma^{\mu\nu}$. Then $\Gamma^{\mu\nu} = \Gamma^{\mu\nu}_\mathrm{tree} - \Pi^{\mu\nu}$ is
\begin{equation}
\begin{split}
\label{eq:gluonSErenormalization}
\Gamma^{\mu\nu}_\mathrm{tree}(v,\xi; p) - \Pi^{\mu\nu} ={}&
\Gamma^{\mu\nu}_\mathrm{tree}(v,\xi; p) - \Pi^{\mu\nu}_\LB
- \delta Z_I (g^{\mu\nu} p^2 - p^\mu p^\nu) - \delta Z_h h^{\mu\nu}p^2
\\&
+  \left[\frac{\xi + 1}{2\xi}\delta Z_h
+ \frac{v^2}{\xi}\,\frac{1}{v^2-1}\,\delta Z_v
\right]
(p^\mu \tilde p^\nu + \tilde p^\mu p^\nu)
\\&
-\left[\frac{v^2}{\xi} \delta Z_I + \frac{v^2+1}{\xi} \delta Z_h
- \frac{v^2}{\xi}\,\delta Z_\xi
+\frac{v^2}{\xi}\,\frac{v^2+1}{v^2-1}\,\delta Z_v
\right]
\tilde p^\mu\tilde p^\nu 
+\cO(\as^2)\;.
\end{split}
\end{equation}
\end{widetext}
We note that the terms $\Gamma^{\mu\nu}_\mathrm{tree}(v,\xi; p)$ cancel in Eq.~(\ref{eq:gluonSErenormalization}). The renormalized one loop gluon self-energy, $\Pi^{\mu\nu}$, should not have ultraviolet poles.  We calculate the ultraviolet poles in $\Pi^{\mu\nu}_\LB(p)$ at order $\as$ from the diagrams in Fig.~\ref{fig:GluonSE}, giving
\begin{equation}
\begin{split}
\label{eq:gluonSEUV}
\Pi^{\mu\nu}_\LB(p) ={}&
\frac{\as}{4\pi}\,\frac{S_\epsilon}{\epsilon}
\Bigg\{
-c_A(v,\xi)
\big[p^2\,g^{\mu\nu} - p^\mu p^\nu\big]
\\&
- \tilde c_A(v,\xi)\,h^{\mu\nu} p^2
+\frac{\tilde c_A(v,\xi)}{2}\,
\big[p^\mu \tilde{p}^\nu + \tilde{p}^\mu p^\nu\big]
\Bigg\}
\\ &
\times \left(1 + \cO(\epsilon)\right) + \cO(\as^2)
\;,
\end{split}
\end{equation}
where $c_A(v,\xi)$ and $\tilde c_A(v,\xi)$ are given in Eq.~(\ref{eq:cAtildecA}). Evidently, the pole proportional to $p^2 g^{\mu\nu} - p^\mu p^\nu$ is removed if we choose
\begin{equation}
\delta Z_I =  \frac{\as}{4\pi}\,\frac{S_\epsilon}{\epsilon}\,c_A(v,\xi)
\;.
\end{equation}
The pole proportional to $h^{\mu\nu} p^2$ is removed if we choose
\begin{equation}
\delta Z_h =  \frac{\as}{4\pi}\,\frac{S_\epsilon}{\epsilon}\,\tilde c_A(v,\xi)
\;.
\end{equation}
Then the pole proportional to $p^\mu \tilde{p}^\nu + \tilde{p}^\mu p^\nu$ is removed if we choose 
\begin{equation}
\delta Z_v = -\frac{\as}{4\pi}\,\frac{S_\epsilon}{\epsilon}\,
\frac{v^2 - 1}{2 v^2}\,\tilde c_A(v,\xi)
\;.
\end{equation}
There is no pole proportional to $\tilde p^\mu \tilde p^\nu$. This can be arranged if we choose 
\begin{equation}
\delta Z_\xi = \frac{\as}{4\pi}\,\frac{S_\epsilon}{\epsilon}\,
\left[ c_A(v,\xi) 
+\frac{v^2 + 1}{2 v^2}\,\tilde c_A(v,\xi)
\right]
\;.
\end{equation}
This calculation gives the order $\as$ results given in Eq.~(\ref{eq:ZA}) and (\ref{eq:ZvZxi}). 

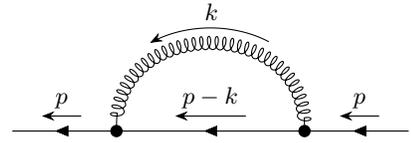
\begin{figure}[t]
   \begin{tikzpicture}[baseline=(a.base)]
      \begin{feynman}
        \vertex [dot] (a) {};
        \vertex [left=of a](c) {};
        \vertex at ($(a) + (2.5cm, 0cm)$)[dot] (d) {};
        \vertex [right=of d](e) {};  
        \diagram*{
          (d) -- [gluon, half right, rmomentum={[arrow shorten=0.7]$k$}](a),
          (e) -- [fermion, with arrow=0.5, rmomentum={$p$}](d) 
          -- [fermion, with arrow=0.5, rmomentum={[arrow shorten=0.7]$p-k$}](a) 
          -- [fermion, with arrow=0.5, rmomentum={$p$}](c),
        };	
      \end{feynman}
    \end{tikzpicture}
\caption{
Quark self-energy.
\label{fig:quarkSE}}
\end{figure}

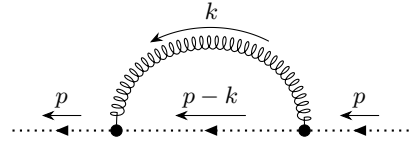
\begin{figure}[t]
  \begin{tikzpicture}[baseline=(a.base)]
      \begin{feynman}
        \vertex [dot] (a) {};
        \vertex [left=of a](c) {};
        \vertex at ($(a) + (2.5cm, 0cm)$)[dot] (d) {};
        \vertex [right=of d](e) {};  
        \diagram*{
          (d) -- [gluon, half right, rmomentum={[arrow shorten=0.7]$k$}](a),
          (e) -- [ghost, with arrow=0.5, rmomentum={$p$}](d) 
          -- [ghost, with arrow=0.5, rmomentum={[arrow shorten=0.7]$p-k$}](a) 
          -- [ghost, with arrow=0.5, rmomentum={$p$}](c),
        };	
      \end{feynman}
    \end{tikzpicture}
\caption{
Ghost self-energy.
\label{fig:ghostSE}}
\end{figure}

\begin{figure}[t]
  \begin{tikzpicture}[baseline={([yshift=-.5ex]current bounding box.center)}]
    \begin{feynman}
      \vertex [dot] (a) {};
      \vertex [dot, below left=of a](b) {};
      \vertex [dot, below right=of a](c) {};
      \vertex [above=of a] (a1) {};
      \vertex [below left=of b](b1) { };
      \vertex [below right=of c](c1) { };
      \diagram*{
        (a1)--[gluon,](a)
        -- [fermion,](b)
        -- [gluon,](c)
        -- [fermion,](a),
        (b1)--[anti fermion,](b),
        (c1)--[fermion,](c);
      };	
    \end{feynman}
  \end{tikzpicture}
  \begin{tikzpicture}[baseline={([yshift=-.5ex]current bounding box.center)}]
    \begin{feynman}
      \vertex [dot] (a) {};
      \vertex [dot, below left=of a](b) {};
      \vertex [dot, below right=of a](c) {};
      \vertex [above=of a] (a1) {};
      \vertex [below left=of b](b1) {};
      \vertex [below right=of c](c1) {};
      \diagram*{
        (a1)--[gluon, ](a)
        -- [gluon,](b)
        -- [anti fermion,](c)
        -- [gluon, ](a),
        (b1)--[anti fermion,](b),
        (c1)--[fermion,](c);
      };	
    \end{feynman}
  \end{tikzpicture}
\caption{
Quark-gluon vertex at one loop. There are two graphs.
\label{fig:quarkgluonvertex}}
\end{figure}
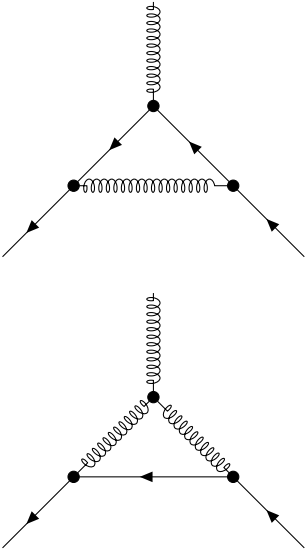

The result in Eq.~(\ref{eq:Zqresult}) for the quark field renormalization can be derived from the quark self-energy, Fig.~\ref{fig:quarkSE}, as outlined in Appendix \ref{sec:quarkpole}. The result in Eq.~(\ref{eq:Zghostresult}) for the ghost field renormalization can be derived from the ghost self-energy, Fig.~\ref{fig:ghostSE}, as outlined in Appendix \ref{sec:ghostpole}. The result in Eq.~(\ref{eq:Zgresult}) for the renormalization of $\as$ can be determined by calculating the one loop correction to the quark-gluon vertex, Fig.~\ref{fig:quarkgluonvertex}, as outlined in Appendix \ref{sec:quarkgluonvertex}.

These calculations determine all of the renormalization factors $Z$. As a check on these calculations, in Appendix \ref{sec:threegluonvertex} we calculate the $1/\epsilon$ poles in the one-loop three-gluon vertex function and verify that the factors $Z_A^{\mu\nu}$ and $Z_g$ provide the needed counterterms to cancel these poles.

\section{The gluon self-energy}
\label{sec:gluonselfenergy}

In Appendix \ref{sec:gluonSE}, we use the diagrams in Fig.~\ref{fig:GluonSE} to calculate the one loop gluon self-energy $\Pi^{\mu\nu}(p)$ with $\MSbar$ renormalization with $\xi = 1$ and $p^2 < 0$. We express the unrenormalized $\Pi^{\mu\nu}_\LU(p)$ as an integral over Feynman parameters and as an integral in $d = 4 - 2 \epsilon$ dimensions over the loop momentum. The integral over the loop momentum can then be performed analytically. This gives terms with UV poles $1/\epsilon$. The pole terms are reported in Eq.~(\ref{eq:gluonSEUV}). (In Eq.~(\ref{eq:gluonSEUV}), we have added a calculation of the UV poles for $\xi \ne 1$.) The UV poles are removed by $\MSbar$ renormalization, as described in Sec.~\ref{sec:renorm}. This gives the renormalized $\Pi^{\mu\nu}(p)$.

The gluon self-energy has the decomposition
\begin{equation}
\begin{split}
\label{eq:Pimunustructure}
\Pi^{\mu\nu}(p) ={}& \frac{\as}{4\pi}
\big\{
A_1 p^2 g^{\mu\nu}
+ A_2 p^2 h^{\mu\nu}
+ A_3   p^\mu p^\nu
\\&\quad
+ A_4   \tilde p^\mu \tilde p^\nu
+ A_5   (p^\mu \tilde p^\nu + \tilde p^\mu  p^\nu)
\big\}
\;.
\end{split}
\end{equation}
The coefficients $A_i$ are given as integrals over Feynman parameters. Using the Feynman rules to construct the graphs for the first order gluon self-energy function, we see that it obeys the identity
\begin{equation}
p_\mu \Pi^{\mu\nu}(p)\, p_\nu = 0
\;.
\end{equation}
This implies that there is a linear relation among the five coefficients $A_i$. When the $A_i$ are computed numerically, this relation provides a check on the calculation.

\begin{figure}
\begin{tikzpicture}
  \begin{axis}[
    title=\textsc{$\Pi^{\mu\nu}$ coefficients},
    xlabel={$p\cdot \tilde p/p^2$}, ylabel={$A_i$},
    legend cell align=left,
    every axis legend/.append style = {
      at={(0.02,0.98)},
      anchor=north west
    },
    xmin=1.0, xmax=3.0,
    ymin=-3.8, ymax=5.0
    ]
    
\pgfplotstableread{
1.0 -0.722265
1.2  0.0382404
1.4  0.725926
1.6  1.36084
1.8  1.95487
2.0  2.50909
2.2  3.03815
2.4  3.52888
2.6  3.99728
2.8  4.44519
3.0  4.85619
}\gmunu

\pgfplotstableread{
1.0  2.8739
1.2  2.46687
1.4  2.11078
1.6  1.78185
1.8  1.49086
2.0  1.21694
2.2  0.992492
2.4  0.759374
2.6  0.554099
2.8  0.36142
3.0  0.181543
}\pmupnu

\pgfplotstableread{
1.0  2.42389
1.2  2.36444
1.4  2.31148
1.6  2.26565
1.8  2.22689
2.0  2.19183
2.2  2.15731
2.4  2.12957
2.6  2.10402
2.8  2.08126
3.0  2.05782
}\hmunu

\pgfplotstableread{
1.0  1.15842
1.2  0.974602
1.4  0.839358
1.6  0.73872
1.8  0.659328
2.0  0.590411
2.2  0.542878
2.4  0.491678
2.6  0.449356
2.8  0.416384
3.0  0.386856
}\pTmupTnu

\pgfplotstableread{
1.0 -2.86875
1.2 -2.81294
1.4 -2.7638
1.6 -2.71022
1.8 -2.67137
2.0 -2.63057
2.2 -2.59324
2.4 -2.55565
2.6 -2.51767
2.8 -2.4821
3.0 -2.449
}\pmupTnu

    \addplot[black,semithick] table {\gmunu};
    \addplot[red,semithick] table {\pmupnu};
    \addplot[blue,semithick] table {\hmunu};
    \addplot[darkgreen,semithick] table {\pTmupTnu};
    \addplot[purple,semithick] table {\pmupTnu};

\node[anchor=west, fill=none] (source) at (axis cs:2.1,3.6)
{$p^2 g^{\mu\nu}$};

\node[anchor=west, fill=none] (source) at (axis cs:2.1,2.5)
{\textcolor{blue}{$p^2 h^{\mu\nu}$}};

\node[anchor=west, fill=none] (source) at (axis cs:2.1,1.4)
{\textcolor{red}{$p^\mu p^\nu$}};

\node[anchor=west, fill=none] (source) at (axis cs:2.1,0.2)
{\textcolor{darkgreen}{$\tilde p^\mu \tilde p^\nu$}};

\node[anchor=west, fill=none] (source) at (axis cs:2.1,-2.9)
{\textcolor{purple}{$p^\mu \tilde p^\nu + \tilde p^\mu p^\nu$}};

  \end{axis}
\end{tikzpicture}
\caption{
Coefficients $A_i(p\cdot\tilde p/p^2)$ in $\Pi^{\mu \nu}$ with $v = 2$, $\xi = 1$. We take $p^2 < 0$, so that also $p \cdot \tilde p < 0$. The renormalization scale is set to $\mu^2 = |p\cdot \tilde p|$. The number of quark flavors is $n_\Lf = 3$. 
\label{fig:Pimunu}}
\end{figure}
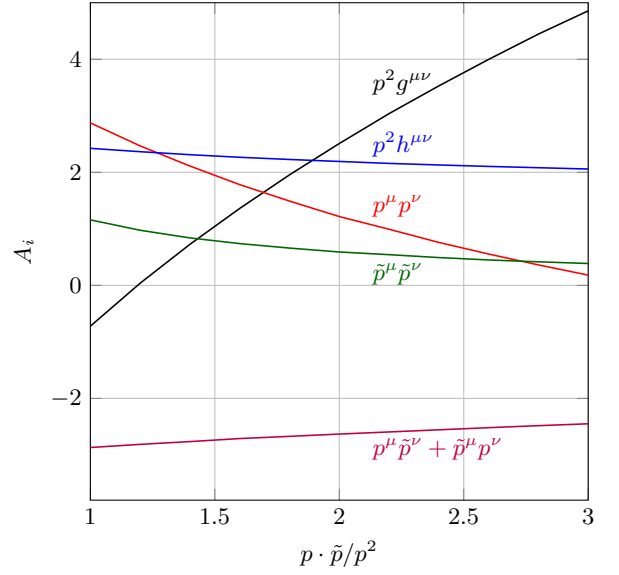

These integrals can be performed by numerical integration. We choose $p^2 < 0$. Since $p\cdot \tilde p < p^2$, we also have $p\cdot \tilde p < 0$. We set the renormalization scale to $\mu^2 = p\cdot \tilde p$. The coefficients $A_i$ are dimensionless, so they are functions of $p\cdot \tilde p/p^2$. We show results in Fig.~\ref{fig:Pimunu} for the choice $v = 2$ and $n_\Lf = 3$.

Of special importance are S-matrix elements involving an initial or final state T gluon. In this case, the final gluon propagator is amputated and we take $p^2 \to 0$ and multiply by a polarization vector $\varepsilon^\mu(p,s)$.  According to Eq.~(\ref{eq:Smatrix}), in the case that there is a self-energy insertion on the gluon line, we multiply the one particle irreducible subgraph by the LSZ factor $\sqrt{R_\Lg}$. Using Eq.~(\ref{eq:Rgdef}) for  $R_\Lg$, we derive
\begin{equation}
\sqrt{R_\Lg(p)} - 1 = 
\left[\frac{\varepsilon_\mu \Pi^{\mu\nu}\varepsilon_\nu}
{2p^2}\right]_{p^2 \to 0}
+ \cO(\as^2)
\;.
\end{equation}
This quantity has IR divergences. We define $R_\Lg$ from the renormalized $\Pi^{\mu \nu}(p)$ by taking the $p^2 \to 0$ limit with dimensional regulation to control the IR divergences. Then we take $\epsilon \to 0$, obtaining $1/\epsilon^2$ and $1/\epsilon$ IR poles plus finite terms. The calculation, with $\xi = 1$, is described in Appendix \ref{sec:gluonSEP0}. The result is given in Eq.~(\ref{eq:PiTRresult}). The result for the pole terms, from Eq.~(\ref{eq:PiTRpolesresult}), is
\begin{equation}
\begin{split}
\label{eq:PiTRpolesbis}
\left[\frac{\varepsilon_\mu \Pi^{\mu\nu}\varepsilon_\nu}
{2p^2}\right]_{p^2 \to 0}\hskip - 1.5 cm &\\
 ={}&-\frac{\as}{4\pi}\, \frac{S_\epsilon}{\epsilon^2}\,C_\LA
 -\frac{\as}{4\pi}\, \frac{S_\epsilon}{\epsilon}\,\gamma_\Lg
 \\&
 -\frac{\as}{4\pi}\, \frac{S_\epsilon}{\epsilon}\,C_\LA
\Bigg[
\frac{v-1}{v}
- \log\!\left(\frac{v-1}{v+1}\right)
\\&\quad
+ \log\!\left(\frac{\mu^2}{4(p\cdot n)^2}\right)
\Bigg]
+ \cO(\epsilon^0)
\;.
\end{split}
\end{equation}
Here $\gamma_\Lg$ is the standard coefficient given in Eq.~(\ref{eq:gammag}). We will combine this result with results from gluon exchange graphs and from quark self-energy graphs in Sec.~\ref{sec:Spoles}.

\section{The quark self-energy}
\label{sec:quarkselfenergy}

The one-loop quark self-energy $-\mi \Sigma(p)$ is given by the Feynman diagram shown in Fig.~\ref{fig:quarkSE}. It can be calculated using using similar techniques to those used for the gluon self-energy, although the calculation is simpler than for the gluon case. One can first find the UV poles of the unrenormalized one-loop self-energy, $\Sigma_\LU(p)$, with the result (Eq.~(\ref{eq:quarkSEpole}))
\begin{equation}
\begin{split}
\Sigma_\LU(p) ={}& -\frac{\as}{4\pi}\,\frac{S_\epsilon}{\epsilon}\,\s{p}\,C_\LF 
\left[\frac{(v-1)^2}{v(v+1)}
+ \frac{\xi}{v}\right]
+ \cO(\epsilon^0)
\;.
\end{split}
\end{equation}
This is the pole that is removed by renormalization and gives the renormalization factor $Z_\Lq$ given in Eq.~(\ref{eq:Zqresult}).

For S-matrix elements involving an initial or final state quark or antiquark, we need the LSZ factor $\sqrt{R_\Lq}$. For this, we need $\Sigma(p)$ in the limit $p^2 \to 0$. Using Eq.~(\ref{eq:Rqdef2}) and the accompanying definition of the Dirac spinor factors, we have
\begin{equation}
\sqrt{R_\Lq(p)} - 1 = \left[\frac{\bar u(p,s) \Sigma(p)
u(p,s)}{2 p^2}\right]_{p^2 \to 0}\!\! +\, \cO(\as^2)
\;.
\end{equation}
We calculate this with $\xi = 1$ with a calculation that is similar to the calculation in Appendix \ref{sec:gluonSEP0} for gluons, but somewhat simpler. As for gluons, the result contains infrared double and single poles. The result for the pole terms is
\begin{equation}
\begin{split}
\label{eq:quarkSEpoles}
\left[\frac{\bar u(p,s) \Sigma(p)
u(p,s)}{2 p^2}\right]_{p^2 \to 0}\hskip - 3 cm &
\\
={}& -\frac{\as}{4\pi}\,
\frac{S_\epsilon}{\epsilon^2}\,
C_\LF
-\frac{\as}{4\pi}\,
\frac{S_\epsilon}{\epsilon}\,
\gamma_\Lq
\\& -\frac{\as}{4\pi}\, \frac{S_\epsilon}{\epsilon}\,C_\LF
\Bigg[
\frac{v-1}{v}
- \log\!\left(\frac{v-1}{v+1}\right)
\\&\quad
+\log\!\left(\frac{\mu^2}{4(p\cdot n)^2}\right)
\Bigg]
+ \cO(\epsilon^0)
\;.
\end{split}
\end{equation}
Here $\gamma_\Lq$ is the standard coefficient
\begin{equation}
\label{eq:gammaq}
\gamma_\Lq = \frac{3 C_\LF}{2}
\;.
\end{equation}
We will combine this result with results from gluon exchange graphs and from gluon self-energy graphs in Sec.~\ref{sec:Spoles}.

\section{Poles of the S matrix}
\label{sec:Spoles}

We have seen that the S matrix for initial or final state external partons has infrared poles in $\epsilon$ when we include one virtual loop. Each external parton has a label $l$ and can be a T gluon or a massless quark or antiquark. One source of poles is the exchange of a gluon between two of the external partons. From Eq.~(\ref{eq:Sexchange}), this contribution is
\begin{equation}
\begin{split}
\label{eq:Sexchanges}
\frac{1}{2}\sum_l &\sum_{k \ne l} V_{lk}\,
\bm T_l \cdot \bm T_k
\\={}& -\sum_l \sum_{k \ne l} 
\bm T_l \cdot \bm T_k\,
\frac{\as}{4\pi}\,\frac{S_\epsilon}{\epsilon}
\log\!\left(\!\frac{-2p_l\cdot p_k + \mi 0}{\mu^2}\!\right)
\\ &
+ \sum_l \bm T_l^2\,
\frac{\as}{4\pi}\,\frac{S_\epsilon}{\epsilon}
\Bigg\{\frac{v-1}{v}
- \log\!\left(\frac{v-1}{v+1}\right)
\\&\quad
+ \log\!\left(\frac{\mu^2}{4\, (p_l\cdot n)^2}\!\right)
\Bigg\}
+ \cO(\epsilon)
\;.
\end{split}
\end{equation}
The other source of poles is the self-energy graphs, each of which contributes $\sqrt{R_l} - 1$ at order $\as$. These contributions are given in Eq.~(\ref{eq:PiTRpolesbis}) for gluons and Eq.~(\ref{eq:quarkSEpoles}) for quarks. They each include a constant $\gamma_l$, with $\gamma_l = \gamma_\Lg$, Eq.~(\ref{eq:gammag}), if $l$ is a gluon and $\gamma_l = \gamma_\Lq$, Eq.~(\ref{eq:gammaq}), if $l$ is a quark. Otherwise, they have the same form, with $\bm T_l^2 = C_\LA$ if $l$ is a gluon and $\bm T_l^2 = C_\LF$ if $l$ is a quark. The result is
\begin{equation}
\begin{split}
\label{eq:SselfEs}
\sum_l (\sqrt{R_l}& - 1)\\
 ={}&- \frac{\as}{4\pi}\, 
 \sum_l\left[\bm T_l^2\,\frac{S_\epsilon}{\epsilon^2}
 + \frac{S_\epsilon}{\epsilon}\gamma_l\right]
\\ &
- \sum_l \bm T_l^2\,
\frac{\as}{4\pi}\,\frac{S_\epsilon}{\epsilon}
\Bigg\{\frac{v-1}{v}
- \log\!\left(\frac{v-1}{v+1}\right)
\\&\quad
+ \log\!\left(\frac{\mu^2}{4\, (p_l\cdot n)^2}\!\right)
\Bigg\}
+ \cO(\epsilon)
\;.
\\ \\ \\
\end{split}
\end{equation}

In the sum of these contributions, the terms involving $v$ and $p_l\cdot n$ cancel, leaving
\begin{equation}
\begin{split}
\label{eq:SselfEs2}
\sum_l &(\sqrt{R_l} - 1) + \frac{1}{2}\sum_l \sum_{k \ne l} V_{lk}\,
\bm T_l \cdot \bm T_k
\\
 ={}&
 - \frac{\as}{4\pi}\, 
 \sum_l\left[\bm T_l^2\,\frac{S_\epsilon}{\epsilon^2}
 + \frac{S_\epsilon}{\epsilon}\gamma_l\right]
 \\&
-\sum_l \sum_{k \ne l} 
\bm T_l \cdot \bm T_k\,
\frac{\as}{4\pi}\,\frac{S_\epsilon}{\epsilon}
\log\!\left(\!\frac{-2p_l\cdot p_k + \mi 0}{\mu^2}\!\right)
\\&
+ \cO(\epsilon^0) + \cO(\as^2)
\;.
\end{split}
\end{equation}
This is the standard result that one finds in Feynman gauge.

\section{Conclusions}
\label{sec:conclusions}

We have investigated the features of interpolating gauge in QCD or other gauge field theories. This gauge was proposed by Doust \cite{Doust} and Baulieu and Zwanziger \cite{BaulieuZwanziger} as a way to interpolate between a covariant gauge and Coulomb gauge, with the aim of making Coulomb gauge better defined. The attraction of this gauge for us is that it may be useful for QCD constructions in which the infrared singularities of the theory are of paramount interest.

In any gauge, QCD with massless quarks has infrared singularities that appear when a gluon with momentum $q$ attaches to an on-shell parton with momentum $p$. First, there is a soft singularity when $q \to 0$. Second, there is a collinear singularity when $q \to \lambda p$ with a fixed value of $\lambda$. There are also soft$\times$collinear singularities when $q \to \lambda p$ with $\lambda \to 0$. These singularities are important for the subtractions needed in the calculation of cross sections for infrared safe observables in high energy processes. The same singularities are important for the analysis of large logarithms that appear when a process has a hard momentum scale and a much smaller soft momentum scale. The infrared singularities control the large logarithms that appear when one takes the soft scale toward zero. One way of summing these large logarithms is through the use of a parton shower algorithm. It is the extension of parton shower algorithms to use higher order parton splitting functions that provides our principle motivation for investigating interpolating gauge.

As presented in Sec.~\ref{sec:interpolatinggauge}, interpolating gauge depends on a vector $n$ that defines a preferred reference frame and on two parameters, $v$ and $\xi$, with $1 \le v < \infty$. With $v = 1$, we have a standard covariant gauge, Feynman gauge for $\xi = 1$ and Lorenz gauge for $\xi = 0$. We obtain Coulomb gauge for $v \to \infty$ with any $\xi$. We always choose $v > 1$ except when we want to connect to Feynman gauge. We mostly use $\xi = 1$ in this paper.

QCD calculations typically use Feynman gauge because of its simplicity. However, in Feynman gauge, there is a problem with collinear singularities. A collinear singularity and a soft$\times$collinear singularity can appear in graphs in which the gluon couples to a far off-shell internal line in a graph. After integrating over the gluon momentum in $4 - 2 \epsilon$ dimensions, one obtains poles $1/\epsilon^2$ and $1/\epsilon$. These poles are unphysical in the sense that they result from the gluon polarization being proportional to its momentum $q$, while a physical gluon has polarization $\varepsilon$ that is orthogonal to $q$. Graphs in which a gluon is exchanged between two external partons also have a double pole in Feynman gauge. Again, the double pole is unphysical in that it results from the gluon polarization being proportional to its momentum. One must use Ward identities to sort out these effects, as outlined in Appendix \ref{sec:FeynmanGauge}. After using Ward identities, the collinear singularities are effectively associated with self-energy insertions on the external lines.

One could eliminate the collinear singularity problem by choosing a physical gauge, for instance Coulomb gauge. However, as emphasized in Ref.~\cite{BaulieuZwanziger}, one then faces a problem with an ambiguity in defining the gauge. To remove this ambiguity in Coulomb gauge, Ref.~\cite{BaulieuZwanziger} takes the $v \to \infty$ limit of interpolating gauge.

For the uses that we have in mind, there is no need to take a limit $v \to \infty$. As long as $v > 1$, the problem with collinear singularities is removed. Any finite value of $v$ that is not too close to 1 will do. For instance, one can choose $v = 2$.

The attraction of interpolating gauge with $v > 1$ compared to Feynman gauge is that {\em for each graph} the collinear singularity problem does not occur. One does not need to apply Ward identities and sum over graphs to bring the collinear singularities into the form of self-energy insertions on the external lines. They have this form from the beginning. We verified this explicitly at one loop order in Secs.~\ref{sec:gluonselfenergy} and \ref{sec:quarkselfenergy}.

The properties of the $v>1$ theory that are important for us can be easily understood, as we found in Sec.~\ref{sec:TandLgluons}. The four component gluon field $A^\mu(x)$ describes two sorts of gluons, each with two components: T gluons and L gluons. The T gluons are transversely polarized and propagate with the speed of light, $c=1$. The L gluons carry the remaining two polarizations and, in the tree-level propagator, propagate with speed $v$ in a frame with $n = (1,0,0,0)$. In such a frame, an on-shell L gluon with momentum $q$ has $|q^0| = v |\vec q\,|$. This makes it impossible for an on-shell L gluon to be collinear with an on-shell lightlike particle, with $|p^0| = |\vec p\,|$. This property eliminates the collinear singularities for L gluons.
 
We have provided analysis and calculations for some of the important features of interpolating gauge:

\begin{itemize}

\item Ref.~\cite{BaulieuZwanziger} argues that the renormalization program works at all orders of perturbation theory in this gauge. In Sec.~\ref{sec:renorm}, we define the needed renormalization factors $Z$ and calculate their order $\as$ contributions from the ultraviolet divergences of one loop graphs.

\item BRST invariance leads to identities for Green functions in interpolating gauge. Ref.~\cite{BaulieuZwanziger} used BRST invariance to analyze the renormalization program. In Sec.~\ref{sec:brst}, we derive identities for the change in Green functions induced by changing the vector $n$ and the parameters $v$ and $\xi$. We then use these identities to show that the S matrix for quarks and T gluons is invariant under changes of $n$, $v$, and $\xi$.

\item The gluon propagator in interpolating gauge is not Lorentz covariant and has several terms. Consequently, calculations of loop diagrams are not as simple as in Feynman gauge. Nevertheless, we found in Appendix \ref{sec:gluonSE} that results can be obtained using Feynman parameterization and numerical integration. Results for the one loop gluon self-energy were shown in Fig.~\ref{fig:Pimunu}.

\item Poles $1/\epsilon$ and $1/\epsilon^2$ arising from infrared singularities are of particular interest in this paper. Results for gluon exchange between external on-shell partons were presented in Sec.~\ref{sec:exchange}. Results for the self-energy of an on-shell T gluon were presented in Appendix \ref{sec:gluonSEP0} and Sec.~\ref{sec:gluonselfenergy}. Results for the self-energy of an on-shell quark were presented in Sec.~\ref{sec:quarkselfenergy}.

\end{itemize}

We conclude that there may be practical uses for interpolating gauge.

\acknowledgments{ 
This work was supported in part by the United States Department of Energy under grant DE-SC0011640.  It was also supported in part by the Munich Institute for Astro-, Particle and BioPhysics, which is funded by the Deutsche Forschungsgemeinschaft under Germany's Excellence Strategy -- EXC-2094--390783311.
}

\appendix 

\section{Infrared singularities in Feynman gauge}
\label{sec:FeynmanGauge}

In interpolating gauge, all the collinear singularities are located in the self-energy graphs on the external legs. In Feynman gauge, these collinear singularities appear in many graphs and one has to utilize the Ward identity to be able to
factorize them out. This prevents us from being able to define infrared singular functions that match the soft and collinear singularities with one singular function for each graph. In this section we demonstrate these complications at one-loop level.

In interpolating gauge it was natural to decompose the gluon propagator as sum of T and L gluons. In Feynman gauge, we use a different decomposition. We try to separate the pure soft modes, labelled S, from the collinear and soft-collinear modes, labelled C. We write

\begin{equation}
  \label{eq:gluon-prop-feynman}
  \begin{split}
    D^{\mu\nu}(q) = {}& -\frac{g^{\mu\nu}}{q^2+\mi0}
    = \frac{P_\LS^{\mu\nu}(q)}{q^2+\mi0}  + \frac{P_\LC^{\mu\nu}(q)}{q^2+\mi0}
  \,.
  \end{split}
\end{equation}
Here the $P_\LS^{\mu\nu}(q)$ and $P_\LC^{\mu\nu}(q)$ tensors are defined as
\begin{equation}
  \begin{split}
    P_\LS^{\mu\nu}(q) = {}& P_T^{\mu\nu}(q) 
    + \frac{q^\mu q^\nu + q^2 n^\mu n^\nu}{(q\!\cdot\!n)^2-q^2}
    \\
    = {}& -g^{\mu\nu} + \frac{q\!\cdot\!n\,(q^\mu n^\nu 
    + n^\mu q^\nu)}{(q\!\cdot\!n)^2-q^2}
    \;,
    \\
    P_\LC^{\mu\nu}(q) = {}&P_L^{\mu\nu}(q) 
    - \frac{q^\mu q^\nu + q^2 n^\mu n^\nu}{(q\!\cdot\!n)^2-q^2}
    \\
    = {}& -\frac{q\!\cdot\!n\,(q^\mu n^\nu + n^\mu q^\nu)}
    {(q\!\cdot\!n)^2-q^2}
    \,.
  \end{split}
\end{equation}
Here $(q\cdot n)^2-q^2 = \vec{q}^{\,2}$ if we work in a frame in which $n = (1,0,0,0)$. When an internal gluon line becomes on-shell, $q^2\to 0$, both parts of the propagator become singular. However, as we shall see,  $P_\LS^{\mu\nu}(q)$ and $ P_\LC^{\mu\nu}(q)$ contribute differently to soft and to collinear singularities.

\begin{figure}[t]
\begin{equation*}
    \sum_l \int\frac{d^d q}{(2\pi)^d}\sum_G\;\;
    \begin{tikzpicture}[baseline=(a.base)]
      \begin{feynman}[]
        \vertex[ellipse blob,fill=none, minimum width=1cm, minimum height=2cm] (b)
        {\rotatebox[origin=c]{90}{$R_l(G)$}};
        \vertex [dot] at ($(b) + (0:1.5cm)$)  (a) {};
        \vertex [right=1cm of a] (c) {};
        \diagram*{
          (a)--[gluon, quarter left, momentum'={$q$}](b.south east);
          (b.north east) --[cut={[]0.4},rmomentum'={$p_l-q$}](a)--[-|,rmomentum={$p_l$}](c);
        };
      \end{feynman}
    \end{tikzpicture}
\end{equation*}
\caption{
Illustration of the right hand side of Eq.~(\ref{eq:1-loop-singular-graphs}). The external line terminated by ``$\;|\,$'' indicates a polarization vector $\epsilon$ or a Dirac spinor $u$, $\bar u$, $v$ or $\bar v$. The double terminations ``$\;||\,$'' indicate that the line is approximated as being on shell, with two polarization vectors or Dirac spinor factors.
\label{fig:Feynman1}}
\end{figure}
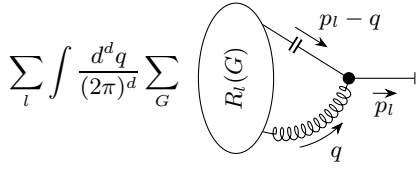

At 1-loop level the possible singular graphs with $m$ external partons can be written in the form illustrated in Fig.~\ref{fig:Feynman1},
\begin{equation}
  \label{eq:1-loop-singular-graphs}
  \begin{split}
    \big|M^{(1)}(\{p,&{}f\}_m)\big\rangle \\
    = {}&\mi g\mu^{\epsilon}
    \sum_l \int\frac{d^dq}{(2\pi)^d}\,
    \frac{\bm{T}^a_l J_l^\mu(q)\, D_{\mu\nu}(q)}
    {(p_l-q)^2+\mi0}
    \\
    &\quad\times\sum_{G}\brax{\nu, a}\ket{R_l(G; \{p,f\}_m)}
    \\
    &+ \cdots
    \;.
  \end{split} 
\end{equation}
The ellipsis ``$\cdots$'' here stands for contributions that do not have leading infrared singularities. In the singular part, external parton $l$, which can be either a quark or a gluon, absorbs a gluon with momentum $q$ and color index $a$. Its propagator is $D_{\mu\nu}(q)$. We sum over $l$ and integrate over $q$. The gluon couples to a color matrix $\bm{T}^a_l$ and an effective current $J_l^\mu(q)$, described below. The propagator for parton $l$ before the emission has a factor $1/[(p_l-q)^2 + \mi 0]$, which is singular when $q$ is collinear to $p_l$ or soft. The current $J_l^\mu(q)$ includes either a polarization vector or a Dirac spinor appropriate for the limit $(p_l-q)^2 \to 0$. The complementary polarization vector or spinor is included in the rest of the graph. These on-shell polarization vectors or spinors are indicated by the orthogonal lines on parton line $l$.

The amplitude $\brax{\nu,a}\ket{R_l(G;  \{p,f\}_m)}$ represents the rest of the Feynman graph that contributes to the given process. This is an $m+1$ point graph in Feynman gauge. The extra leg is a gluon line with color $a$ and polarization index $\nu$. This extra gluon carries the momentum $q^\mu$. The external leg $l$ of $|R_l(G; \{p,f\}_m)\rangle$ caries a momentum of $p_l-q$, which is treated as being an on-shell external parton with its appropriate polarization vector or spinor. The dependence on the polarizations or spins of parton $l$ is suppressed in the notation.

In $\brax{\nu,a}\ket{R_l(G; \{p,f\}_m)}$, the gluon with momentum $q$ is connected to any internal line in the graph or to any external leg of the graph except for the leg of parton $l$.

The currents $J_l(q)$ are matrices $J_l(q)_{\hat s,s}$ in the spin space for parton $l$. They also depend on the timelike reference momentum $n$ and the momentum and flavor of the parton, $p_l,f_l$, but for the sake of simplicity we hide those arguments and the spin indices. We define 
\begin{equation}
  \label{eq:soft-current}
  J_l^\mu(q)_{\hat s,s} = 2p_l^\mu \delta_{\hat s,s} 
  - A^{\mu\nu}_{f_l}(p_l)_{\hat s,s}\, q_\nu
  \;.
\end{equation}
Here the first term appears in the eikonal approximation for soft gluon absorption, while the second term is flavor dependent and has non-trivial dependence on the spins. 

For a final state quark, we define \cite{NSI}
\begin{equation}
  A^{\mu\nu}_\Lq(p)_{\hat s,s} =
  \frac{\bar u(p,s)\gamma^\mu\gamma^\nu\s{n}\,u(p,\hat s)}{2p\!\cdot\!n}
  \;.
\end{equation}
Analogous formulas apply for antiquarks and for initial state quarks and antiquarks. 

When parton $l$ is a gluon, an extra step is needed \cite{NSsubtractions}. We have a gluon with momentum that we can call $p_\La$ and Lorentz index $\alpha$ and a gluon with momentum that we can call $p_\Lb$ and Lorentz index $\beta$ combining at a three-gluon vertex to make an on-shell gluon with momentum $p = p_\La + p_\Lb$ and polarization vector $\varepsilon(p,s)$. There are three terms in the three-gluon vertex, Eq.~(\ref{eq:threegluonvertex}). One term is proportional to $g_{\alpha\beta} (p_\Lb - p_\La)\cdot \varepsilon(p,s)$. Since $p\cdot \varepsilon(p,s) = 0$, this term is nonsingular when $p_\La$ and $p_\Lb$ are collinear or either of them is soft. We neglect this term. One term is proportional to $(p + p_\La)_\beta\, \varepsilon_\alpha(p,s)$. In this term, we identify $q$ in Fig.~\ref{fig:Feynman1} with $p_\Lb$. The remaining term is proportional to $-(p + p_\La)_\alpha\, \varepsilon_\beta(p,s)$. In this term, we identify $q$ in Fig.~\ref{fig:Feynman1} with $p_\La$. With this identification, we have
\begin{equation}
  \begin{split}
   A^{\mu\nu}_\Lg(p)_{\hat s,s} = {}& g^{\mu\nu} \delta_{s\hat s}
    \;.
  \end{split}
\end{equation}
%
%

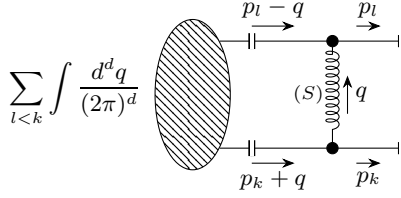
\begin{figure}[t]
\begin{equation*}
   \sum_{l<k} \int\frac{d^d q}{(2\pi)^d}\;\;
    \begin{tikzpicture}[baseline=(b.base)]
      \begin{feynman}[]
        \vertex[ellipse blob, minimum width=1cm, minimum height=2cm] (b){};
        \vertex [dot] at ($(b.north east) + (0:1.5cm)$)  (a1) {};
        \vertex [dot] at ($(b.south east) + (0:1.5cm)$)  (a2) {};
        \vertex [right=1cm of a1] (c1) {};
        \vertex [right=1cm of a2] (c2) {};
        \diagram*{
          (a1)--[gluon, momentum'={$q$},edge label'={\rotatebox[origin=c]{0}{\scriptsize $(S)$}}](a2);
          (b.north east) --[cut={[]0.3},rmomentum'={$p_l-q$}](a1)--[-|,rmomentum'={$p_l$}](c1);
          (b.south east) --[cut={[]0.3},rmomentum={$p_k+q$}](a2)--[-|,rmomentum={$p_k$}](c2);
        };
      \end{feynman}
    \end{tikzpicture}
\end{equation*}
\caption{
Illustration of the first term on the right-hand-side of Eq.~(\ref{eq:1-loop-singular-graphs-decomposed1}), in which a gluon with propagator  $P_S^{\mu\nu}(q)/[q^2 + \mi 0]$ is exchanged between two partons, $l$ and $k$.
\label{fig:Feynman2}}
\end{figure}
%

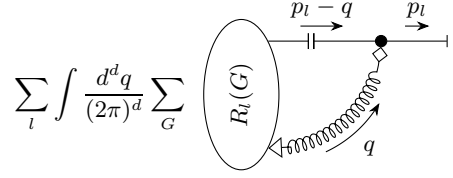
\begin{figure}[t]
\begin{equation*}
   \sum_l \int\frac{d^d q}{(2\pi)^d}\sum_G\;\;
    \begin{tikzpicture}[baseline=(b.base)]
      \begin{feynman}[]
        \vertex[ellipse blob,fill=none, minimum width=1cm, minimum height=2cm] (b)
        {\rotatebox[origin=c]{90}{$R_l(G)$}};
        \vertex [dot] at ($(b.north east) + (0:1.5cm)$)  (a) {};
        \vertex [right=1cm of a] (c) {};
        \diagram*{
          (b.south east)--[{Triangle[angle=75:6.5pt,open,flex]}-{Turned Square[length=6.pt,open,flex]},
          gluon, decoration={pre length=3mm, post length=2.5mm}, quarter right, rmomentum={$q$}](a);
          (b.north east) --[cut={[]0.4},rmomentum'={$p_l-q$}](a)--[-|,rmomentum'={$p_l$}](c);
        };
      \end{feynman}
    \end{tikzpicture}
\end{equation*}
\caption{
Illustration of the second term on the right-hand-side of Eq.~(\ref{eq:1-loop-singular-graphs-decomposed1}), in which a gluon with propagator  $- q\cdot n\, n^\mu q^\nu/[((q\cdot n)^2 - q^2)(q^2 + \mi 0)]$ is exchanged between parton $l$ and the rest of the graph.
\label{fig:Feynman3}}
\end{figure}
%

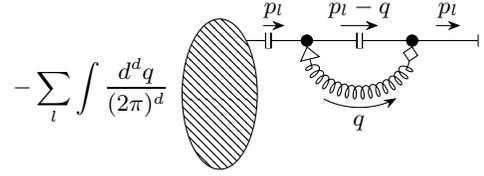
\begin{figure}[t]
\begin{equation*}
- \sum_l \int\frac{d^d q}{(2\pi)^d}\;\;
    \begin{tikzpicture}[baseline=(b.base)]
      \begin{feynman}[]
        \vertex[ellipse blob, minimum width=1cm, minimum height=2cm] (b){};
        \vertex [dot] at ($(b.north east) + (0:0.8cm)$)  (a1) {};
        \vertex [dot] at ($(b.north east) + (0:2.2cm)$)  (a2) {};
        \vertex [right=1cm of a2] (c) {};
        \diagram*{
          (a1)--[{Triangle[angle=75:6.5pt,open,flex]}-{Turned Square[length=6.pt,open,flex]},
          gluon, decoration={pre length=3mm, post length=2.5mm},
           half right, rmomentum={$q$}](a2);
          (b.north east) --[cut={[]0.4},rmomentum'={$p_l$}](a1)
          --[cut={[]0.5},rmomentum'={$p_l-q$}](a2)--[-|,rmomentum'={$p_l$}](c);
        };
      \end{feynman}
    \end{tikzpicture}
\end{equation*}
\caption{
Illustration of the second term on the right-hand-side of Eq.~(\ref{eq:1-loop-singular-graphs-decomposed2}). This is the negative of the one term that was not included in Fig.~\ref{fig:Feynman3}.
\label{fig:Feynman4}}
\end{figure}
%

The expression in Eq.~\eqref{eq:1-loop-singular-graphs} is not immediately usable if we want to define subtraction scheme, since it does not have a factorized form. To obtained such a form we should separate the contributions that apply in the soft and collinear limits. Using the decomposition of the gluon propagator in Eq.~\eqref{eq:gluon-prop-feynman} we
have
\begin{equation}
  \label{eq:1-loop-singular-graphs-decomposed}
  \begin{split}
    \big|M^{(1)}&{}(\{p,f\}_m)\big\rangle \\
    = {}&\mi g\mu^{\epsilon}
    \sum_l  \int\frac{d^dq}{(2\pi)^d}\,
    \frac{\bm{T}^a_l J_l^\mu(q)\,[P_\LS(q)]_{\mu\nu} }
    {[q^2+\mi0][(p_l-q)^2+\mi0]}
    \\
    &\quad\times\sum_{G}\brax{\nu, a}\ket{R_l(G; \{p,f\}_m}
    \\
    &+ \mi g\mu^{\epsilon}
    \sum_l  \int\frac{d^dq}{(2\pi)^d}\,
    \frac{\bm{T}^a_l  J_l^\mu(q)\,[P_\LC(q)]_{\mu\nu}}
    {[q^2+\mi0][(p_l-q)^2+\mi0]}
    \\
    &\quad\times\sum_{G}\brax{\nu, a}\ket{R_l(G; \{p,f\}_m)}
    \\
    &+ \cdots
    \;.
  \end{split} 
\end{equation}
We note that the first term cannot be singular in the collinear limit, in which $q$ becomes parallel to $p_l$, because in this limit $J_l(q)^\mu \propto q^\mu$ and $q^\mu [P_\LC(q)]_{\mu\nu} = 0$. However, this term can be singular in the soft limit if the other end of the gluon is connected to another external leg, say a leg with index $k$. Since we have only soft singularity, the current can be simplified: it is enough to keep the $2p_l^\mu$ part in Eq.~\eqref{eq:soft-current} since the term $A^{\mu\nu} q_\nu$ is suppressed by the factor $q$. This leads to the eikonal approximation in the first term of Eq.~\eqref{eq:1-loop-singular-graphs-decomposed}, depicted in Fig.~\ref{fig:Feynman2}.

The second term in Eq.~\eqref{eq:1-loop-singular-graphs-decomposed} can be singular in both the soft and collinear limits. In the limit in which $q$ is collinear with $p_l$, $J_l(q)^\mu$ is proportional to $q^\mu$. Then the term in $[P_\LC(q)]_{\mu\nu}$ proportional to $n_\mu q_\nu$ gives a leading collinear singularity, but the term proportional to $q_\mu n_\nu$ does not contribute because it is suppressed by a factor $q^2$. In the limit in which $q$ is soft, $q \to 0$, there are two cases to consider. When the soft gluon connects the external parton $l$ to an internal line of the graph, there are not enough powers of $1/q$ to give a leading singularity.\footnote{For an attachment to an internal line, there is a collinear singularity but no soft singularity. However, in one way of organizing the calculation, there is a sum of two soft singularities that cancel. Thus we sometimes speak of having both soft and collinear singularities.} When the soft gluon connects two external partons, there is a leading soft singularity. In this case, there is one graph, but two ends of the gluon line, one with a factor $n_\mu$ and the other with a factor $q_\nu$. We can use this symmetry to let index of the external line connected to $n_\mu$ be denoted by $l$ and the index of the other external line be denoted by something else.

The arguments presented above give for the singular terms in Eq.~\eqref{eq:1-loop-singular-graphs-decomposed},
\begin{equation}
   \label{eq:1-loop-singular-graphs-decomposed1}
   \begin{split}
    \big|M^{(1)}&{}(\{p,f\}_m)\big\rangle \\
    = {}&4\mi\pi\,\as\mu^{2\epsilon}
    \sum_{l<k}  \bm{T}_l\!\cdot\!\bm{T}_k\ket{M^{(0)}(\{p,f\}_m)}
    \\
    &\times\int\frac{d^dq}{(2\pi)^d}\,
    \frac{4p_l\!\cdot\!P_\LS(q)\!\cdot\!p_k}
    {[q^2+\mi0][(p_l-q)^2+\mi 0][(p_k+q)^2+\mi0]}
    \\
    &- \mi g\mu^{\epsilon}
    \sum_l  \int\frac{d^dq}{(2\pi)^d}\,
    \frac{\bm{T}^a_l\,n_\mu J_l^\mu(q)}{[q^2+\mi0][(p_l-q)^2+\mi0]}
    \\
    &\quad\times\sum_{G}\frac{q\!\cdot\!n\, q^\nu}{(q\!\cdot\!n)^2-q^2}
    \brax{\nu, a}\ket{R_l(G;  \{p,f\}_m)}
    \\
    &+ \cdots
    \;.
  \end{split} 
\end{equation}
Here the first term, illustrated in Fig.~\ref{fig:Feynman2}, has a factorized form with one contribution for each graph. In this term, $\ket{M^{(0)}(\{p,f\}_m)}$ is the amplitude without the soft gluon exchange. The integration over the loop momentum can be performed analytically. The second term, illustrated in Fig.~\ref{fig:Feynman3}, contains a sum over almost every $m+1$ parton graph, where the extra parton is a gluon carrying the momentum $q$. The factor $n_\mu$ is indicated by a diamond shape where the gluon line joins line $l$. The factor $q^\nu$ is indicated by the open arrow at the end of the gluon line. This extra gluon can connect to every internal or external line in the graph except to parton $l$. Parton $l$ carries a momentum of $p_l-q$. It becomes on shell either in the soft or the collinear limit. In the singular regions we can apply the Ward identity, as
\begin{equation}
  \label{eq:1-loop-singular-ward-id}
  \begin{split}
    \sum_{G}q^\nu{}&\brax{\nu, a}\ket{R_l(G;  \{p,f\}_m,\{q,\Lg\})}
   \\
    ={}&  g\mu^{\epsilon} \bm{T}_l^a\ket{M^{(0)}(\{p,f\}_m)} + \cO(q^2)
    \;.
  \end{split}
\end{equation}
This is the negative of the contribution that one would have obtained from an attachment to parton $l$. After using the Ward identity, we can write the singular part of the one-loop amplitudes in factorized form as
\begin{equation}
  \label{eq:1-loop-singular-graphs-decomposed2}
  \begin{split}
    \big|M^{(1)}&{}(\{p,f\}_m)\big\rangle \\
    = {}&4\mi\pi\,\as\mu^{2\epsilon}
    \sum_{l<k}  \bm{T}_l\!\cdot\!\bm{T}_k\ket{M^{(0)}(\{p,f\}_m)}
    \\
    &\times\int\frac{d^dq}{(2\pi)^d}\,
    \frac{4p_l\!\cdot\!P_{S}(q)\!\cdot\!p_k}
    {[q^2+\mi0][(p_l-q)^2+\mi0][(p_k+q)^2+\mi0]}
    \\
    &+ 4\mi\pi\,\as\mu^{2\epsilon}
    \sum_l \bm{T}_l^2\ket{M^{(0)}(\{p,f\}_m)}
    \\
    &\quad\times\int\frac{d^dq}{(2\pi)^d}\,
    \frac{n\!\cdot\!J_l(q)\,q\!\cdot\!n}
    {[(q\!\cdot\!n)^2-q^2]\,[q^2+\mi0][(p_l-q)^2+\mi0]}
    \\
    &+ \cdots
    \;.
  \end{split} 
\end{equation}
The first term in Eq.~(\ref{eq:1-loop-singular-graphs-decomposed2}) is illustrated in Fig.~\ref{fig:Feynman2}. The second term is illustrated in Fig.~\ref{fig:Feynman4}.

We note that by taking the limits in Eq.~\eqref{eq:1-loop-singular-ward-id} and using the Ward identity we might have introduced some spurious UV poles. We have to keep in mind that these UV singularities are fake and they need to be removed.

Now we can perform the integrals, leading to 
\begin{equation}
  \label{eq:1-loop-singular-graphs-decomposed3}
  \begin{split}
    \big|M^{(1)}(\{p&{},f\}_m)\big\rangle \\
    = {}&\frac{\as}{4\pi}\frac{S_\epsilon}{\epsilon}
    \sum_{l\neq k}  \bm{T}_l\!\cdot\!\bm{T}_k\ket{M^{(0)}(\{p,f\}_m)}
    \\
    &\quad\times
    \left\{
      \frac{1}{\epsilon}\left[\frac{\mu^2}{-2p_l\!\cdot\!p_k}\right]^\epsilon
      -\frac{1}{2\epsilon}\left[\frac{\mu^2}{(2p_l\!\cdot\!n)^2}\right]^\epsilon
    \right.
    \\
    &\quad\qquad\left.
      -\frac{1}{2\epsilon}\left[\frac{\mu^2}{(2p_k\!\cdot\!n)^2}\right]^\epsilon
    \right\}
    \\
    &-\frac{\as}{4\pi}\frac{S_\epsilon}{\epsilon}
    \sum_l \bm{T}_l^2\ket{M^{(0)}(\{p,f\}_m)}
    \\
    &\quad\times
    \left\{
      \frac{1}{\epsilon}\left[\frac{\mu^2}{(2p_l\!\cdot\!n)^2}\right]^\epsilon
      +\frac{2}{1+\delta_l}
    \right\}
    \\
    &+ \cdots
    \;,
  \end{split} 
\end{equation}
where $\delta_l=1$ if $l$ is a gluon otherwise $\delta_l = 0$. Using color conservation, we can simplify the result even further and obtain the result
\begin{align}
\label{eq:1-loop-singular-graphs-decomposed4}
    \big|{}&M^{(1)}(\{p,f\}_m)\big\rangle
    = \frac{\as}{4\pi}\frac{S_\epsilon}{\epsilon}
    \notag\\
    &\quad\times\bigg\{ \sum_{l\neq k}  \bm{T}_l\!\cdot\!\bm{T}_k
      \frac{1}{\epsilon}\left[\frac{\mu^2}{-2p_l\!\cdot\!p_k}\right]^\epsilon
      -\sum_l \frac{2}{1+\delta_l}\, \bm{T}_l^2
      \bigg\}
      \notag\\
      &\quad\times\ket{M^{(0)}(\{p,f\}_m)}
   + \cdots
   \;. 
\end{align}

This derivation shows that it is quite practical to use Feynman gauge to extract the infrared singularities of amplitudes when we add one gluon exchange to graphs that are either tree graphs or have only hard momenta in loops. However, one faces complications if one wants to extend this method to include more loops with potentially soft or collinear momenta of the partons in the loops. In this case, there are more gluons with propagators $P_\LC^{\mu\nu}(q)/[q^2 + \mi 0]$. These gluons can couple to each other, so that there are now multiple special cases to be considered when applying the needed Ward identities \cite{BabisCalcs, AnastasiouSterman}.

\section{Calculation of the gluon self-energy}
\label{sec:gluonSE}

We have seen that interpolating gauge offers the advantage that collinear singularities are eliminated in virtual exchange graphs. However, manifest covariance is lost and there are several terms in the gluon propagator. Thus calculations of loop diagrams are not as simple as in Feynman gauge. We have in mind using numerical integration to perform calculations, so perhaps this lack of simplicity in not a crucial problem. In this appendix, we investigate this issue by performing a numerical calculation in interpolating gauge.

We discuss the calculation of the gluon self-energy function $\Pi^{\mu\nu}(p)$ with a space-like value of $p$, using interpolating gauge with $\xi = 1$. The function $\Pi^{\mu\nu}(p)$ has an ultraviolet pole, which we eliminate by renormalization. It has no infrared poles. Infrared poles will emerge if we take $p^2 \to 0$.

\begin{widetext} 
The unrenormalized (``U'') version of the self-energy function that we wish to calculate has the form
\begin{equation}
\begin{split}
\label{eq:gselfE}
-\mi\,\Pi^{\mu\nu}_\LU(p) 
={}& 
4\pi\as   
\mu^{2\epsilon}\int \frac{d^d q}{(2\pi)^d}\,
\frac{N^{\mu\nu}(q,p)}
{[q^2+\mi 0][k^2+\mi 0]
[q\cdot \tilde q +\mi 0] [k\cdot \tilde k +\mi 0]}
\;.
\end{split}
\end{equation}
Here $d = 4 - 2 \epsilon$ and the loop momentum is $q$, with $k = p-q$. The numerator function is
\begin{equation}
\begin{split}
\label{eq:Nmunufull}
N^{\mu\nu}(q,p) ={}& 
- \frac{C_\LA}{2}\,\Gamma_{\mu \alpha \beta}(-p,q,k)
\Gamma_{\nu \gamma \delta}(p,-q,-k)
N^{\alpha \gamma}(q)\,
N^{\beta \delta}(k) 
+ \frac{C_\LA}{2}\,q^2 k^2 
\left[\tilde q^\mu \tilde k^\nu + \tilde k^\mu \tilde q^\nu \right]
\\ &
-\frac{C_\LA}{2}
\left[
N^{\mu\nu}(k) 
- g^{\mu\nu} N^\alpha_\alpha(k) 
\right]
q^2 q\cdot \tilde q
-\frac{C_\LA}{2}
\left[
N^{\mu\nu}(q) 
- g^{\mu\nu} N^\alpha_\alpha(q)
\right] k^2\, k\cdot \tilde k
\\ &
+ 4 T_\LR n_\Lf
\left[k^\mu q^\nu + q^\mu k^\nu - g^{\mu\nu} q\cdot k\right]
q\cdot\tilde q\,k\cdot\tilde k
\;.
\end{split}
\end{equation}
The second term is for the ghost loop. The third and fourth terms are from the tadpole diagram, symmetrized under $q \leftrightarrow k$. The last term is the quark loop. These contributions are depicted in Fig.~\ref{fig:GluonSE}. We have defined $\Gamma_{\alpha \beta \gamma}(p_a, p_b, p_c)$ using
\begin{equation}
\label{eq:Gammatree}
\Gamma_{\alpha \beta \gamma}^{abc}(p_a, p_b, p_c)
= g f_{abc}\Gamma_{\alpha \beta \gamma}(p_a, p_b, p_c)
\;.
\end{equation}
Then, from Eq.~(\ref{eq:threegluonvertex}),
\begin{equation}
\Gamma_{\alpha \beta \gamma}(p_a, p_b, p_c) =
-\big\{ 
g_{\alpha\beta}(p_{a} - p_{b})_\gamma
+ g_{\beta\gamma}(p_{b} - p_{c})_\alpha
+ g_{\gamma\alpha} (p_{c} - p_{a})_\beta
\big\}
\;.
\end{equation}
We have also defined $N^{\mu\nu}(q)$ as 
\begin{equation}
N^{\mu\nu}(q) = q^2\,q\cdot\tilde q\,D^{\mu\nu}(q)
\end{equation}
with $\xi = 1$. From Eq.~(\ref{eq:gluonpropagator3}), this is
\begin{equation}
\label{eq:gluonpropagator2B}
N^{\mu\nu}(q)
=   - g^{\mu\nu} q\cdot \tilde q
    + q^\mu\,\tilde q^\nu + \tilde q^\mu\,q^\nu
    -  \frac{v^2+1}{v^2}\, q^\mu\,q^\nu
\;.
\end{equation}
The numerator factor omits a factor $4\pi\as \delta_{a a'}$, where $a$ and $a'$ are the gluon color indices. The $\delta_{a a'}$ has been removed from $\Pi^{\mu\nu}(p)$ and $4\pi\as$ has been factored out of the numerator function.

It is possible to simplify some parts of the integrand. First, any factors of $q^2$, $q\cdot \tilde q$, $k^2$, or $k\cdot \tilde k$ in the numerator can be used to cancel the corresponding factors in the denominator, giving a simpler denominator. Second, if the resulting denominator depends only on $q$ but not on $k$ or depends only on $k$ but not on $q$, then we have a scaleless integral. With dimensional regularization, a scaleless integral consists of an ultraviolet pole that exactly cancels an infrared pole. Such scaleless integrals can simply be dropped. However, it is feasible to proceed without implementing any of these simplifications. Then one applies the same general method to all contributions to the integrand. This is the method that we explain below.

We can combine the four denominator factors using three Feynman parameters $x$, $y$ and $z$:
\begin{equation}
\begin{split}
\label{eq:gSEFeynman}
\Pi^{\mu\nu}_\LU(p) ={}& 
\mi\mu^{2\epsilon}\int \frac{d^d q}{(2\pi)^d}\,
N^{\mu\nu}(q,p)
\int_0^1\!dx\ J(x) \int_0^1\!dy \int_0^1\!dz
\\&\times
\frac{1}
{[x(1-y) q^2 + x y k^2
+ (1-x) (1-z)\,q\cdot \tilde q
+ (1-x) z\,k\cdot \tilde k
+ \mi 0]^4}
\;,
\end{split}
\end{equation}
\end{widetext} 
where 
\begin{equation}
\label{eq:Jdef}
J(x) = 4\pi\as\,6\,x(1-x)
\;.
\end{equation}

To proceed, we manipulate the denominator in Eq.~(\ref{eq:gSEFeynman}). We use the projection tensors $P_+$ and $P_-$, Eq.~(\ref{eq:Pplusminus}), and indicate contractions of Lorentz indices with a dot, as in $A\cdot B = C$ for $A^\mu_{\ \alpha} B^\alpha_{\ \nu} = C^\mu_{\ \nu}$. The denominator is
\begin{equation}
\begin{split}
D ={}& x(1-y) q^2 + xy (p-q)^2
+ (1-x) (1-z)\,q\cdot \tilde q
\\&
+ (1-x) z\,(p-q)\cdot (\tilde p - \tilde q)
\\
={}& q\cdot A(x) \cdot q - 2 q \cdot A(x) \cdot  w(x,y,z) 
\\&
+ xy p^2 + (1-x) z p\cdot \tilde p
\;,
\end{split}
\end{equation}
where
\begin{equation}
\begin{split}
\label{eq:Axdef}
A(x) ={}& \frac{1-x + v^2 x}{v^2}\, P_+ + P_-
\;,
\\
A(x)^{-1} ={}& \frac{v^2}{1-x + v^2 x}\, P_+ + P_-
\;,
\\
w(x,y,z) ={}& 
A^{-1}(x)
\left[
xy p +  (1-x) z \tilde p
\right]
\;.
\end{split}
\end{equation}
Notice that we have combined the terms proportional to $q^2$ and $q\cdot \tilde q$ into one term $q \cdot A(x) \cdot q$. Completing the square in the denominator gives
\begin{equation}
\begin{split}
D ={}&(q - w(x,y,z))\cdot A(x) \cdot (q - w(x,y,z))
\\&
- \Lambda^2(x,y,z)
\;,
\end{split}
\end{equation}
where
\begin{equation}
\begin{split}
\label{eq:Lambdaxyz0}
\Lambda^2(x,y,z) ={}& 
w(x,y,z)\cdot A(x) \cdot w(x,y,z)
\\&
- xy p^2 - (1-x)z p\cdot \tilde p
\;.
\end{split}
\end{equation}
This may be better expressed as
\begin{equation}
\begin{split}
\label{eq:Lambdaxyz}
\Lambda^2(x,y,z) ={}& 
-\lambda_0(x,y,z)\  p^2
-\lambda_1(x,y,z)\ p\cdot \tilde p
\;,
\end{split}
\end{equation}
where
\begin{equation}
\begin{split}
\label{eq:lambda0lambda1}
\lambda_0(x,y,z) ={}& \frac{x(1-x)^2}{1-x + v^2 x}\, (y - z)^2
 + x y (1-y)
\;,
\\
\lambda_1(x,y,z) ={}& \frac{x^2(1-x) v^2}{1-x + v^2x}\, (y - z)^2
\\&
 + (1-x) z (1-z)
 \;.
\end{split}
\end{equation}
Notice that $\lambda_0(x,y,z)$ and $\lambda_1(x,y,z)$ are positive for $0<x<1$, $0<y<1$, $0<z<1$.

In the integral (\ref{eq:gSEFeynman}), we can shift the integration variable from $q$ to $q'= q-w$, so that
\begin{equation}
\begin{split}
\Pi^{\mu\nu}_\LU(p) ={}& 
\int_0^1\!dx\ J(x) \int_0^1\!dy \int_0^1\!dz\
\mi\,\mu^{2\epsilon}\int \frac{d^d q'}{(2\pi)^d}
\\&\times
\frac{N^{\mu\nu}(q' + w(x,y,z),p)}
{[q'\cdot A(x) \cdot q'
- \Lambda^2(x,y,z)
+ \mi 0]^4}
\;.
\end{split}
\end{equation}
Then we can change variables again to $\ell = A^{1/2}(x) q'$, with a new jacobian,
\begin{equation}
\label{eq:newjacobian}
J'(x) = [(1-x + v^2 x)/v^2]^{-1/2}
\;,
\end{equation}
so that
\begin{equation}
\begin{split}
\Pi^{\mu\nu}_\LU(p) ={}& 
\!\!\int_0^1\!dx J(x)J'(x) \int_0^1\!dy \int_0^1\!dz\,
\mi \mu^{2\epsilon}\!\int\! \frac{d^d \ell}{(2\pi)^d}
\\&\times
\frac{N^{\mu\nu}(A^{-1/2}(x) \ell + w(x,y,z),p)}
{[\ell^2
- \Lambda^2(x,y,z)
+ \mi 0]^4}
\;.
\end{split}
\end{equation}

We write $N^{\mu\nu}(A^{-1/2} \ell + w,p)$ as a function of $\ell$ and $p$. We can eliminate terms that are odd in $\ell$ because they will integrate to zero. There are then terms in the numerator proportional to 6, 4, 2, and 0 powers of $\ell$. That is, there are terms proportional to $\ell^\alpha \ell^\beta \ell^\gamma \ell^\delta \ell^\rho \ell^\sigma$, $\ell^\alpha \ell^\beta \ell^\gamma \ell^\delta$, $\ell^\alpha \ell^\beta$ and 1. We define Lorentz invariant symmetric tensors by
\begin{equation}
\begin{split}
T_2^{\alpha\beta} ={}& g^{\alpha\beta}
\;,
\\
T_4^{\alpha\beta\gamma\delta} ={}& g^{\alpha\beta}T_2^{\gamma\delta}
+ g^{\alpha\gamma}T_2^{\beta\delta}
+ g^{\alpha\delta}T_2^{\beta\gamma}
\;,
\\
T_6^{\alpha\beta\gamma\delta\rho\sigma} ={}& 
g^{\alpha\beta}T_4^{\gamma\delta\rho\sigma}
+ g^{\alpha\gamma}T_4^{\beta\delta\rho\sigma}
+ g^{\alpha\delta}T_4^{\beta\gamma\rho\sigma}
\\&
+ g^{\alpha\rho}T_4^{\beta\gamma\delta\sigma}
+ g^{\alpha\sigma}T_4^{\beta\gamma\delta\rho}
\;.
\end{split}
\end{equation}
After integration, we obtain terms proportional to $T_6^{\alpha\beta\gamma\delta\rho\sigma}$, $T_4^{\alpha\beta\gamma\delta}$, $T_2^{\alpha\beta}$ and 1. We can use these to define coefficients $A_0$, $A_2$, $A_4$, and $A_6$:
\begin{equation}
\begin{split}
A_{0}(\Lambda^2) ={}&
\mi\,\mu^{2\epsilon}\int\!\frac{d^d \ell}{(2\pi)^d}
\frac{1}{[\ell^2 - \Lambda^2 + \mi 0]^4}
\;,
\\
A_{2}(\Lambda^2) T_2^{\alpha\beta} ={}&
\mi\,\mu^{2\epsilon}\int\!\frac{d^d \ell}{(2\pi)^d}
\frac{\ell^\alpha \ell^\beta}{[\ell^2 - \Lambda^2 + \mi 0]^4}
\;,
\\
A_{4}(\Lambda^2) T_4^{\alpha\beta\gamma\delta}
={}&
\mi\,\mu^{2\epsilon} \int\!\frac{d^d \ell}{(2\pi)^d}
\frac{\ell^\alpha \ell^\beta \ell^\gamma \ell^\delta}
{[\ell^2 - \Lambda^2 + \mi 0]^4}
\;,
\\
A_{6}(\Lambda^2) T_6^{\alpha\beta\gamma\delta\rho\sigma}
={}&
\mi\,\mu^{2\epsilon} \int\!\frac{d^d \ell}{(2\pi)^d}
\frac{\ell^\alpha \ell^\beta \ell^\gamma \ell^\delta \ell^\rho \ell^\sigma}
{[\ell^2 - \Lambda^2 + \mi 0]^4}
\;.
\end{split}
\end{equation}
The coefficients $A_J$ are
\begin{equation}
\begin{split}
A_{0}(\Lambda^2) ={}&
-\frac{(\epsilon + 1)}{6}\,
\frac{1}{(\Lambda^2)^2}\,
I_\Ls(\Lambda^2)
\;,
\\
A_{2}(\Lambda^2) ={}&
\frac{1}{12}\,
\frac{1}{\Lambda^2}\,
I_\Ls(\Lambda^2)
\;,
\\
A_{4}(\Lambda^2)
={}&
-\frac{1}{24}\,
\frac{1}{\epsilon}\,
I_\Ls(\Lambda^2)
\;,
\\
A_{6}(\Lambda^2)
={}&
\frac{1}{48}\,
\frac{1}{\epsilon (\epsilon - 1)}\,
\Lambda^2\,
I_\Ls(\Lambda^2)
\;,
\end{split}
\end{equation}
where the standard factor $I_\Ls(\Lambda^2)$ is
\begin{equation}
\begin{split}
\label{eq:Isdef}
I_\Ls(\Lambda^2)
={}&  \frac{\Gamma(1+\epsilon)}{(4\pi)^{2 - \epsilon}}\,
\left(\frac{\mu^2}{\Lambda^2}\right)^\epsilon
\;.
\end{split}
\end{equation}
After integration over $\ell$, we have a result of the form
\begin{equation}
\begin{split}
\Pi^{\mu\nu}_\LU(p) ={}& 
\int_0^1\!dx\ J(x)\,J'(x) \int_0^1\!dy \int_0^1\!dz
\\&\times
I^{\mu\nu}_\LU(p;x,y,z)
\;.
\end{split}
\end{equation}
The $(\ell)^6$ and $(\ell)^4$ contributions to $I^{\mu\nu}(p;x,y,z)$ are proportional to $1/\epsilon$. The $1/\epsilon$ pole is to be removed by renormalization. The $(\ell)^2$ and $(\ell)^0$ contributions are finite when $\epsilon \to 0$.

The ultraviolet pole terms are defined by
\begin{equation}
I^{\mu\nu}_\mathrm{pole}(p;x,y,z) =
\left[\epsilon\,I^{\mu\nu}_\LU(p;x,y,z)\right]_{\epsilon \to 0}
\;.
\end{equation}
We remove the poles according to the $\MSbar$ prescription by subtracting
\begin{equation}
\label{eq:ImunuCT}
I^{\mu\nu}_\mathrm{CT}(p;x,y,z) = 
\frac{S_\epsilon}{\epsilon}\,
I^{\mu\nu}_\mathrm{pole}(p;x,y,z)
\;,
\end{equation}
where $S_\epsilon$ is the standard factor defined in Eq.~(\ref{eq:epsilonMSbar}).
The renormalized version of $I^{\mu\nu}$ is then
\begin{equation}
\begin{split}
I^{\mu\nu}(p;x,&y,z) \\={}& 
\big[I^{\mu\nu}_\LU(p;x,y,z) - I^{\mu\nu}_\mathrm{CT}(p;x,y,z)\big]_{\epsilon \to 0}
\;.
\end{split}
\end{equation}
The integrand $I^{\mu\nu}_\LU$ has the form
\begin{equation}
\begin{split}
I^{\mu\nu}_\LU(p;x,y,z;\epsilon) ={}& 
\frac{1}{\epsilon}
I_\Ls(\Lambda^2)\widetilde I^{\mu\nu}(p;x,y,z;\epsilon)
\\={}&
\frac{1}{(4\pi)^2}
\frac{(4\pi)^\epsilon \Gamma(1+\epsilon)}{\epsilon}
\left(\frac{\mu^2}{\Lambda^2}\right)^\epsilon
\\&\times
\widetilde I^{\mu\nu}(p;x,y,z;\epsilon)
\;.
\end{split}
\end{equation}
This gives us
\begin{equation}
\begin{split}
I^{\mu\nu}(p;x,y,z) ={}& 
\frac{1}{(4\pi)^2}
\widetilde I^{\mu\nu}(p;x,y,z;0)
\log\!\left(\frac{\mu^2}{\Lambda^2}\right)
\\&
+ \frac{1}{(4\pi)^2}
\left[\frac{d \widetilde I^{\mu\nu}(p;x,y,z;\epsilon)}{d\epsilon}
\right]_{\epsilon = 0}
\;.
\end{split}
\end{equation}
For the parts of $I^{\mu\nu}$ that did not have a pole $1/\epsilon$, this prescription gives simply the contribution to $I^{\mu\nu}$ evaluated at $\epsilon = 0$.

The renormalized version of $\Pi^{\mu\nu}$ is then obtained by integrating $I^{\mu\nu}$ over the Feynman parameters:
\begin{equation}
\begin{split}
\label{eq:PimunuRdef}
\Pi^{\mu\nu}(p) ={}& 
\int_0^1\!dx\ J(x)\,J'(x) \int_0^1\!dy \int_0^1\!dz
\\&\times
I^{\mu\nu}(p;x,y,z)
\;,
\end{split}
\end{equation}
This integral can be performed by numerical integration. Typical results are reported in Fig.~\ref{fig:Pimunu}.

The pole term in $\Pi^{\mu\nu}$ is
\begin{equation}
\begin{split}
\Pi^{\mu\nu}_\mathrm{pole}(p) ={}& 
\int_0^1\!dx\ J(x)\,J'(x) \int_0^1\!dy \int_0^1\!dz
\\&\times
I^{\mu\nu}_\mathrm{pole}(p;x,y,z)
\;.
\end{split}
\end{equation}
The integrals over the Feynman parameters can be performed analytically, giving
\begin{equation}
\begin{split}
\label{eq:Pimunupole}
\Pi^{\mu\nu}_\mathrm{pole}(p) ={}&
C_1(v)\left\{
g^{\mu\nu} p^2-p^\mu p^\nu
\right\}
\\&
+ C_2(v)\left\{
2 h^{\mu\nu} p^2 - \tilde p^\mu p^\nu - p^\mu \tilde p^\nu
\right\}
\;,
\end{split}
\end{equation}
where
\begin{equation}
\begin{split}
\label{eq:Z1Z2}
C_1(v) ={}& \frac{C_\LA \as}{4 \pi}\,
\frac{- 11 v^3 - 16 v^2 - 7 v + 2}{3 v (v+1)^2}
\\&
+ \frac{\as}{4 \pi}\,\frac{4}{3}\,T_\LR n_\Lf
\;,
\\
C_2(v) ={}& \frac{C_\LA \as}{4 \pi}\,
\frac{2 v(2 v + 1)}{3(v+1)^2}
\;.
\end{split}
\end{equation}
This calculation has been for $\xi = 1$. This result for the ultraviolet pole matches the more general result in Eq.~(\ref{eq:gluonSEUV}) at $\xi = 1$.

\section{The gluon self-energy with $p^2 \to 0$}
\label{sec:gluonSEP0}

We now investigate the one-loop self-energy for a T gluon that is an external particle in the S matrix. This means that we need not the full $\Pi^{\mu\nu}(p)$ but only $\Pi^{\mu\nu}(p) \varepsilon_\nu(p,s)$. Given the tensor structure (\ref{eq:Pimunustructure}) of $\Pi^{\mu\nu}(p)$, we see that $\Pi^{\mu\nu}(p) \varepsilon_\nu(p,s)$ is proportional to $\varepsilon^\mu(p,s)$. Thus it suffices to consider $\varepsilon_\mu(p,s) \Pi^{\mu\nu}(p) \varepsilon_\nu(p,s)$. Since an external T gluon is on its mass shell, and is accompanied by a tree level propagator proportional to $1/p^2$, the quantity that we need for the S matrix, Eq.~(\ref{eq:Smatrix}), is the first order contribution to $\sqrt{R_\Lg}$, the square root of the residue of the T-gluon propagator at $p^2 = 0$. This is
\begin{equation}
\label{eq:PiIR}
\Pi_\mathrm{IR}=
\lim_{p^2 \to 0}
\frac{\varepsilon_\mu(p,s) \Pi^{\mu\nu}(p) \varepsilon_\nu(p,s)}{2 p^2}
\;.
\end{equation}
The factor $1/2$ gives us the order $\as$ contribution to $\sqrt{R_\Lg}$ instead of $R_\Lg$. In taking the limit $p^2 \to 0$, we start with $p^2 < 0$.

Some care is required in calculating $\Pi_\mathrm{IR}$ because taking the limit $p^2 \to 0$ leads to infrared divergences. We maintain the dimensional regulation of the calculation with $d = 4 - 2\epsilon$, with $\epsilon < 0$ so that we regulate an infrared divergence. We take the limit $p^2 \to 0$ in Eq.~(\ref{eq:PiIR}).  Then we let $\epsilon \to 0$, giving poles $1/\epsilon^2$ and $1/\epsilon$ plus a finite $\epsilon^0$ contribution. The finite contribution then contains logarithms of the dimensional regularization scale $\mu^2$. Knowing the IR pole terms and the scale dependence is just what one wants for use in a parton shower algorithm. 

Fortunately, $\varepsilon_\mu \Pi^{\mu\nu} \varepsilon_\nu$ is much simpler than the full $\Pi^{\mu\nu}$. This simplicity enables us to manipulate the integrand for $\varepsilon_\mu \Pi^{\mu\nu} \varepsilon_\nu$ so that all of the IR poles can be extracted, leaving only one integral for the coefficient of a finite part that remains for numerical integration.

\begin{widetext} 
The unrenormalized self-energy function that we wish to calculate is given by the form that we used in Eq.~(\ref{eq:gselfE}),
\begin{equation}
\begin{split}
\label{eq:gselfET}
\varepsilon_\mu(p,s)\Pi^{\mu\nu}_\LU(p)\varepsilon_\nu(p,s)
={}& 
4\pi\as\,  
\mi\mu^{2\epsilon}\int \frac{d^d q}{(2\pi)^d}\,
\frac{\varepsilon_\mu(p,s) N^{\mu\nu}(q,p)\varepsilon_\nu(p,s)}
{[q^2+\mi 0][k^2+\mi 0]
[q\cdot \tilde q +\mi 0] [k\cdot \tilde k +\mi 0]}
\;.
\end{split}
\end{equation}
\end{widetext} 

After some manipulation, we can write the numerator in the form
\begin{equation}
\begin{split}
\varepsilon_\mu  N^{\mu\nu} \varepsilon_\nu ={}& 
N_1 + N_2 + N_3 + N_4 + N_5 
+\mathrm{scaleless}
\,,
\end{split}
\end{equation}
where ``scaleless'' denotes contributions that give scaleless integrals, which vanish in dimensional regularization, and where
\begin{equation}
\begin{split}
N_1 ={}& 
\left\{c_{1,1} (q\cdot \varepsilon)^2
+ c_{1,2}\, p^2 \right\} q \cdot \tilde q\  k\cdot \tilde k
\;,
\\
N_2 ={}& c_2\,p^2\,(q\cdot \varepsilon)^2\, 
k \cdot \tilde k
\;,
\\
N_3 ={}& 
c_3\, p^2\, q\cdot \tilde p\ k\cdot\tilde k 
\;,
\\
N_4 ={}&
\left\{c_{4,1}\,
p^2\,p\cdot \tilde p
+c_{4,2}\,(p^2)^2\,
\right\}\,(q\cdot \varepsilon)^2
\;,
\\
N_5 ={}&
c_5\,(p^2)^2\,  
k\cdot \tilde k
\;.
\end{split}
\end{equation}
Here we have defined coefficients
\begin{equation}
\begin{split}
\label{eq:cijcoefficients}
c_{1,1} ={}& 4 C_\LA (1-\epsilon) - 8 T_\LR n_\Lf
\;,
\\
c_{1,2} ={}& - 4 C_\LA + 2 T_\LR n_\Lf
\;,
\\
c_{2} ={}& - 2C_\LA \, \frac{v^2-1}{v^2}
\;,
\\
c_{3} ={}&  4 C_\LA
\;,
\\
c_{4,1} ={}&  C_\LA\, \frac{v^2-1}{v^2}
\;,
\\
c_{4,2} ={}&    -C_\LA\, \frac{v^4-1}{2v^4}
\;,
\\
c_{5} ={}& - C_\LA\frac{v^2+1}{v^2}
\;.
\end{split}
\end{equation}

We analyze each of these contributions in turn.

We begin with the contribution from $N_1$:
\begin{equation}
\begin{split}
\label{eq:gselfET1}
\varepsilon_\mu\Pi^{\mu\nu}_{\LU,1}\varepsilon_\nu\!
={}& 
4\pi \as\,
\mi\mu^{2\epsilon}\!\int\! \frac{d^d q}{(2\pi)^d}\,
\frac{c_{1,1} (q\cdot \varepsilon)^2\! + c_{1,2}\, p^2}
{[q^2+\mi 0][k^2+\mi 0]}
.
\end{split}
\end{equation}
We can write this as an integral over a Feynman parameter $x$ as
\begin{align}
\varepsilon_\mu\Pi^{\mu\nu}_{\LU,1}\varepsilon_\nu
={}& 
4\pi \as\,  
\mi \mu^{2\epsilon}\int \frac{d^d q}{(2\pi)^d}\,
\int_0^1\!dx
\notag
\\&\times
\frac{c_{1,1} (q\cdot \varepsilon)^2 + c_{1,2}\, p^2}
{[(1-x) q^2 + x k^2+\mi 0]^2}
\;.
\end{align}
Using $k = p - q$, this is
\begin{align}
\varepsilon_\mu\Pi^{\mu\nu}_{\LU,1}\varepsilon_\nu
={}& 
\int_0^1\!dx\,
4\pi \as\,
\mi \mu^{2\epsilon}\int \frac{d^d q}{(2\pi)^d}\,
\notag
\\&\times
\frac{c_{1,1} (q\cdot \varepsilon)^2 + c_{1,2}\, p^2}
{[(q - x p)^2 + x(1-x) p^2+\mi 0]^2}
\;.
\end{align}
We change variables to $\ell = q - x p$ and use $\varepsilon \cdot p = 0$ to obtain
\begin{align}
\varepsilon_\mu\Pi^{\mu\nu}_{\LU,1}\varepsilon_\nu
={}& 
\int_0^1\!dx\,
4\pi \as\,
\mi \mu^{2\epsilon}\int \frac{d^d \ell}{(2\pi)^d}
\notag
\\&\times
\frac{c_{1,1} (\ell\cdot \varepsilon)^2 + c_{1,2}\, p^2}
{[\ell^2 + x(1-x) p^2+\mi 0]^2}
\;.
\end{align}
We can perform the integration over $\ell$ using
\begin{align}
\label{eq:integrals2}
\mi\,\mu^{2\epsilon}\!\int \frac{d^d \ell}{(2\pi)^d}
\frac{1}{[\ell^2 - \Lambda^2 + \mi 0]^2}
={}& 
-
\frac{1}{\epsilon}\,
I_\Ls(\Lambda^2)
\;,
\notag\\
\mi\,\mu^{2\epsilon}\!\int \frac{d^d \ell}{(2\pi)^d}
\frac{\ell^\mu \ell^\nu}{[\ell^2 - \Lambda^2 + \mi 0]^2}
={}& -
\frac{\Lambda^2 
I_\Ls(\Lambda^2)}{2\epsilon(1-\epsilon)}\, g^{\mu\nu}
,
\end{align}
where $I_\Ls(\Lambda^2)$ was defined in Eq.~(\ref{eq:Isdef}). This gives
\begin{align}
\frac{\varepsilon_\mu\Pi^{\mu\nu}_{\LU,1}\varepsilon_\nu}{2 p^2}
={}& 
-\frac{\as}{4\pi}\!\int_0^1\!\!dx\,
\frac{x(1-x)c_{1,1} + (1-\epsilon)c_{1,2}}
{2\varepsilon (1-\varepsilon)}
\notag
\\&\times
\Gamma(1+\epsilon)
\left(\frac{- x(1-x) p^2}{4\pi\,\mu^2}\right)^{-\epsilon}
\;.
\end{align}
We take $-p^2 > 0$ and take the limit $p^2 \to 0$ at a fixed value of $\epsilon$ with $\epsilon < 0$. This gives us
\begin{equation}
\begin{split}
\label{eq:Pi1result}
\lim_{p^2 \to 0}\frac{\varepsilon_\mu\Pi^{\mu\nu}_{\LU,1}\varepsilon_\nu}{2 p^2}
={}& 
0
\;.
\end{split}
\end{equation}

\begin{widetext} 
Next, we turn to the contribution proportional to $N_2$:
\begin{equation}
\begin{split}
\label{eq:gselfET2}
\frac{\varepsilon_\mu \Pi^{\mu\nu}_{\LU,2}\varepsilon_\nu}{2p^2}
={}& 
\frac{c_2}{2}\,
4\pi \as\,
\mi \mu^{2\epsilon}\int \frac{d^d q}{(2\pi)^d}\,
\frac{(q\cdot \varepsilon)^2}
{[q^2+\mi 0][k^2+\mi 0][q\cdot \tilde q+\mi 0]}
\;.
\end{split}
\end{equation}
We can write this as an integral over a Feynman parameters $x$ and $y$ as
\begin{equation}
\begin{split}
\frac{\varepsilon_\mu \Pi^{\mu\nu}_{\LU,2}\varepsilon_\nu}{2p^2}
={}& 
\frac{c_2}{2}\,4\pi \as\,
\mi \mu^{2\epsilon}\int \frac{d^d q}{(2\pi)^d}\,
\int_0^1\!dx\,2 x\int_0^1\!dy\,
\frac{(q\cdot \varepsilon)^2}
{[x(1-y)q^2 + xy k^2 + (1-x) q\cdot \tilde q + \mi 0]^3}
\;.
\end{split}
\end{equation}
With $k = p - q$, this is
\begin{equation}
\begin{split}
\frac{\varepsilon_\mu \Pi^{\mu\nu}_{\LU,2}\varepsilon_\nu}{2p^2} ={}&  
c_2\,4\pi \as 
\int_0^1\!dx\  x \int_0^1\!dy\ 
\mi\mu^{2\epsilon}\int \frac{d^d q}{(2\pi)^d}\,
\frac{(\varepsilon\cdot q)^2}
{[(q - w(x,y,0))\cdot A(x) \cdot (q - w(x,y,0)) 
- \Lambda^2(x,y,0) + \mi 0]^3}
\;.
\end{split}
\end{equation}
Here $A(x)$ and $w(x,y,z)$ were defined in Eq.~(\ref{eq:Axdef}) and $\Lambda^2(x,y,z)$ was defined in Eq.~(\ref{eq:Lambdaxyz}). We change variables to $q' = q - w$, noting that $\varepsilon \cdot w = 0$ since $\varepsilon \cdot p = 0$. Then we change variables to $\ell = A^{1/2}(x) q'$, with a Jacobian $J'(x)$ given in Eq.~(\ref{eq:newjacobian}). Since $\varepsilon\cdot n = 0$ we have $\varepsilon\cdot q' = \varepsilon\cdot \ell$. This gives
\begin{equation}
\begin{split}
\frac{\varepsilon_\mu \Pi^{\mu\nu}_{\LU,2} \varepsilon_\nu }{2 p^2}={}& 
c_2 \int_0^1\!dx\  x 
\left[\frac{v^2}{1-x + v^2 x}\right]^{1/2}
\int_0^1\!dy\  
4\pi\mi\as\,
\mu^{2\epsilon}\int \frac{d^d \ell}{(2\pi)^d}\,
\frac{(\varepsilon\cdot \ell)^2}
{[\ell^2 
- \Lambda^2(x,y,0) + \mi 0]^3}
\;.
\end{split}
\end{equation}

We can perform the $\ell$ integration using
\begin{equation}
\begin{split}
\mi\,\mu^{2\epsilon}\int \frac{d^d \ell}{(2\pi)^d}
\frac{\ell^\mu \ell^\nu}{[\ell^2 - \Lambda^2 + \mi 0]^3}
={}& - g^{\mu\nu}\ \frac{1}{4}\,
\frac{1}{\epsilon}\,
I_\Ls(\Lambda^2)
\;.
\end{split}
\end{equation}
This gives us
\begin{equation}
\begin{split}
\frac{\varepsilon_\mu \Pi^{\mu\nu}_{\LU,2} \varepsilon_\nu }{2 p^2}={}& 
c_2 \int_0^1\!dx\  x 
\left[\frac{v^2}{1-x + v^2 x}\right]^{1/2}
\int_0^1\!dy\  
\frac{\as}{4\pi}
\frac{\Gamma(1+\epsilon)}{4\epsilon}
\left(\frac{4\pi\mu^2}{\Lambda^2(x,y,0)}\right)^\epsilon
\;.
\end{split}
\end{equation}
It is straightforward to take the limit $p^2 \to 0$. At $p^2 = 0$ we have from Eq.~(\ref{eq:Lambdaxyz}),
\begin{equation}
\label{eq:Lambdaxy0}
\Lambda^2(x,y,0)= - \frac{x^2(1-x)y^2 v^2}{1-x + v^2x}\,p\cdot \tilde p
\;.
\end{equation}
Thus
\begin{equation}
\begin{split}
\lim_{p^2 \to 0}\frac{\varepsilon_\mu \Pi^{\mu\nu}_{\LU,2} \varepsilon_\nu }{2 p^2}
={}& 
\frac{\as}{4\pi}
\frac{\Gamma(1+\epsilon)}{4\epsilon}\,
\left(\frac{4\pi\mu^2}{- p\cdot \tilde p}\right)^\epsilon
c_2 \int_0^1\!dx\  x 
\int_0^1\!dy\  
\left[\frac{v^2}{1-x + v^2 x}\right]^{1/2}
\left(\frac{1-x + v^2x}{x^2(1-x)y^2 v^2}\right)^{\epsilon}
\;.
\end{split}
\end{equation}
This has a $1/\epsilon$ pole and a finite part as $\epsilon \to 0$. After performing the integrals over $y$ and $x$, we find
\begin{equation}
\begin{split}
\lim_{p^2 \to 0}\frac{\varepsilon_\mu \Pi^{\mu\nu}_{\LU,2} \varepsilon_\nu }{2 p^2} ={}& \frac{\as}{4\pi}\,\frac{c_2}{4}\,
\left(\frac{\mu^2}{-p\cdot\tilde p}\right)^\epsilon
\left\{
\frac{S_\epsilon}{\epsilon}\,I_{2,1}(v)
+ 2I_{2,1}(v) +  I_{2,2}(v)
\right\}
+ \cO(\epsilon^0)
\;.
\end{split}
\end{equation}
The factor $(\mu^2/(-p\cdot\tilde p))^\epsilon$  can be expanded to give a contribution proportional to $\log(\mu^2/(-p\cdot\tilde p))$. The needed integrals are
\begin{equation}
\begin{split}
\label{eq:I21I22}
I_{2,1}(v) ={}& \frac{2 v (v+2)}{3 (v+1)^2}
\;,
\\
I_{2,2}(v) ={}& \!\frac{4 v}{9(v^2-1)^2}\bigg((5-3\log(2)) v^3 
- 3 v^2
-\! (12 - 9 \log(2))v + 10
\\&\quad
+ 3 v (v^2 - 3)\log\!\left[\frac{v+1}{v}\right]
+ 6\log\!\left[\frac{(v+1)^2}{4v}\right]
\!\bigg)
.
\end{split}
\end{equation}
For $v=2$, $I_{2,1}(2)=0.592593$ and $I_{2,2}(2)=1.28204$.

We are now ready to consider the contribution proportional to $N_3$: 
\begin{equation}
\begin{split}
\label{eq:gselfET3}
\frac{\varepsilon_\mu \Pi^{\mu\nu}_{\LU,3}\varepsilon_\nu}{2p^2}
={}& 
4\pi \as\,\frac{c_3}{2}\,
 \mi\mu^{2\epsilon}\int \frac{d^d q}{(2\pi)^d}\,
\frac{q\cdot\tilde p}
{[q^2+\mi 0][k^2+\mi 0][q\cdot \tilde q+\mi 0]}
\;.
\end{split}
\end{equation}
We can write this as an integral over a Feynman parameters $x$ and $y$ and rearrange the denominator as for $\varepsilon_\mu \Pi^{\mu\nu}_{\LU,2} \varepsilon_\nu$, giving
\begin{equation}
\begin{split}
\frac{\varepsilon_\mu \Pi^{\mu\nu}_{\LU,3}\varepsilon_\nu}{2p^2} ={}&   
c_3\,\int_0^1\!dx\ x \int_0^1\!dy\ 
4\pi\as\,
\mi\mu^{2\epsilon}\int \frac{d^d q}{(2\pi)^d}\,
\frac{q\cdot\tilde p}
{[(q - w(x,y,0))\cdot A(x) \cdot (q - w(x,y,0)) 
- \Lambda^2(x,y,0) + \mi 0]^3}
\;.
\end{split}
\end{equation}
We change variables to $q' = q - w$. Then in the numerator, $q\cdot\tilde p = q'\cdot\tilde p + w\cdot\tilde p$, but we can drop $q'\cdot\tilde p$ since it is odd under $q' \to - q'$. Then we change variables to $\ell = A^{1/2}(x) q'$, with a Jacobian $J'(x)$. This gives
\begin{equation}
\begin{split}
\frac{\varepsilon_\mu \Pi^{\mu\nu}_{\LU,3} \varepsilon_\nu }{2 p^2}={}& 
c_3 \int_0^1\!dx\  x 
\left[\frac{v^2}{1-x + v^2 x}\right]^{1/2}
\int_0^1\!dy\  
4\pi\mi\as\,
\mu^{2\epsilon}\int \frac{d^d \ell}{(2\pi)^d}\,
\frac{w(x,y,0)\cdot\tilde p}
{[\ell^2 - \Lambda^2(x,y,0) + \mi 0]^3}
\;.
\end{split}
\end{equation}

We can perform the $\ell$ integration using
\begin{equation}
\begin{split}
\mi\,\mu^{2\epsilon}\int \frac{d^d \ell}{(2\pi)^d}
\frac{1}{[\ell^2 - \Lambda^2 + \mi 0]^3}
={}&  \frac{I_\Ls(\Lambda^2)}{2 \Lambda^2}\,
\;.
\end{split}
\end{equation}
This gives us
\begin{equation}
\begin{split}
\frac{\varepsilon_\mu \Pi^{\mu\nu}_{\LU,3} \varepsilon_\nu }{2 p^2}
={}& 
\frac{c_3}{2} \frac{\as}{4\pi}\Gamma(1+\epsilon)\! \int_0^1\!dx\,  x 
\left[\frac{v^2}{1-x + v^2 x}\right]^{1/2}
\int_0^1\!dy\,  
\frac{w(x,y,0)\cdot\tilde p}{\Lambda^2(x,y,0)}
\left(\frac{4\pi\mu^2}{\Lambda^2(x,y,0)}\right)^\epsilon
.
\end{split}
\end{equation}
It is straightforward to take the limit $p^2 \to 0$. At $p^2 = 0$, $\Lambda^2(x,y,0) $ is given by Eq.~(\ref{eq:Lambdaxy0}). For  $w(x,y,0)\cdot \tilde p$ at $p^2 = 0$, we use Eqs.~(\ref{eq:Axdef}), (\ref{eq:tildedef}), (\ref{eq:hdef}), and (\ref{eq:Pplusminus}) to derive
\begin{equation}
\label{eq:wxy0}
w(x,y,0)\cdot \tilde p = -\frac{\Lambda^2(x,y,0)}{(1-x)y}
\;.
\end{equation}
This gives us
\begin{equation}
\begin{split}
\lim_{p^2 \to 0}\frac{\varepsilon_\mu \Pi^{\mu\nu}_{\LU,3} \varepsilon_\nu }{2 p^2}
={}& 
-\frac{c_3}{2} \frac{\as}{4\pi}\Gamma(1+\epsilon)\left(\frac{4\pi\mu^2}{-p\cdot \tilde p}\right)^\epsilon
\int_0^1\!\frac{dy}{y^{1+2\epsilon}} \int_0^1\!\frac{dx}{(1-x)^{1+\epsilon}}\,
\left[\frac{v^2 x^2}{1-x + v^2 x}\right]^{1/2-\epsilon}
.
\end{split}
\end{equation}
The $y$-integration is trivial and gives a factor $-1/(2\epsilon)$. The $x$-integration produces another $1/\epsilon$ from the endpoint $x \to 1$. This gives a double pole, a single pole, and a finite contribution as $\epsilon \to 0$. The integrals over $x$ and $y$ needed for these contributions can be performed analytically, giving
\begin{equation}
\begin{split}
\lim_{p^2 \to 0}\frac{\varepsilon_\mu \Pi^{\mu\nu}_{\LU,3} \varepsilon_\nu }{2 p^2}
={}& 
-\frac{c_3}{4} \frac{\as}{4\pi}\frac{S_\epsilon}{\epsilon}
[1 + \frac{\pi^2}{6} \epsilon^2]
\left(\frac{\mu^2}{-p\cdot \tilde p}\right)^\epsilon
\left\{
\frac{1}{\epsilon}
+ I_{3,1} + \epsilon \left(I_{3,2} - \frac{\pi^2}{6}\right)
\right\}
+\cO(\epsilon)
\;.
\end{split}
\end{equation}
The integrals $I_{3,1}$ and $I_{3,2}$ are given below. We have included the standard factor $S_\epsilon$ and compensated by multiplying by
\begin{equation}
\Gamma(1+\epsilon)\Gamma(1-\epsilon) = 1 + \frac{\pi^2}{6} \epsilon^2 + \cO(\epsilon^4)
\;.
\end{equation}
This gives
\begin{equation}
\begin{split}
\lim_{p^2 \to 0}\frac{\varepsilon_\mu \Pi^{\mu\nu}_{\LU,3} \varepsilon_\nu }{2 p^2}
={}& 
-\frac{c_3}{4} \frac{\as}{4\pi} \left(\frac{\mu^2}{-p\cdot \tilde p}\right)^\epsilon
\left\{
\frac{S_\epsilon}{\epsilon^2}
+ \frac{S_\epsilon}{\epsilon} I_{3,1}(v)  
+ I_{3,2}(v) 
\right\}
+\cO(\epsilon)
\;.
\end{split}
\end{equation}
The factor $(\mu^2/(-p\cdot\tilde p))^\epsilon$  can be expanded to give contributions proportional to powers of $\log(\mu^2/(-p\cdot\tilde p))$. The needed integrals are
\begin{equation}
\begin{split}
\label{eq:I31I32}
I_{3,1}(v) ={}& \frac{2 v}{v+1} - 2\log\left(\frac{2 v}{v+1}\right)
\\
I_{3,2}(v) ={}&  \frac{8 v}{v+1}
+\frac{4 v}{v^2-1}\,\log\left(\frac{2}{v+1}\right)
-\frac{4 v}{v + 1}\,\log\left(\frac{2v}{v+1}\right)
+ \log(2)\log\left(\frac{2 v^2 (v-1)^2}{(v+1)^4}\right)
+ 2\left[\log\left(\frac{v}{v+1}\right)\right]^2
\\&
+ \left[\log\left(v\right)\right]^2
+  \log(v+1)\log\left(\frac{v+1}{(v-1)^2}\right)
- 2\, \mathrm{Li}_2\left( -\frac{2}{v-1}\right)
+ 2\, \mathrm{Li}_2\left( -\frac{v+1}{v-1}\right)
+ 2\, \mathrm{Li}_2\left( -\frac{1}{v}\right)
\\&
- 2\, \mathrm{Li}_2\left( \frac{v-1}{2v}\right)
+ 2\, \mathrm{Li}_2\left( \frac{v-1}{v}\right)
+ 2\, \mathrm{Li}_2\left( -\frac{v-1}{v+1}\right)
- 2\, \mathrm{Li}_2\left( \frac{v-1}{v+1}\right)
\;.
\end{split}
\end{equation}
For $v = 2$, $I_{3,1}(2) = 0.757969$ and $I_{3,2}(2) = 1.27310$

The contribution proportional to $N_4$ is quite simple. We start with
\begin{equation}
\begin{split}
\label{eq:gselfET4}
\frac{\varepsilon_\mu \Pi^{\mu\nu}_{\LU,4}\varepsilon_\nu}{2 p^2}
={}& 
\frac{4\pi\as}{2}\,
\mi\mu^{2\epsilon}\int \frac{d^d q}{(2\pi)^d}\,
\frac{[c_{4,1}\,  p \cdot \tilde p + c_{4,2}\, p^2 ](q\cdot \varepsilon)^2}
{[q^2+\mi 0][k^2+\mi 0][q\cdot \tilde q+\mi 0][k\cdot \tilde k +\mi 0]}
\;.
\end{split}
\end{equation}
We can evaluate this using the method of Appendix~\ref{sec:gluonSE}. We introduce integrations over Feynman parameters $x$, $y$, and $z$. Then we change integration variables from $q$ to $\ell = A^{1/2}(x)(q - w(x,y,z))$. The analysis is simpler than for the general contribution in Appendix~\ref{sec:gluonSE} because, since $\varepsilon\cdot p = \varepsilon\cdot n = 0$, we have $q\cdot \varepsilon = \ell\cdot \varepsilon$. After performing the $\ell$-integration, we have
\begin{equation}
\begin{split}
\frac{\varepsilon_\mu \Pi^{\mu\nu}_{\LU,4}\varepsilon_\nu}{2 p^2}
={}& 
\frac{\as}{4\pi}\,\frac{\Gamma(1+\epsilon)}{4}
\int_0^1\!dx\ x(1-x) \left[\frac{v^2}{1-x + v^2 x}\right]^{1/2}
\int_0^1\!dy \int_0^1\!dz\
\frac{c_{4,1}\,  p \cdot \tilde p + c_{4,2}\, p^2}{\Lambda^2(x,y,z)}
\left(\frac{4\pi \mu^2}{\Lambda^2(x,y,z)}\right)^\epsilon
\;.
\end{split}
\end{equation}
We take the $p^2 \to 0$ limit of this by simply setting $p^2 = 0$ in the integral. The integral over the Feynman parameters does not produce any poles in $\epsilon$, so we can simply set $\epsilon = 0$. This gives 
\begin{equation}
\begin{split}
\frac{\varepsilon_\mu \Pi^{\mu\nu}_{\LU,4}\varepsilon_\nu}{2 p^2}
={}& 
\frac{\as}{4\pi}\,\frac{c_{4,1}}{4}\,
I_4(v) + \cO(\epsilon)
\;,
\end{split}
\end{equation}
where
\begin{equation}
\begin{split}
I_4(v) ={}& \int_0^1\!dx\ x(1-x) \left[\frac{v^2}{1-x + v^2 x}\right]^{1/2}
\\&\times
\int_0^1\!dy \int_0^1\!dz
\left[
\frac{v^2 x}{1-x + v^2x}\,x(1-x) (y - z)^2
 + (1-x) z (1-z)\right]^{-1}
\;.
\end{split}
\end{equation}
As far as we can see, this integral cannot be expressed in closed form using logarithms and dilogarithms. For $v = 2$, it is $I_4(2) = 3.67765$.

The contribution proportional to $N_5$ vanishes. To see why, we begin with
\begin{equation}
\begin{split}
\label{eq:gselfET5}
\frac{\varepsilon_\mu \Pi^{\mu\nu}_{\LU,5}\varepsilon_\nu}{2 p^2}
={}& 
4\pi\as\,\frac{c_5}{2}\,
\mi\mu^{2\epsilon}\int \frac{d^d q}{(2\pi)^d}\,
\frac{p^2}
{[q^2+\mi 0][k^2+\mi 0][q\cdot \tilde q+\mi 0]}
\;.
\end{split}
\end{equation}
We introduce an integral over Feynman parameters $x$ and $y$ and then integrate over the loop momentum as for $\varepsilon_\mu \Pi^{\mu\nu}_{\LU,3}\varepsilon_\nu$. This leads to
\begin{equation}
\begin{split}
\frac{\varepsilon_\mu \Pi^{\mu\nu}_{\LU,5}\varepsilon_\nu}{2 p^2} ={}& 
\frac{\as}{4\pi}\frac{c_5}{2}\, 
\Gamma(1+\epsilon) 
\int_0^1\!dx  
\left[\frac{v^2 x^2}{1-x + v^2 x}\right]^{1/2}
\int_0^1\!dy\ 
\frac{p^2}{\Lambda^2(x,y,0)}
\left(\frac{4\pi \mu^2}{\Lambda^2(x,y,0)}\right)^\epsilon
\;.
\end{split}
\end{equation}
We want the limit of this as $p^2 \to 0$. There is a factor of $p^2$ in the numerator, but this factor multiplies an integral that diverges if we set $\epsilon$ to zero, so we should be cautious about taking the $p^2 \to 0$ limit inside the integral. The function $\Lambda^2$ is, from Eq.~(\ref{eq:Lambdaxyz}),
\begin{equation}
\begin{split}
\Lambda^2(x,y,0) ={}& 
-\left[\frac{x(1-x)^2}{1-x + v^2 x}\, y^2
 + x y (1-y)\right] p^2
- \frac{v^2 x^2(1-x)}{1-x + v^2x}\, y^2\, p\cdot \tilde p
\;.
\end{split}
\end{equation}
In the limit $p^2 \to 0$, the integral over $y$ is dominated by very small $y$. To capture this behavior, we integrate over $y$ from 0 to $\infty$ instead of from 0 to 1 and approximate $\Lambda^2$ by
\begin{equation}
\begin{split}
\Lambda^2(x,y,0) \approx{}& 
-  x y\,  p^2
- \frac{v^2 x^2(1-x)}{1-x + v^2x}\, y^2\, p\cdot \tilde p
\;.
\end{split}
\end{equation}
We change integration variables from $y$ to $y'$ defined by
\begin{equation}
y'\, p^2 = y \, p\cdot \tilde p
\;.
\end{equation}
Then
\begin{equation}
\begin{split}
\Lambda^2(x,y,0) \approx{}& 
=\frac{-(p^2)^2}{p\cdot \tilde p}
\left[x y'
+\frac{v^2 x^2 (1-x)}{1-x + v^2x}\, y^{\prime\, 2}\right]
\;.
\end{split}
\end{equation}
This gives us
\begin{equation}
\begin{split}
\frac{\varepsilon_\mu \Pi^{\mu\nu}_{\LU,5}\varepsilon_\nu}{2p^2} \approx{}& 
-\frac{\as}{4\pi}\frac{c_5}{2} 
\left[\frac{-(p^2)^2}{4\pi \mu^2\,p\cdot\tilde p}\right]^{-\epsilon}\!
\Gamma(1+\epsilon)
\int_0^1\!dx 
\left[\frac{v^2 x^2}{1-x + v^2 x}\right]^{1/2}\!
\int_0^\infty\!dy'  
\left[x y'
+\frac{v^2 x^2 (1-x)}{(1-x) + v^2x}\,  y^{\prime\, 2}\right]^{-1-\epsilon}
\!.
\end{split}
\end{equation}
We take the limit $p^2 \to 0$ with $\epsilon < 0$. The factor $[-p^2]^{-2\epsilon}$ vanishes while the integrals over $x$ and $y$ are finite. Thus
\begin{equation}
\left[\frac{\varepsilon_\mu \Pi^{\mu\nu}_{\LU,5}\varepsilon_\nu}
{2p^2}\right]_{p^2 \to 0} = 0
\;.
\end{equation}

The contributions derived above give us the unrenormalized $\varepsilon_\mu \Pi^{\mu\nu}_\LU\varepsilon_\nu/(2p^2)$. We have to subtract the ultraviolet counterterms, which we can obtain from Eq.~(\ref{eq:gluonSEUV}):
\begin{equation}
\begin{split}
\label{eq:transverseCT}
\frac{\varepsilon_\mu \Pi^{\mu\nu}_\mathrm{CT}\varepsilon_\nu}
{2p^2}
={}& 
\frac{\as}{4\pi}\,
\frac{S_\epsilon}{\epsilon}\,\frac{c_\LA(v,1) + \tilde c_\LA(v,1)}{2}
=\frac{\as}{4\pi}\,
\frac{S_\epsilon}{\epsilon}
\left\{ C_\LA 
\frac{3 v^3 + 12 v^2 + 7 v - 2}{6 v (v+1)^2}
- 
\frac{2}{3}\,T_\LR n_\Lf
\right\}
\;.
\end{split}
\end{equation}
The sum of these contributions is the renormalized $\varepsilon_\mu \Pi^{\mu\nu}\varepsilon_\nu/(2p^2)$:
\begin{equation}
\begin{split}
\label{eq:PiTRresult}
\left[\frac{\varepsilon_\mu \Pi^{\mu\nu}\varepsilon_\nu}
{2p^2}\right]_{p^2 \to 0}
 ={}&\frac{\as}{4\pi}\Bigg\{
\frac{c_2}{4}\,
\left(\frac{\mu^2}{-p\cdot\tilde p}\right)^\epsilon
\left\{
\frac{S_\epsilon}{\epsilon}\,I_{2,1}(v)
+ 2I_{2,1}(v) +  I_{2,2}(v)
\right\}
\\&
-\frac{c_3}{4}\left(\frac{\mu^2}{-p\cdot \tilde p}\right)^\epsilon
\left\{
\frac{S_\epsilon}{\epsilon^2}
+ \frac{S_\epsilon}{\epsilon} I_{3,1}(v)  
+ I_{3,2}(v) 
\right\} 
+\frac{c_{4,1}}{4}\,I_4(v)
\\&
- \frac{S_\epsilon}{\epsilon}\,\frac{c_\LA(v,1) + \tilde c_\LA(v,1)}{2}
\Bigg\}
+ \cO(\epsilon)
\;.
\end{split}
\end{equation}

It is of interest to collect the single and double pole terms. We write $-p\cdot\tilde p$ at $p^2 = 0$ using Eq.~(\ref{eq:qdottildeq}) as $(v^2 - 1)/v^2](p\cdot n)^2$. This gives
\begin{equation}
\begin{split}
\left[\frac{\varepsilon_\mu \Pi^{\mu\nu}\varepsilon_\nu}
{2p^2}\right]_{p^2 \to 0}
 ={}&-\frac{\as}{4\pi}\, \frac{S_\epsilon}{\epsilon^2}\,\frac{c_3}{4}
 \\&
 +\frac{\as}{4\pi}\, \frac{S_\epsilon}{\epsilon}
\Bigg[
\frac{c_2}{4}\,I_{2,1}(v)
-\frac{c_3}{4}\,I_{3,1}(v) 
-\frac{c_3}{4} \log\!\left(\frac{v^2 \mu^2}
{(v^2 - 1)(p\cdot n)^2}\right)
-\frac{c_\LA(v,1) + \tilde c_\LA(v,1)}{2}
\Bigg]
\\&
+ \cO(\epsilon^0)
\;.
\end{split}
\end{equation}
Using Eqs.~(\ref{eq:cAtildecA}), (\ref{eq:cijcoefficients}), (\ref{eq:I21I22}),  and (\ref{eq:I31I32}) for the values for the coefficients here, we obtain
\begin{equation}
\begin{split}
\label{eq:PiTRpolesresult}
\left[\frac{\varepsilon_\mu \Pi^{\mu\nu}\varepsilon_\nu}
{2p^2}\right]_{p^2 \to 0}
 ={}&-\frac{\as}{4\pi}\, \frac{S_\epsilon}{\epsilon^2}\,C_\LA
 -\frac{\as}{4\pi}\, \frac{S_\epsilon}{\epsilon}\,\gamma_\Lg
 \\&
 +\frac{\as}{4\pi}\, \frac{S_\epsilon}{\epsilon}
\Bigg[
- \frac{v-1}{v}\,C_\LA
+ C_\LA \log\!\left(\frac{v-1}{v+1}\right)
- C_\LA \log\!\left(\frac{\mu^2}{4(p\cdot n)^2}\right)
\Bigg]
\\&
+ \cO(\epsilon^0)
\;.
\end{split}
\end{equation}
Here $\gamma_\Lg$ is the standard coefficient defined in Eq.~(\ref{eq:gammag}). The double and single pole terms in the first line of Eq.~(\ref{eq:PiTRpolesresult}) match the standard result for the pole terms in the S matrix in Feynman gauge. The terms in the second line will cancel against the gluon contributions to the $\bm T_l\cdot \bm T_l$ terms from gluon exchange graphs.

\end{widetext} 

\section{Renormalization calculations}
\label{sec:UVpoles}

In Sec.~\ref{sec:renorm}, we saw how the UV pole terms in the gluon self-energy diagrams in interpolating gauge lead to the one-loop contributions to the renormalization factors $Z_A$, $Z_v$ and $Z_\xi$. In this appendix, we outline the structure of some of the other UV divergent one loop diagrams and see how the UV pole terms lead to the remaining renormalization factors $Z$ reported in Sec.~\ref{sec:renorm}.

We expand each renormalization factor in powers of $\as$ as $Z = 1 + \delta Z + \cO(\as^2)$ and $(Z_A)^\mu_\nu = g^\mu_\nu + (\delta Z_A)^\mu_\nu + \cO(\as^2)$, where $\delta Z$ contains one power of $\as$. We compute one loop quantities like the renormalized quark self-energy function $\Sigma(p)$ that are proportional to one power of $\as$ and depend on gauge parameters $v$ and $\xi$. The unrenormalized version is denoted by $\Sigma_\LU(p)$. This is the same as this quantity in the bare theory, $\Sigma_\LB(p)$, except that we use the renormalized $g$, $v$ and $\xi$. Since $g = g_\LB (1 + \cO(\as))$, $v = v_\LB(1  + \cO(\as))$, and $\xi = \xi_\LB (1+ + \cO(\as))$, changing from bare to renormalized $g$, $v$ and $\xi$ does not affect the result at order $\as$.

\subsection{Quark self-energy}
\label{sec:quarkpole}

A straightforward calculation of the one loop diagram in Fig.~\ref{fig:quarkSE} for the unrenormalized quark self-energy $- \mi \Sigma_\LU$ gives the UV pole
\begin{equation}
\begin{split}
\label{eq:quarkSEpole}
    \Sigma_\LU(p) = {}& -\frac{\as}{4\pi}
    \frac{S_\epsilon}{\epsilon}\,\s{p}\, C_\LF
    \left[\frac{(v-1)^2}{v(v+1)} + \frac{\xi}{v}\right]
    +\cO(\epsilon^0)
    \;.
\end{split}
\end{equation}
The inverse of the quark propagator is $\s{p} - \Sigma(p)$ at order $\as$, where $\Sigma(p)$ is the renormalized quark self-energy. This gives us the relation
\begin{equation}
\left[\s{p} - \Sigma_\LU(p)\right] = Z_\psi^{-1}\left[\s{p} - \Sigma(p)\right]
+ \cO(\as^2)
\;.
\end{equation}
Then
\begin{equation}
\s{p} - \Sigma_\LU(p) = \s{p} - \Sigma(p) - \delta Z_\psi\, \s{p}
\;.
\end{equation}
This gives us
\begin{equation}
\delta Z_\psi\, \s{p} = \Sigma_\LU(p) - \Sigma(p)
\;.
\end{equation}
We use our result Eq.~(\ref{eq:quarkSEpole}) for the pole term in $\Sigma_\LU(p)$ and note that by definition the renormalized $\Sigma(p)$ has no pole. This gives
\begin{equation}
\delta Z_\psi = -\frac{\as}{4\pi}
    \frac{S_\epsilon}{\epsilon}\, C_\LF
    \left[\frac{(v-1)^2}{v(v+1)} + \frac{\xi}{v}\right]
\;.
\end{equation}
This is the result reported in Eq.~(\ref{eq:Zqresult}).

\subsection{Ghost self-energy}
\label{sec:ghostpole}

The determination of $\delta Z_\eta$ for the ghost field is slightly more subtle. A straightforward calculation of the one loop diagram in Fig.~\ref{fig:ghostSE} for the unrenormalized ghost self-energy $- \mi \Pi^\mathrm{ghost}_\LU$ gives the UV pole
\begin{equation}
\begin{split}
\label{eq:Pighostpole}
\Pi^\mathrm{ghost}_\LU(p)\hskip - 0.6 cm &
\\  ={}& 
\frac{\as}{4\pi}\,\frac{S_\epsilon}{\epsilon}\, C_\LA
\Big\{
\left[\frac{16 v^2 + v + 1}{12 v (v+1)} - \frac{\xi}{4v}\right]\,p \cdot \tilde p
\\&\quad
-\frac{2 (2v+1)(v-1)}{3 v^3 (v+1)}\,(p \cdot n)^2
\Big\}
+ \cO(\epsilon^0)
\;.
\end{split}
\end{equation}
The inverse of the ghost propagator is $p\cdot\tilde p - \Pi^\mathrm{ghost}(p)$ at order $\as$. This gives us
\begin{equation}
\begin{split}
p\cdot \tilde p_\LB - \Pi^\mathrm{ghost}_\LU(p) ={}& 
Z_\eta^{-1} (p\cdot \tilde p - \Pi^\mathrm{ghost}(p))
+ \cO(\as^2)
\;.
\end{split}
\end{equation}
Here $\tilde p_\LB = (h_\LB)^\mu_\nu p^\nu$ is defined using a factor $1/v_\LB^2$ in $(h_\LB)^\mu_\nu$, with $h^\mu_\nu$ as in Eq.~(\ref{eq:hdef}). Since $1/v_\LB^2 = Z_v^{-1}/v^2$, we have
\begin{equation}
\begin{split}
p\cdot \tilde p_\LB ={}& \frac{Z_v^{-1} - 1}{v^2}
(p\cdot n)^2
+ p\cdot \tilde p
\;.
\end{split}
\end{equation}
This gives us
\begin{equation}
\begin{split}
p\cdot \tilde p +  \frac{Z_v^{-1}-1}{v^2}\, (p\cdot n)^2 
- \Pi^\mathrm{ghost}_\LU(p) \hskip - 3 cm &
\\
={}& 
Z_\eta^{-1} (p\cdot \tilde p - \Pi^\mathrm{ghost}(p))
+ \cO(\epsilon^0)
\;.
\end{split}
\end{equation}
Expanding the renormalization factors to order $\as$ gives
\begin{equation}
\begin{split}
p\cdot \tilde p - \frac{\delta Z_v}{v^2}\,& (p\cdot n)^2 - \Pi^\mathrm{ghost}_\LU(p) 
\\
={}& 
p\cdot \tilde p
- \delta Z_\eta\, p\cdot \tilde p - \Pi^\mathrm{ghost}(p)
\;.
\end{split}
\end{equation}
Thus
\begin{equation}
\begin{split}
\delta Z_\eta\, p\cdot \tilde p ={}& 
\Pi^\mathrm{ghost}_\LU(p) 
-\Pi^\mathrm{ghost}(p) 
+ \frac{\delta Z_v}{v^2}\, (p\cdot n)^2
\;.
\end{split}
\end{equation}
We can use Eq.~(\ref{eq:Pighostpole}) for the pole term in $\Pi^\mathrm{ghost}_\LU(p)$. We already know $\delta Z_v$ from the gluon self-energy. It is given in Eq.~(\ref{eq:ZvZxi}). Using these results, we have
\begin{equation}
\begin{split}
\delta Z_\eta\,  p\cdot \tilde p ={}&
\frac{\as}{4\pi}\,\frac{S_\epsilon}{\epsilon}
C_\LA
\bigg\{
\left[\frac{16 v^2 + v + 1}{12 v (v+1)} - \frac{\xi}{4v}\right] p \cdot \tilde p
\\&\quad
-\frac{2 (2v+1)(v-1)}{3 v^3 (v+1)}\,(p \cdot n)^2
\bigg\}
\\&
+\frac{\as}{4\pi}\,\frac{S_\epsilon}{\epsilon}\,
C_\LA  \frac{2 (2v+1)(v-1) }{3v^3(v+1)} (p\cdot n)^2
.
\end{split}
\end{equation}
The terms proportional to $(p\cdot n)^2$ cancel, leaving
\begin{equation}
\begin{split}
\delta Z_\eta 
={}&
\frac{\as}{4\pi}\,\frac{S_\epsilon}{\epsilon}\,
C_\LA
\left[\frac{16 v^2 + v + 1}{12 v (v+1)} - \frac{\xi}{4v}\right] 
\;.
\end{split}
\end{equation}
This is the result reported in Eq.~(\ref{eq:Zghostresult}).

\subsection{Quark-gluon vertex}
\label{sec:quarkgluonvertex}

We use the graphs in Fig.~\ref{fig:quarkgluonvertex} to calculate the unrenormalized one-loop quark gluon vertex function, $-\mi g \Gamma^\mu_\LU(p)$,
\begin{equation}
\begin{split}
\label{eq:Gammamupole}
     \Gamma^\mu_{\LU} ={}&
     \frac{\as}{4\pi}\frac{S_\epsilon}{\epsilon}\,\gamma_\mu
     \bigg[C_\LF \left(\frac{(v-1)^2}{v(v+1)} +  \frac{\xi}{v}\right)
     \\&\quad
       + C_\LA\frac{3v^2+ 2 v + 1}{3 v (v+1)^2} + C_\LA
       \frac{\xi-1}{4v}\bigg]
     \\
     &\quad
     + \frac{\as}{4\pi}\frac{S_\epsilon}{\epsilon}\, \tilde{\gamma}^\mu \,C_\LA
    \frac{2v (1 + 2v)}{3 (v+1)^2}
     + \cO(\epsilon^0)
\;.
\end{split}
\end{equation}
The (renormalized) quark-gluon amputated three point function divided by $g$ is $\gamma^\mu + \Gamma^\mu(p) + \cO(\as^2)$. To apply renormalization, we use,
\begin{equation}
\gamma^\mu +   \Gamma_\LU^\mu 
= [Z_A^{-1/2}]^\mu_\nu Z^{-1}_\psi Z^{-1}_g (\gamma^\nu + \Gamma^\nu)
+ \cO(\as^2)
\;.
\end{equation}
Writing $Z = 1 + \delta Z$ in each case, this is
\begin{equation}
\begin{split}
\gamma^\mu + \Gamma_\LU^\mu 
={}&\gamma^\mu + \Gamma^\mu
\\&
- \frac{1}{2}\,[\delta Z_A]^\mu_\nu \gamma^\nu
- \delta Z_\psi \gamma^\mu
- \delta Z_g \gamma^\mu
\;.
\end{split}
\end{equation}
That is
\begin{equation}
\begin{split}
\Gamma_\LU^\mu - \Gamma^\mu ={}& 
- \frac{1}{2}\,[\delta Z_A]^\mu_\nu \gamma^\nu
- \delta Z_\psi \gamma^\mu
- \delta Z_g \gamma^\mu
\;.
\end{split}
\end{equation}
Given our result (\ref{eq:Gammamupole}) for $\Gamma^\mu_\LU$ and using our previous results (\ref{eq:ZA}) for $Z_A$ and (\ref{eq:Zqresult}) $Z_\psi$, we have
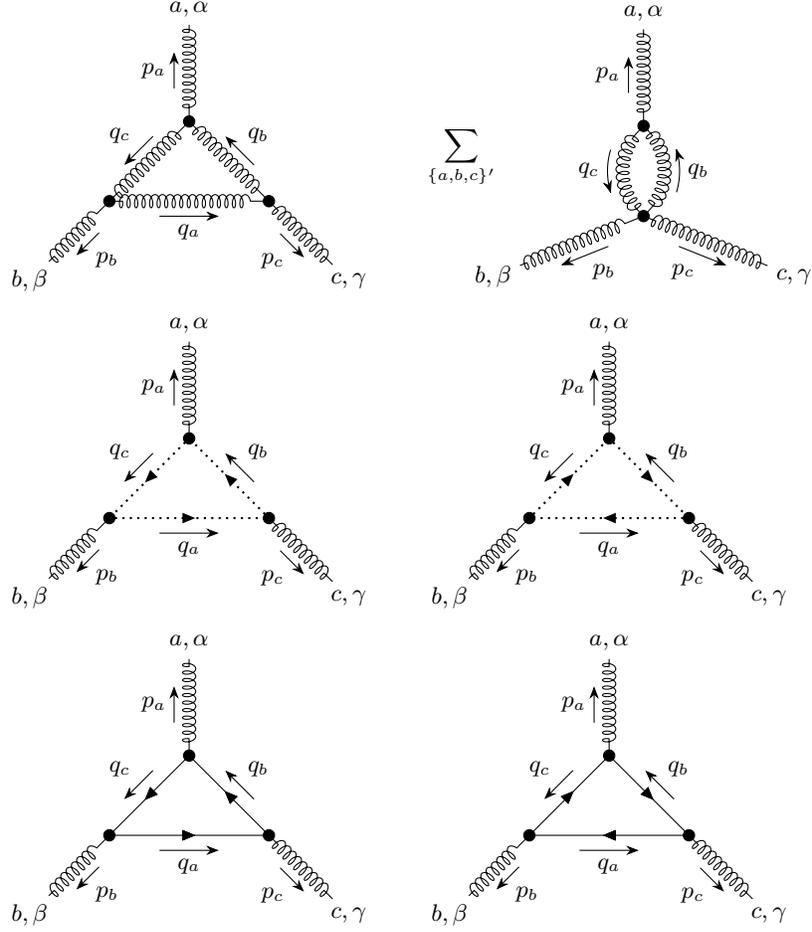
\begin{figure*}[t]
\begin{equation*}
  \begin{split}
    &
    \begin{tikzpicture}[baseline={([yshift=-.5ex]current bounding box.center)}]
      \begin{feynman}
        \vertex [dot] (a) {};
        \vertex [dot, below left=of a](b) {};
        \vertex [dot, below right=of a](c) {};
        \vertex [above=of a] (a1) {$a,\alpha$};
        \vertex [below left=of b](b1) {$b,\beta$};
        \vertex [below right=of c](c1) {$c,\gamma$};
        \diagram*{
          (a1)--[gluon, momentum={[arrow shorten=0.7]$p_a$}](a),
          (b)-- [gluon, momentum'={[arrow shorten=0.7]$q_c$}](a),
          (b)-- [gluon, rmomentum={[arrow shorten=0.7]$q_a$}](c),
          (a)-- [gluon, momentum'={[arrow shorten=0.7]$q_b$}](c),
          (b1)--[gluon, momentum={[arrow shorten=0.7]$p_b$}](b),
          (c)--[gluon, rmomentum={[arrow shorten=0.7]$p_c$}](c1);
        };	
      \end{feynman}
    \end{tikzpicture}
    \quad\quad\sum_{\{a,b,c\}'}\!\!\!\!\!\!\!\!
    \begin{tikzpicture}[baseline={([yshift=-.5ex]current bounding box.center)}]
      \begin{feynman}
        \vertex [dot] (a) {};
        \vertex [dot, below=1.2cm of a](b) {};
        \vertex [above=of a] (a1) {$a,\alpha$};
        \vertex [] at ($(b)+(-2cm,-0.8cm)$) (b1) {$b,\beta$};
        \vertex [] at ($(b)+(2cm,-0.8cm)$) (c1) {$c,\gamma$};
        \diagram*{
          (a1)--[gluon, momentum={[arrow shorten=0.7]$p_a$}](a),
          (b)-- [gluon, quarter left, momentum'={[arrow shorten=0.7]$q_c$}](a),
          (a)-- [gluon, quarter left, momentum'={[arrow shorten=0.7]$q_b$}](b),
          (b1)--[gluon, momentum={[arrow shorten=0.7]$p_b$}](b),
          (b)--[gluon, rmomentum={[arrow shorten=0.7]$p_c$}](c1);
        };	
      \end{feynman}
    \end{tikzpicture}
    \\& 
    \begin{tikzpicture}[baseline={([yshift=-.5ex]current bounding box.center)}]
      \begin{feynman}
        \vertex [dot] (a) {};
        \vertex [dot, below left=of a](b) {};
        \vertex [dot, below right=of a](c) {};
        \vertex [above=of a] (a1) {$a,\alpha$};
        \vertex [below left=of b](b1) {$b,\beta$};
        \vertex [below right=of c](c1) {$c,\gamma$};
        \diagram*{
          (a1)--[gluon, momentum={[arrow shorten=0.7]$p_a$}](a),
          (a)-- [ghost, with arrow=0.5, rmomentum={[arrow shorten=0.7]$q_c$}](b),
          (b)-- [ghost, with arrow=0.5, rmomentum={[arrow shorten=0.7]$q_a$}](c),
          (c)-- [ghost, with arrow=0.5, rmomentum={[arrow shorten=0.7]$q_b$}](a),
          (b1)--[gluon, momentum={[arrow shorten=0.7]$p_b$}](b),
          (c)--[gluon, rmomentum={[arrow shorten=0.7]$p_c$}](c1);
        };	
      \end{feynman}
    \end{tikzpicture}
   \quad\quad  
    \begin{tikzpicture}[baseline={([yshift=-.5ex]current bounding box.center)}]
      \begin{feynman}
        \vertex [dot] (a) {};
        \vertex [dot, below left=of a](b) {};
        \vertex [dot, below right=of a](c) {};
        \vertex [above=of a] (a1) {$a,\alpha$};
        \vertex [below left=of b](b1) {$b,\beta$};
        \vertex [below right=of c](c1) {$c,\gamma$};
        \diagram*{
          (a1)--[gluon, momentum={[arrow shorten=0.7]$p_a$}](a),
          (b)-- [ghost, with arrow=0.5, momentum'={[arrow shorten=0.7]$q_c$}](a),
          (c)-- [ghost, with arrow=0.5, momentum'={[arrow shorten=0.7]$q_a$}](b),
          (a)-- [ghost, with arrow=0.5, momentum'={[arrow shorten=0.7]$q_b$}](c),
          (b1)--[gluon, momentum={[arrow shorten=0.7]$p_b$}](b),
          (c)--[gluon, rmomentum={[arrow shorten=0.7]$p_c$}](c1);
        };	
      \end{feynman}
    \end{tikzpicture}
    \\&
    \begin{tikzpicture}[baseline={([yshift=-.5ex]current bounding box.center)}]
      \begin{feynman}
        \vertex [dot] (a) {};
        \vertex [dot, below left=of a](b) {};
        \vertex [dot, below right=of a](c) {};
        \vertex [above=of a] (a1) {$a,\alpha$};
        \vertex [below left=of b](b1) {$b,\beta$};
        \vertex [below right=of c](c1) {$c,\gamma$};
        \diagram*{
          (a1)--[gluon, momentum={[arrow shorten=0.7]$p_a$}](a),
          (a)-- [fermion, with arrow=0.5, rmomentum={[arrow shorten=0.7]$q_c$}](b),
          (b)-- [fermion, with arrow=0.5, rmomentum={[arrow shorten=0.7]$q_a$}](c),
          (c)-- [fermion, with arrow=0.5, rmomentum={[arrow shorten=0.7]$q_b$}](a),
          (b1)--[gluon, momentum={[arrow shorten=0.7]$p_b$}](b),
          (c)--[gluon, rmomentum={[arrow shorten=0.7]$p_c$}](c1);
        };	
      \end{feynman}
    \end{tikzpicture}
    \quad\quad   
    \begin{tikzpicture}[baseline={([yshift=-.5ex]current bounding box.center)}]
      \begin{feynman}
        \vertex [dot] (a) {};
        \vertex [dot, below left=of a](b) {};
        \vertex [dot, below right=of a](c) {};
        \vertex [above=of a] (a1) {$a,\alpha$};
        \vertex [below left=of b](b1) {$b,\beta$};
        \vertex [below right=of c](c1) {$c,\gamma$};
        \diagram*{
          (a1)--[gluon, momentum={[arrow shorten=0.7]$p_a$}](a),
          (b)-- [fermion, with arrow=0.5, momentum'={[arrow shorten=0.7]$q_c$}](a),
          (c)-- [fermion, with arrow=0.5, momentum'={[arrow shorten=0.7]$q_a$}](b),
          (a)-- [fermion, with arrow=0.5, momentum'={[arrow shorten=0.7]$q_b$}](c),
          (b1)--[gluon, momentum={[arrow shorten=0.7]$p_b$}](b),
          (c)--[gluon, rmomentum={[arrow shorten=0.7]$p_c$}](c1);
        };	
      \end{feynman}
    \end{tikzpicture}
  \end{split}
\end{equation*}
\caption{
Diagrams for the one loop contributions to the three gluon vertex function.
\label{fig:threegluonfunction}}
\end{figure*}

\begin{widetext}
\begin{equation}
\begin{split}
\frac{\as}{4\pi}\frac{S_\epsilon}{\epsilon}\,&\gamma^\mu
     \left[C_\LF \left(\frac{(v-1)^2}{v(v+1)} +  \frac{\xi}{v}\right)
       + C_\LA\frac{3 v^2 + 2 v + 1}{3 v (v+1)^2} + C_\LA
       \frac{\xi-1}{4v}\right]
     + \frac{\as}{4\pi}\frac{S_\epsilon}{\epsilon}\, \tilde{\gamma}^\mu \,C_\LA
    \frac{2v (1 + 2v)}{3 (v+1)^2}
\\ ={}& 
+ \frac{\as}{4\pi} \frac{S_\epsilon}{\epsilon}\,\gamma^\mu
\left\{
- C_\LA \left[\frac{22 v^3+35 v^2+20 v-1}{12 v(v+1)^2}-\frac{\xi}{4 v}\right]
+ \frac{2}{3}\,T_\LR n_\Lf
\right\}
+  \frac{\as}{4\pi} \frac{S_\epsilon}{\epsilon}\,\tilde\gamma^\mu\,
C_\LA\,\frac{2 v (2 v + 1) }{3(v+1)^2}
\\&
+ \frac{\as}{4\pi} \frac{S_\epsilon}{\epsilon}\,\gamma^\mu\,
 C_\LF\left(\frac{(v-1)^2}{v(v+1)}
+ \frac{\xi}{v}\right)
- \delta Z_g \gamma^\mu 
\;.
\end{split}
\end{equation}
\end{widetext}
The coefficients of $\tilde \gamma^\mu$ and the coefficients of $C_\LF$ match. We can solve for $\delta Z_g$:
\begin{equation}
\delta Z_g = 
\frac{\as}{4\pi} \frac{S_\epsilon}{\epsilon}
\left[-\frac{11}{6}\,C_\LA + \frac{2}{3}\,T_\LR n_\Lf
\right]
\;.
\end{equation}
Note that this is independent of $v$ and $\xi$. This is the result reported in Eq.~(\ref{eq:Zgresult}).

\subsection{Three gluon vertex}
\label{sec:threegluonvertex}

We can check how the renormalization program is working by calculating the three gluon vertex function at one loop order. We call the one-particle-irreducible three-gluon Green function $\Gamma^{abc}_{\alpha \beta \gamma}(p_a,p_b,p_c)$, where the momenta are directed out of the vertex function. We define $\Gamma_{\alpha \beta \gamma}(p_a,p_b,p_c)$ by
\begin{equation}
\Gamma^{abc}_{\alpha \beta \gamma}(p_a,p_b,p_c)
= g f_{abc} \Gamma_{\alpha \beta \gamma}(p_a,p_b,p_c)
\;.
\end{equation}
Our aim is to calculate the ultraviolet pole terms in $\Gamma_{\alpha \beta \gamma}(p_a,p_b,p_c)$. In this appendix, we denote the zero loop version, $\Gamma_{\alpha \beta \gamma}(p_a,p_b,p_c)$, Eq.~(\ref{eq:Gammatree}), by $\Gamma^\mathrm{tree}_{\alpha \beta \gamma}(p_a,p_b,p_c)$.

The graphs needed for $\Gamma^{abc}_{\alpha \beta \gamma}(p_a,p_b,p_c)$ are illustrated in Fig.~\ref{fig:threegluonfunction}. We write the integral for the graph with a single gluon loop as
\begin{equation}
\begin{split}
\label{eq:threegluonloopgraph}
g f_{abc}&
\Gamma_{\alpha \beta \gamma}(p_\La, p_\Lb, p_\Lc) 
\\={}& 
g^3\,f_{a\bar c\bar b}f_{b \bar a\bar c}f_{c\bar b\bar a}\,
\mu^{2\epsilon}\int\!\frac{d^d q}{(2\pi)^d}
\\&\times
\Gamma^\mathrm{tree}_{\alpha \alpha_1\alpha_2}(p_\La, q_\Lc, -q_\Lb)
\\
&\times
\Gamma^\mathrm{tree}_{\beta \beta_1 \beta_2}(p_\Lb, q_\La, -q_\Lc)
\Gamma^\mathrm{tree}_{\gamma \gamma_1 \gamma_2}(p_\Lc, q_\Lb, -q_\La)
\\
&\times
\mi D^{\alpha_1\beta_2}(q_\Lc)\,
\mi D^{\beta_1\gamma_2}(q_\La)\,
\mi D^{\gamma_1\alpha_2}(q_\Lb)
\;.
\end{split}
\end{equation}
Here the momenta on the three sides of the loop are related to the loop momentum $q$ by \cite{NSsubtractions}
\begin{align}
q_\La ={}& q + \frac{p_\Lc - p_\Lb}{3}
\;,\notag
\\
q_\Lb ={}& q + \frac{p_\La - p_\Lc}{3}
\;,
\\
q_\Lc ={}& q + \frac{p_\Lb - p_\La}{3}
\;.\notag
\end{align}

We sum over the three graphs that have a four-gluon vertex. The graph in which gluons b and c join at the four-gluon vertex is illustrated in Fig.~\ref{fig:threegluonfunction}. This graph is
\begin{align}
g f_{abc}&
\Gamma_{\alpha \beta \gamma}(p_\La, p_\Lb, p_\Lc)
\notag
\\ ={}& 
\frac{1}{2}\,
\mu^{2\epsilon}\int\!\frac{d^d q}{(2\pi)^d}
\notag\\
&\times
g f_{a\bar c \bar b}
\Gamma^\mathrm{tree}_{\alpha \alpha_1\alpha_2}(p_\La, q_\Lc, -q_\Lb)\,
\mi D^{\alpha_1 \bar\gamma}(q_\Lc)
\mi D^{\bar \beta \alpha_2}(q_\Lb)
\notag\\&\times
(-\mi g^2)\Big\{
f_{\bar a b c}f_{\bar a \bar b \bar c}
[g_{\beta\bar \beta}g_{\gamma \bar\gamma} 
- g_{\beta\bar\gamma} g_{\gamma\bar\beta}]
\notag\\&\quad\quad
+ f_{\bar a b \bar c}f_{\bar a c \bar b }
[g_{\beta \gamma}g_{\bar\beta \bar\gamma} 
- g_{\beta \bar\beta} g_{\gamma\bar\gamma}]
\notag\\&\quad\quad
+ f_{\bar a b \bar b}f_{\bar a c \bar c}
[g_{\beta \gamma}g_{\bar\beta \bar\gamma} 
- g_{\beta\bar\gamma} g_{\gamma\bar\beta}]
\Big\}
\;.
\end{align}
The momenta on the two sides of the loop are related to the loop momentum $q$ by \cite{NSsubtractions}
\begin{equation}
\begin{split}
q_\Lb ={}& q + \frac{1}{2}\,p_\La
\;,
\\
q_\Lc ={}& q - \frac{1}{2}\,p_\La
\;.
\end{split}
\end{equation}

We express the integrals needed for the graphs in Fig.~\ref{fig:threegluonfunction} using Feynman parameter representations, along the lines of Appendix \ref{sec:gluonSE}. This allows us to extract the ultraviolet pole terms in the form
\begin{widetext}
\begin{equation}
\begin{split}
\label{eq:3gpolestructure}
\Gamma_{\alpha \beta \gamma}(p_\La, p_\Lb, p_\Lc) 
={}& \frac{\as}{4\pi}
\frac{S_\epsilon}{\epsilon}
\Big\{
A \big[ 
g_{\alpha\beta}(p_{\La} - p_{\Lb})_\gamma
+ g_{\beta\gamma}(p_{\Lb} - p_{\Lc})_\alpha
+ g_{\gamma\alpha} (p_{\Lc} - p_{\La})_\beta
\big]
\\& \quad\quad
+ B \big[ 
h_{\alpha\beta}(p_{\La} - p_{\Lb})_\gamma
+ h_{\beta\gamma}(p_{\Lb} - p_{\Lc})_\alpha
+ h_{\gamma\alpha} (p_{\Lc} - p_{\La})_\beta
\big]
\\& \quad\quad
+ C \big[ 
g_{\alpha\beta}(\tilde p_{\La} - \tilde p_{\Lb})_\gamma
+ g_{\beta\gamma}(\tilde p_{\Lb} - \tilde p_{\Lc})_\alpha
+ g_{\gamma\alpha} (\tilde p_{\Lc} - \tilde p_{\La})_\beta
\big]
\\& \quad\quad
+ D \big[ 
h_{\alpha\beta}(\tilde p_{\La} - \tilde p_{\Lb})_\gamma
+ h_{\beta\gamma}(\tilde p_{\Lb} - \tilde p_{\Lc})_\alpha
+ h_{\gamma\alpha} (\tilde p_{\Lc} - \tilde p_{\La})_\beta
\big]
\Big\}
\;.
\end{split}
\end{equation}
\end{widetext}
For the gluon loop graph, we find
\begin{align}
A ={}& -C_\LA\,\frac{5 v^4 + 15 v^3 + 9 v^2 + 23 v + 12}
{6 v (v+1)^3}
- C_\LA \frac{9 (\xi-1)}{8 v}
\;,\notag
\\
B ={}& -C_\LA\,\frac{v( 2 v^2 + 3 v + 7)}{12 (v+1)^3}
\;,\notag
\\
C ={}& -C_\LA\,\frac{v( 10 v^2 - 3 v + 5)}{12 (v+1)^3}
\;,\notag
\\
D ={}& -C_\LA\,\frac{v}{24}
\;.
\end{align}
For the ghost loop graph, we find
\begin{align}
A ={}& 0
\;,\notag
\\
B ={}& 0
\;,\notag
\\
C ={}& 0
\;,\notag
\\
D ={}& C_\LA\,\frac{v}{24}
\;.
\end{align}
For the graphs with a four-gluon vertex, we find
\begin{align}
A ={}& C_\LA\,\frac{(3 v+1)(3 v^3+6 v^2 + 3 v + 2)}
{2 v (v+1)^3}
+ C_\LA \frac{3 (\xi - 1)}{8 v}
\;,\notag
\\
B ={}& -C_\LA\,\frac{v(10 v^2 + 15 v + 3)}{4 (v+1)^3}
\;,\notag
\\
C ={}& -C_\LA\,\frac{v(2 v^2 + 9 v + 1)}{4 (v+1)^3}
\;,\notag
\\
D ={}& 0
\;.
\end{align}
For the graphs with a quark loop, we find
\begin{align}
A ={}& -\frac{4}{3}\, T_\LR n_\Lf 
\;,\notag
\\
B ={}& 0
\;,\notag
\\
C ={}& 0
\;,\notag
\\
D ={}& 0
\;.
\end{align}

For the sum of all graphs, we find
\begin{align}
\label{eq:poletermssummed}
A ={}& C_\LA \left(\frac{11}{3}
- \frac{3 v^2 + 2v + 1}{ v (v+1)^2}
- \frac{3 (\xi - 1)}{4 v}\right)
- \frac{4}{3}\,T_\LR n_\Lf
\;,\notag
\\
B ={}& -C_\LA\,\frac{4 v (2 v + 1)}{3 (v+1)^2}
\;,\notag
\\
C ={}& -C_\LA\,\frac{2 v (2 v+1)}
{3(v+1)^2}
\;,\notag
\\
D ={}& 0
\;.
\end{align}

The renormalization factors $Z$ have already been defined. We now check whether these renormalization factors $Z$ provide the counterterms needed to remove these poles. Renormalizing the gluon field gives
\begin{equation}
\begin{split}
\Gamma^{abc}_{\LB,\alpha\beta\gamma}(g_\LB) ={}& 
(Z_\LA^{-1/2})^{\bar\alpha}_\alpha
(Z_\LA^{-1/2})^{\bar\beta}_\beta
(Z_\LA^{-1/2})^{\bar\gamma}_\gamma
\Gamma^{abc}_{\bar\alpha\bar\beta\bar\gamma}(g)
\;,
\end{split}
\end{equation}
where the subscript B denotes the vertex function in the bare theory and $g_\LB$ is the bare coupling. This gives us
\begin{equation}
\begin{split}
\Gamma^{abc}_{\alpha\beta\gamma}(g) ={}& 
(Z_\LA^{1/2})^{\bar\alpha}_\alpha
(Z_\LA^{1/2})^{\bar\beta}_\beta
(Z_\LA^{1/2})^{\bar\gamma}_\gamma
\\&\times
\left[
g_\LB f_{abc}\Gamma^\mathrm{tree}_{\bar\alpha\bar\beta\bar\gamma}
+\Gamma^{\mathrm{loop},abc}_{\LB,\bar\alpha\bar\beta\bar\gamma}(g)
\right]
\;.
\end{split}
\end{equation}
In the first term, we can use $g_\LB = Z_g g$. In the second term, we can replace $g_\LB$ by $g$ since we work only to order $g^3$. This is then the full one loop three point function but without its renormalization counterterms, which we denote by a subscript U for {\em unrenormalized}. Thus
\begin{equation}
\begin{split}
\Gamma^{abc}_{\alpha\beta\gamma}(g) ={}& 
(Z_\LA^{1/2})^{\bar\alpha}_\alpha
(Z_\LA^{1/2})^{\bar\beta}_\beta
(Z_\LA^{1/2})^{\bar\gamma}_\gamma
Z_\Lg
g f_{abc}\Gamma^\mathrm{tree}_{\bar\alpha\bar\beta\bar\gamma}
\\&
+\Gamma^{\mathrm{loop},abc}_{\LU,\bar\alpha\bar\beta\bar\gamma}(g)
+ \cO(g\as^2)
\;.
\end{split}
\end{equation}
To order $g^3$, this is
\begin{widetext}
\begin{equation}
\begin{split}
\Gamma^{abc}_{\alpha\beta\gamma}(g) ={}& 
\Gamma^{\mathrm{tree},abc}_{\alpha\beta\gamma}(g)
+
\left(\frac{1}{2}\,g^{\bar\alpha}_\alpha g^{\bar\beta}_\beta 
(\delta Z_\LA)^{\bar\gamma}_\gamma
+\frac{1}{2}\,g^{\bar\alpha}_\alpha (\delta Z_\LA)^{\bar\beta}_\beta
g^{\bar\gamma}_\gamma 
+\frac{1}{2}\,(\delta Z_\LA)^{\bar \alpha}_\alpha g^{\bar\beta}_\beta
g^{\bar\gamma}_\gamma 
+ g^{\bar\alpha}_\alpha g^{\bar\beta}_\beta g^{\bar\gamma}_\gamma \delta Z_g
\right)
g f_{abc}\Gamma^\mathrm{tree}_{\bar\alpha\bar\beta\bar\gamma}
\\&
+\Gamma^{\mathrm{loop},abc}_{\LU,\bar\alpha\bar\beta\bar\gamma}(g)
+ \cO(g\as^2)
\;.
\end{split}
\end{equation}
For the renormalization factors, we use Eqs.~(\ref{eq:ZA}) and (\ref{eq:Zgresult})
\begin{equation}
\begin{split}
(\delta Z_\LA)^\mu_\nu ={}& \frac{\as}{4\pi}\,\frac{S_\epsilon}{\epsilon}
\left[c_\LA(v,\xi) g^\mu_\nu + \tilde c_\LA(v,\xi) h^\mu_\nu \right]
\;,
\\
\delta Z_g ={}& -\frac{\as}{4\pi}\,\frac{S_\epsilon}{\epsilon} \gamma_\Lg
\;.
\end{split}
\end{equation}
This gives us
\begin{equation}
\begin{split}
\Gamma^{abc}_{\alpha\beta\gamma}(g) 
-\Gamma^{\mathrm{tree},abc}_{\alpha\beta\gamma}(g)
={}& 
\frac{\as}{4\pi}\,\frac{S_\epsilon}{\epsilon}
\left[\frac{3}{2}\,c_\LA(v,\xi) - \gamma_\Lg\right]
g f_{abc}\Gamma^\mathrm{tree}_{\alpha\beta\gamma}
\\&
+ \frac{\as}{4\pi}\,\frac{S_\epsilon}{\epsilon} \frac{1}{2}\,\tilde c_\LA(v,\xi)
\left(g^{\bar\alpha}_\alpha g^{\bar\beta}_\beta h^{\bar\gamma}_\gamma
+ g^{\bar\alpha}_\alpha h^{\bar\beta}_\beta g^{\bar\gamma}_\gamma 
+ h^{\bar \alpha}_\alpha g^{\bar\beta}_\beta g^{\bar\gamma}_\gamma 
\right)
\Gamma^\mathrm{tree}_{\bar\alpha\bar\beta\bar\gamma}\,
g f_{abc}
\\&
+\Gamma^{\mathrm{loop},abc}_{\LU,\bar\alpha\bar\beta\bar\gamma}(g)
+ \cO(g\as^2)
\;.
\end{split}
\end{equation}

We use Eqs.~(\ref{eq:cAtildecA}) and  (\ref{eq:gammag}),
\begin{equation}
\begin{split}
\frac{3}{2} c_\LA(v,\xi) - \gamma_\Lg
={}& \left(
\frac{11}{3}
-\frac{3 v^2 + 2 v +1}{v (v+1)^2}
- \frac{3 (\xi - 1)}{4 v}\right) C_\LA
- \frac{4}{3}\, T_\LR n_\Lf
\;,
\\
\tilde c_\LA(v,\xi) ={}& - \frac{4 v (2v+1)}{3(v+1)^2}\,C_\LA
\;,
\end{split}
\end{equation}
and the structure of  $\Gamma^\mathrm{tree}$ from Eq.~(\ref{eq:Gammatree}). This gives
\begin{equation}
\begin{split}
\Gamma^{abc}_{\alpha\beta\gamma}(g) 
-\Gamma^{\mathrm{tree},abc}_{\alpha\beta\gamma}(g) 
={}& 
-\frac{\as}{4\pi}\,\frac{S_\epsilon}{\epsilon}
\left[\left(\frac{11}{3}
-\frac{3 v^2 + 2 v +1}{v (v+1)^2} 
- \frac{3 (\xi - 1)}{4 v}\right) C_\LA
- \frac{4}{3}\, T_\LR n_\Lf\right] 
g f_{abc}
\\&\times
\left\{g_{\alpha\beta}(p_\La - p_\Lb)_\gamma
+ g_{\beta\gamma}(p_\Lb - p_\Lc)_\alpha
+ g_{\gamma\alpha} (p_\Lc - p_\La)_\beta
\right\}
\\&
+ \frac{\as}{4\pi}\,\frac{S_\epsilon}{\epsilon}\,
\frac{2 v (2v+1)}{3(v+1)^2}\,C_\LA
g f_{abc}
\Big\{ 
2 h_{\alpha\beta}(p_\La - p_\Lb)_\gamma
+ 2 h_{\beta\gamma}(p_\Lb - p_\Lc)_\alpha
+ 2 h_{\gamma\alpha} (p_\Lc - p_\La)_\beta
\\&\quad\quad
+ g_{\alpha\beta}(\tilde p_\La - \tilde p_\Lb)_\gamma
+ g_{\beta\gamma}(\tilde p_\Lb - \tilde p_\Lc)_\alpha
+ g_{\gamma\alpha} (\tilde p_\Lc - \tilde p_\La)_\beta
\Big\}
\\&
+ \Gamma^{\mathrm{loop},abc}_{\LU,\bar\alpha\bar\beta\bar\gamma}(g)
+ \cO(g\as^2)
\;.
\end{split}
\end{equation}
This is
\begin{equation}
\Gamma^{abc}_{\alpha\beta\gamma}(g) = 
\Gamma^{\mathrm{tree},abc}_{\alpha\beta\gamma}(g)
+\Gamma^{\mathrm{loop},abc}_{\LU,\bar\alpha\bar\beta\bar\gamma}(g)
- \Gamma^{\mathrm{loop},abc}_{\mathrm{CT},\bar\alpha\bar\beta\bar\gamma}(g)
+ \cO(g\as^2)
\;,
\end{equation}
where
\begin{equation}
\begin{split}
\Gamma^{\mathrm{loop},abc}_{\mathrm{CT},\alpha\beta\gamma}(g) ={}& 
\frac{\as}{4\pi}\,\frac{S_\epsilon}{\epsilon}
\left[\left(\frac{11}{3}
-\frac{3 v^2 + 2 v +1}{v (v+1)^2} 
- \frac{3 (\xi - 1)}{4 v}\right) C_\LA
- \frac{4}{3}\, T_\LR n_\Lf\right]
g f_{abc}
\\&\times
\left[ g_{\alpha\beta}(p_\La - p_\Lb)_\gamma
+ g_{\beta\gamma}(p_\Lb - p_\Lc)_\alpha
+ g_{\gamma\alpha} (p_\Lc - p_\La)_\beta
\right]
\\&
- \frac{\as}{4\pi}\,\frac{S_\epsilon}{\epsilon} 
\frac{2 v (2v+1)}{3(v+1)^2}\,C_\LA
g f_{abc}
\Big\{ 
2 h_{\alpha\beta}(p_\La - p_\Lb)_\gamma
+ 2 h_{\beta\gamma}(p_\Lb - p_\Lc)_\alpha
+ 2 h_{\gamma\alpha} (p_\Lc - p_\La)_\beta
\\&\quad\quad
+ g_{\alpha\beta}(\tilde p_\La - \tilde p_\Lb)_\gamma
+ g_{\beta\gamma}(\tilde p_\Lb - \tilde p_\Lc)_\alpha
+ g_{\gamma\alpha} (\tilde p_\Lc - \tilde p_\La)_\beta
\Big\}
\;.
\end{split}
\end{equation}
We can write this as
\begin{equation}
\begin{split}
\Gamma_{\mathrm{CT},\alpha \beta \gamma}(p_\La, p_\Lb, p_\Lc) ={}& \frac{\as}{4\pi}
\frac{S_\epsilon}{\epsilon}
\Big\{
A \big[ 
g_{\alpha\beta}(p_{\La} - p_{\Lb})_\gamma
+ g_{\beta\gamma}(p_{\Lb} - p_{\Lc})_\alpha
+ g_{\gamma\alpha} (p_{\Lc} - p_{\La})_\beta
\big]
\\& \quad\quad
+ B \big[ 
h_{\alpha\beta}(p_{\La} - p_{\Lb})_\gamma
+ h_{\beta\gamma}(p_{\Lb} - p_{\Lc})_\alpha
+ h_{\gamma\alpha} (p_{\Lc} - p_{\La})_\beta
\big]
\\& \quad\quad
+ C \big[ 
g_{\alpha\beta}(\tilde p_{\La} - \tilde p_{\Lb})_\gamma
+ g_{\beta\gamma}(\tilde p_{\Lb} - \tilde p_{\Lc})_\alpha
+ g_{\gamma\alpha} (\tilde p_{\Lc} - \tilde p_{\La})_\beta
\big]
\\& \quad\quad
+ D \big[ 
h_{\alpha\beta}(\tilde p_{\La} - \tilde p_{\Lb})_\gamma
+ h_{\beta\gamma}(\tilde p_{\Lb} - \tilde p_{\Lc})_\alpha
+ h_{\gamma\alpha} (\tilde p_{\Lc} - \tilde p_{\La})_\beta
\big]
\Big\}
\;.
\end{split}
\end{equation}
Then
\begin{equation}
\begin{split}
\label{eq:counterterms}
A ={}& \left(\frac{11}{3}
-\frac{3 v^2 + 2 v +1}{v (v+1)^2}
- \frac{3 (\xi - 1)}{4 v}\right) C_\LA
- \frac{4}{3}\, T_\LR n_\Lf
\;,
\\
B ={}&  -\frac{4 v (2v+1)}{3(v+1)^2}\,C_\LA
\;,
\\
C ={}&  -\frac{2 v (2v+1)}{3(v+1)^2}\,C_\LA
\;,
\\
D ={}& 0
\;.
\end{split}
\end{equation}
We see that the counterterms exactly cancel the pole terms in Eq.~(\ref{eq:poletermssummed}).

\end{widetext}




\end{document}